%% file: paper.tex
\RequirePackage[l2tabu, orthodox]{nag}
\newcommand*{\ATLASLATEXPATH}{}
\documentclass[cernpreprint,texlive=2011,txfonts,USenglish,texmf]{atlasdoc} 
\pdfoutput=1
\usepackage[full]{\ATLASLATEXPATH atlaspackage}
\usepackage{\ATLASLATEXPATH atlasbiblatex}
\usepackage{\ATLASLATEXPATH atlascontribute}
\usepackage{\ATLASLATEXPATH atlasphysics}
\hypersetup{pdftitle={ATLAS document},pdfauthor={The ATLAS Collaboration}}
\graphicspath{{}}
\include{defs}

\newcommand*{\numRF}[2]{\num[round-mode=figures,round-precision=#2]{#1}}
\newcommand*{\numRP}[2]{\num[round-mode=places, round-precision=#2]{#1}}

\AtlasTitle{Measurement of top quark pair differential cross-sections in the dilepton channel in $pp$ collisions at $\sqrt{s}$ = 7 and 8 TeV with ATLAS}

\author{The ATLAS Collaboration}

\AtlasRefCode{TOPQ-2015-07}

\PreprintIdNumber{CERN-EP-2016-144}

\AtlasJournalRef{Phys. Rev. D 94, 092003 (2016)}

\AtlasDOI{10.1103/PhysRevD.94.092003}

\AtlasAbstract{%
Measurements of normalized differential cross-sections of top quark pair ($t\bar t$) production are presented as a function of the mass, the transverse momentum and the rapidity of the $t\bar t$ system
in proton-proton collisions at center-of-mass energies of $\sqrt{s}$ = 7 TeV and 8 TeV. 
The dataset corresponds to an integrated luminosity of 4.6 fb$^{-1}$ at 7 TeV and 20.2 fb$^{-1}$ at 8 TeV, recorded with the ATLAS detector at the Large Hadron Collider. 
Events with top quark pair signatures are selected in the dilepton final state, 
requiring exactly two charged leptons and at least two jets with at least one of the jets identified as likely to contain a $b$-hadron.
The measured distributions are corrected for detector effects and selection efficiency to cross-sections at the parton level. 
The differential cross-sections are compared with different Monte Carlo generators and theoretical calculations of $t\bar t$ production. 
The results are consistent with the majority of predictions in a wide kinematic range.
}

\begin{document}
\maketitle

\tableofcontents

\section{Introduction}
\label{sec:intro}

The top quark is the most massive elementary particle in the Standard Model (SM).
Its mass is close to the scale of electroweak symmetry breaking, implying a unique sensitivity to interactions beyond the SM. 
The production of top quarks at the Large Hadron Collider (LHC) is dominated by pair production of top and antitop quarks ($\ttbar$) via the strong interaction. 
Possible new phenomena beyond the SM can modify the kinematic properties of the $\ttbar$ system. 
Thus measurements of these distributions provide a means of testing the SM prediction at the TeV scale.
In addition, more accurate and detailed knowledge of top quark pair production 
is an essential component of the wide-ranging LHC physics program,
since $\ttbar$ events are the dominant background to many searches for new physics as well as Higgs boson measurements. 

The large $\ttbar$ production cross-section at the LHC leads to a large number of $\ttbar$ pairs, allowing precise inclusive and differential measurements in a wide kinematic range.
The inclusive $\ttbar$ production cross-section ($\sigmattbar$) has been measured
in proton-proton ($pp$) collisions at \rts{} = 7 TeV, 8 TeV and 13 TeV by the ATLAS
and CMS experiments~\cite{TOPQ-2013-04, CMS-TOP-11-005, CMS-TOP-12-007, CMS-TOP-13-004, TOPQ-2015-09, CMS-TOP-15-003},
with a best reported precision of 3.6\% (3.7\%) at 7 (8) TeV~\cite{CMS-TOP-13-004}. 
Measurements of the $\ttbar$ differential cross-section as a function of the kinematic properties of the top quark or the $\ttbar$ pair have also been performed by ATLAS~\cite{TOPQ-2011-07, TOPQ-2012-08, TOPQ-2013-07, TOPQ-2014-15, TOPQ-2015-06}
and CMS~\cite{CMS-TOP-11-013, CMS-TOP-12-028, CMS-TOP-14-012, CMS-TOP-14-018}. 

This paper presents measurements of the normalized differential \ttbar{} cross-sections
as a function of the invariant mass ($\mttbar$), the transverse momentum ($\ptttbar$),
and the rapidity ($\absyttbar$) of the \ttbar{} system
in $pp$ collisions at \rts{} = 7 TeV and 8 TeV
recorded by the ATLAS detector~\cite{PERF-2007-01}.
The dilepton $\ttbar$ decay mode used in this measurement yields a clean signal and thus provides an accurate test for the modeling of $\ttbar$ production. 
This paper complements other ATLAS measurements that use the lepton+jets ($\ell$+jets) $\ttbar$ decay mode~\cite{TOPQ-2011-07, TOPQ-2012-08, TOPQ-2013-07, TOPQ-2014-15, TOPQ-2015-06}.

A top quark pair is assumed to decay into two $W$ bosons and two $b$-quarks with a branching ratio of 100\%. 
The dilepton decay mode of $\ttbar$ used in this analysis refers to the mode
where both $W$ bosons decay into a charged lepton (electron or muon) and a neutrino.
Events in which the $W$ boson decays into an electron or a muon through a $\tau$ lepton decay 
are also included. 

Dileptonic $\ttbar$ events are selected by requiring two leptons (electron or muon) and at least two jets, 
where at least one of the jets is identified as containing a $b$-hadron.
The specific decay modes refer to the $\diel$, $\dimu$, and $\emu$ channels. 
In the 8 TeV measurement, one lepton must be an electron and the other must be a muon (the $\emu$ channel). 
This channel provides a
data sample large enough for the measurement to be limited by systematic uncertainties at 8 TeV.
In the 7 TeV analysis, 
where the integrated luminosity is smaller,
events containing same-flavor electron or muon pairs (the $\diel$ and $\dimu$ channels) are also selected
in order to maximize the size of the available dataset.

\section{ATLAS detector}
\label{sec:detector}

The ATLAS detector\footnote{ATLAS uses a right-handed coordinate system with
its origin at the nominal interaction point (IP) in the center of the detector
and the $z$-axis along the beam pipe.
The $x$-axis points from the IP to the center of the LHC ring,
and the $y$-axis points upward.
Cylindrical coordinates $(r,\phi)$ are used in the transverse plane,
$\phi$ being the azimuthal angle around the beam pipe.
The pseudorapidity is defined in terms of the polar angle $\theta$
as $\eta = -\ln{\tan(\theta/2)}$, 
and transverse momentum and energy are
defined as $\pT = p \sin\theta$ and $\ET = E \sin \theta$. 
Distances in ($\eta$, $\phi$) space are denoted by
$\Delta R = \sqrt{(\Delta\eta)^2 + (\Delta\phi)^2}$.}
is a general-purpose,
cylindrically symmetric detector 
with a barrel and two endcap components.
The inner detector (ID) is closest to the interaction point
and provides precise reconstruction of charged-particle tracks.
It is a combination of high-resolution silicon pixel and strip detectors, and a straw-tube tracking detector.
The ID covers a range of $|\eta| < 2.5$ and is surrounded by
a superconducting solenoid that produces a 2 T axial field within the ID.
Surrounding the ID are electromagnetic and hadronic sampling calorimeters.
The liquid argon (LAr) sampling electromagnetic calorimeter covers the pseudorapidity range of $|\eta| < 3.2$ with high granularity. 
The hadronic sampling calorimeters use steel/scintillator-tiles in $|\eta| < 1.7$ and LAr technology for $1.5 < |\eta| < 4.9$.
The muon spectrometer is the outermost subdetector and is composed
of three layers of chambers.
It is designed for precision measurement and detection of muons
exploiting the track curvature in the toroidal magnetic field.
The trigger system involves a combination of hardware- and software-based triggers at three levels to reduce the raw trigger rate of 20 MHz to 400 Hz.

\section{Data and simulation samples}
\label{sec:samples}

The datasets used in this analysis were collected from LHC $pp$ collisions 
at $\sqrt{s}$ = 7 and 8 TeV in 2011 and 2012.
The total integrated luminosities are
4.6\,\ifb{} with an uncertainty of 1.8\% at \rts{} = 7 TeV and 
20.2\,\ifb{} with an uncertainty of 1.9\% at \rts{} = 8 TeV.
The luminosity was measured using techniques 
described in Refs.~\cite{DAPR-2011-01,DAPR-2013-01}.
The average number of $pp$ interactions per bunch crossing (pileup) 
is about 9 for the 7 TeV dataset and increases to about 21 for the 8 TeV dataset. 
The data sample was collected using single-lepton triggers. 
The $\sqrt{s}=7$ TeV dataset uses a single-muon trigger requiring at least one muon with transverse momentum $\pT$ above 18$\GeV$ and a single-electron trigger requiring at least one electron with a $\pT$ threshold of either 20 or 22$\GeV$, with the $\pT$ threshold being increased during data-taking to cope with increased luminosity. 
In the  $\sqrt{s}=8$ TeV dataset, the logical OR of two triggers is used in order to increase the efficiency for isolated leptons at low transverse momentum, for each lepton type. 
For electrons the two $\pT$ thresholds are 24 GeV and 60 GeV, 
and for muons the thresholds are 24 GeV and 36 GeV,
where only the lower-$\pT$ triggers impose lepton isolation requirements.

Samples of Monte Carlo (MC) simulated events are used
to characterize the detector response and efficiency
for reconstructing \ttbar{} events,
to estimate systematic uncertainties, and to predict the background contributions
from various physics processes. 
The samples were processed through the {\sc Geant4}~\cite{Agostinelli:2002hh}
simulation of the ATLAS detector~\cite{SOFT-2010-01}
and the ATLAS reconstruction software.
For the evaluation of some systematic uncertainties,
generated samples are passed through
a fast simulation using a parameterization of the performance of the ATLAS electromagnetic
and hadronic calorimeters~\cite{ATL-SOFT-PUB-2014-01}.
The simulated events include pileup interactions to emulate the multiple pp interactions in each event present in the data.

The nominal signal \ttbar{} sample, {\sc Powheg+Pythia}, is generated
using the {\sc Powheg}
({\sc Powheg}-hvq patch4, revision 2330, version 3.0)
\cite{Nason:2004rx, Frixione:2007vw, Alioli:2010xd, Frixione:2007nw}
generator,
which is based on next-to-leading-order (NLO) QCD matrix element calculations.
The CT10~\cite{CT10} parton distribution functions (PDF) are
employed and the top quark mass ($m_{t}$) is set to 172.5 GeV.
The $h_{\rm damp}$ parameter in {\sc Powheg}, 
which controls the \pT{} of the first additional emission beyond the Born configuration, 
is set to infinity for the 7 TeV sample and set to $m_{t}$ for the 8 TeV sample.
The main effect of this parameter is to regulate the high-$p_{\rm T}$ emission against which the top quark pair system recoils.
In studies~\cite{ATL-PHYS-PUB-2015-002,ATL-PHYS-PUB-2015-011} 
using data from $\sqrt{s}$ = 7 TeV ATLAS $\ttbar$ differential cross-section
measurements in the $\ell$+jets channel~\cite{TOPQ-2012-08},
$h_{\rm damp}=\mt$ was shown to give a better description of data
than $h_{\rm damp}=\infty$, 
especially in the $\ptttbar$ spectrum~\cite{ATL-PHYS-PUB-2015-002,ATL-PHYS-PUB-2015-011}. 
Thus, the {\sc Powheg} $h_{\rm damp}=\mt$ sample was generated at 8 TeV as the nominal sample.
At 7 TeV, while only the {\sc Powheg} $h_{\rm damp}=\infty$ full MC sample is available,
the generated parton-level distributions with $h_{\rm damp}=\mt$ can be accessed and are used for comparison to the results.
Parton showering and hadronization are simulated with {\sc Pythia}~\cite{Sjostrand:2006za} 
(version 6.427) 
using the Perugia 2011C (P2011C) set of tuned parameters (tune)~\cite{Skands:2010ak}
and the corresponding leading-order (LO) CTEQ6L1 PDF set~\cite{Pumplin:2002vw}.

The effect of
the choice of generators and parton showering models
are studied with predictions from 
{\sc MC@NLO}~\cite{Frixione:2002ik, Frixione:2003ei} 
(version 4.01) 
interfaced to {\sc Herwig}
\cite{Corcella:2000bw} 
(version 6.520) 
for parton showering and hadronization, and to {\sc Jimmy}
\cite{Butterworth:1996zw} 
(version 4.31) 
for modeling multiple parton scattering in the underlying event
using the ATLAS AUET2 tune~\cite{ATL-PHYS-PUB-2011-008} and the CT10 PDFs,
and predictions from {\sc Powheg} interfaced to {\sc Herwig}.
The uncertainties 
in the modeling of extra QCD radiation in $t\bar{t}$ events
are estimated with samples generated using 
{\sc Alpgen} (version 2.14) ~\cite{ALPGEN} with CTEQ5L~\cite{cteq5} PDFs interfaced to {\sc Pythia}
with varied radiation settings 
and {\sc MC@NLO} interfaced to {\sc Herwig} 
with varied renormalization and factorization scales
(\rts{} = 7 TeV),  
or {\sc Powheg} interfaced to {\sc Pythia}
(\rts{} = 8 TeV) in which the parton shower parameters 
are varied to span the ranges compatible with the results of measurements of
\ttbar{} production in association with jets~\cite{TOPQ-2011-21, TOPQ-2012-03, ATL-PHYS-PUB-2015-002}.
All \ttbar{} samples are normalized to the NNLO+NNLL cross-sections
\cite{Cacciari:2011hy,Baernreuther:2012ws,Czakon:2012pz,Czakon:2012zr,Czakon:2013goa,Czakon:2011xx}: 
$\sigmattbar=177.3^{+10}_{-11}$~pb at $\sqrt{s}=7$ TeV and 
$\sigmattbar=253^{+13}_{-15}$~pb at $\sqrt{s}=8$ TeV.

Backgrounds with two real prompt leptons from decays of $W$ or $Z$ bosons (including those produced via leptonic $\tau$ decays) include 
$Wt$ single-top production, 
$Z$+jets production, 
and diboson ($WW$, $WZ$, and $ZZ$)+jets production. 
The largest background in this analysis, $Wt$ production, is modeled
using {\sc Powheg (Powheg}-st\_wtch)~\cite{Re:2010bp}
with the CT10 PDF set
and showered with {\sc Pythia} 
using the Perugia 2011C tune 
and the corresponding CTEQ6L1 PDF set. 
The baseline $Wt$ sample uses the ``diagram removal'' scheme to remove interference terms involving \ttbar{} production, 
and an alternative method using the ``diagram subtraction'' scheme~\cite{Frixione:2008yi}
is used to cross-check the validity of the prediction from the diagram removal scheme
and to assess systematic uncertainties.
The cross-section employed for $Wt$ single-top event generation is 15.7$\pm$1.2~pb (\rts{} = 7 TeV)
and 22.4$\pm$1.5~pb (\rts{} = 8 TeV),
as obtained from NLO+NNLL calculations~\cite{Kidonakis:2010ux}.
The $Z(\rightarrow\ell\ell)$+jets 
background is modeled using {\sc Alpgen} with the CTEQ6L1 PDFs, 
interfaced either to {\sc Herwig} and {\sc Jimmy} with the ATLAS AUET2 tune and the CT10 PDFs (\rts{} = 7 TeV)
or to {\sc Pythia6} with the Perugia P2011C tune and the CTEQ6L1 PDFs, including LO matrix elements for $Zb\bar{b}$ and $Zc\bar{c}$ production (\rts{} = 8 TeV).
Inclusive $Z$ boson cross-sections are set to the NNLO predictions from FEWZ~\cite{fewz}, 
but the normalizations of $Z(\rightarrow\diel/\dimu)$+jets in the \rts{} = 7 TeV analysis are determined from data
using the same procedure used in Refs.~\cite{TOPQ-2010-01,TOPQ-2011-01}.
The diboson background is modeled using 
{\sc Alpgen} with the CTEQ6L1 PDFs interfaced to {\sc Herwig} and {\sc Jimmy} with the AUET2 tune and the CT10 PDFs, and the cross-sections are normalized to NLO QCD calculations~\cite{Campbell:2011}.

Background processes where one or more of the reconstructed lepton candidates are nonprompt or misidentified (referred to as ``fake leptons'')
arise from 
\ttbar{} production,
$W$+jets production, 
and single-top production in the $t$-channel or $s$-channel.
The \rts{} = 7 TeV analysis uses a matrix method~\cite{TOPQ-2010-01} to estimate the fake-lepton background directly from data, while the \rts{} = 8 TeV analysis uses event samples of same-sign leptons in both data and simulations to estimate the fake-lepton contributions in these processes~\cite{TOPQ-2013-04}.
The fake-lepton contributions from $\ttbar$ production are simulated from the same baseline $\ttbar$ signal sample, 
which includes the $\ell$+jets decay channel, 
and $\ttbar$+$V$ samples where $V=W$ or $Z$, modeled by {\sc Madgraph}~\cite{madgraph} interfaced to {\sc Pythia} with the Perugia P2011C tune and the CTEQ6L1 PDFs.
The $W$+jets production is simulated using {\sc Alpgen} with the CTEQ6L1 PDFs interfaced to {\sc Pythia6} with the Perugia P2011C tune and the CTEQ6L1 PDFs, including LO matrix elements for $Wb\bar{b}$, $Wc\bar{c}$, and $Wc$ processes.
The $t$-channel single-top production is modeled using the {\sc AcerMC}~\cite{ACERMC} generator, 
while {\sc Powheg} is used for the production in the $s$-channel, 
and both generators are interfaced to {\sc Pythia6} using the Perugia P2011C tune and the CTEQ6L1 PDFs.
Different methods are used in the two datasets due to the different trigger conditions and because the 7 TeV analysis uses all 3 dilepton channels.
Other backgrounds 
are negligible after the event selections used in this analysis.

Table~\ref{tab:mcsamples} summarizes the baseline signal and background MC simulated samples used in the 7 TeV and 8 TeV analyses. 

\begin{table}
 \begin{center}
  \begin{tabular}{l|cc}
   \toprule
   Physics process                  & 7 TeV analysis               & 8 TeV analysis  \\
   \midrule
   \ttbar{}                         & {\sc Powheg+Pythia} ($h_{\rm damp}=\infty$)  & {\sc Powheg+Pythia} ($h_{\rm damp}=m_{t}$) \\
   $Wt$                  & {\sc Powheg+Pythia}              & {\sc Powheg+Pythia} \\
   $Z(\rightarrow\tau\tau)$+jets    & {\sc Alpgen+Herwig}              & {\sc Alpgen+Pythia} \\
   $Z(\rightarrow\diel/\dimu)$+jets & {\sc Alpgen+Herwig} and data     & - \\
   Diboson+jets                         & {\sc Alpgen+Herwig}              & {\sc Alpgen+Herwig} \\
   Fake leptons                     & Data                             & Various MC samples and data \\
   \bottomrule
  \end{tabular}
  \caption{List of baseline MC samples used in the 7 TeV and 8 TeV analyses.
  The $Z(\rightarrow\diel/\dimu)$+jets process is not included in the 8 TeV analysis as the analysis uses only the $\emu$ channel. 
   }
  \label{tab:mcsamples}
 \end{center}
\end{table}

\section{Object and event selection}
\label{sec:selection}

\subsection{Object definition}
\label{subsec:object}

Electron candidates are reconstructed as charged-particle tracks in the inner detector associated with energy deposits in the electromagnetic calorimeter, and must satisfy tight identification criteria~\cite{PERF-2013-03}.
Electron candidates are required to have transverse energy $\Et > 25 \GeV$ and pseudorapidity $|\eta|<2.47$, while excluding the transition region between the barrel and the endcap calorimeters ($1.37 < | \eta | < 1.52$). 
Isolation requirements on calorimeter and tracking variables are used to reduce
the background from nonprompt electrons. 
The calorimeter isolation variable is based on the energy sum
of cells within a cone of size
$\Delta R=0.2$ around the direction of each electron candidate. 
This energy sum excludes cells associated with the electron cluster and is corrected for leakage from the electron
cluster itself and for energy deposits from pileup. 
The tracking isolation variable is based on the track
$\pT$ sum around the electron in a cone of size
$\Delta R=0.3$, excluding the electron track.  
In every $\pT$ bin, 
both requirements are chosen to result separately in a 90\% (98\%) electron selection efficiency for prompt electrons
from $Z\rightarrow\diel$ decays in the 7 TeV (8 TeV) analysis.

Muon candidates are identified by matching track segments in the muon spectrometer with tracks in the
inner detector, and are required to be in the region $|\eta| < 2.5$ and have $\pT > 20 (25)\GeV$ in the 7 TeV (8 TeV) analysis.
To reduce the background from muons originating from heavy-flavor decays inside jets, 
muons are required to be separated by
$\Delta R=0.4$ from the nearest jet, and to be isolated.
In the 7 TeV analysis, 
the isolation of muons requires the calorimeter transverse energy within a cone of fixed size $\Delta R = 0.2$ and the sum of track $\pT$ within a cone of fixed size $\Delta R=0.3$ around the muon, except the contribution from the muon itself, to be less than 4$\GeV$ and 2.5$\GeV$, respectively. 
In the 8 TeV analysis, muons are required to satisfy
$I^{\ell}<0.05$ where the isolation variable is the ratio of
the sum of $\pT$ of tracks, excluding the muon, in a cone of variable size
$\Delta R = 10 \GeV / \pT(\mu)$ to the $\pT$ of the muon~\cite{Rehermann:2010vq}. 
Both isolation requirements result in an efficiency of about 97\% for prompt muons from $Z\rightarrow\dimu$ decays.

Jets are reconstructed by the \AKT{} algorithm~\cite{akt2} 
with a radius parameter $R=0.4$ using 
calorimeter energy clusters~\cite{topocluster}, 
which are calibrated at the electromagnetic energy scale for the \rts{} = 7 TeV dataset, 
or using the local cluster weighting method for \rts{} = 8 TeV~\cite{PERF-2011-03}.
The energies of jets are then calibrated 
using an energy- and $\eta$-dependent simulation-based
calibration scheme
with {\it in situ} corrections based on data.
Different calibration procedures were used for the 7 TeV and 8 TeV datasets due to the different pileup conditions.
The effects of pileup on the jet energy calibration at 8 TeV are further reduced using the jet area method as described in Ref.~\cite{ATLAS-CONF-2013-083}.
Jets with $\pT$ > 25 GeV and $|\eta|$ < 2.5 are accepted. 
To suppress jets from pileup, 
a requirement on the jet vertex fraction (JVF), 
the ratio of the sum of the \pT{} of tracks associated with both the jet and the primary vertex
to the sum of the \pT{} of all tracks associated with the jet, 
is imposed based on the different pileup conditions in the \rts{} = 7 TeV and \rts{} = 8 TeV~\cite{TOPQ-2013-04}. At 7 TeV, jets are required to satisfy |JVF| > 0.75 while at 8 TeV, jets with $\pT$ < 50 GeV and $|\eta|$ < 2.4 are required to satisfy |JVF| > 0.5.
To prevent double-counting of electron energy deposits as jets, 
the closest jet lying $\Delta R < 0.2$ from
a reconstructed electron is removed; and finally, 
a lepton lying $\Delta R < 0.4$ from a selected jet is discarded
to reject leptons from heavy-flavor decays.

The purity of $\ttbar$ events in the selected sample is improved by tagging jets containing $b$-hadrons (``$b$-tagging''). 
Information from the track impact parameters, secondary vertex position, and decay topology 
is combined in a multivariate discriminant (MV1)~\cite{ATLAS-CONF-2012-043,ATLAS-CONF-2014-046}. 
Jets are defined to be $b$-tagged if the MV1 discriminant value is larger than a threshold (operating point) 
corresponding to an average 70\% efficiency for tagging $b$-quark jets from top quark decays in $\ttbar$ events, 
with about 1\% and 20\% probability of misidentifying light-flavor jets and charm-jets, respectively.

The missing transverse momentum $\Etmiss$ is derived
from the vector sum of calorimeter cell energies within $|\eta|<4.9$
associated with physics objects (electrons, muons, and jets) and corrected with their dedicated calibrations, 
as well as the transverse energy deposited in the calorimeter cells not associated with these objects~\cite{PERF-2011-07}.

\subsection{Event selection}
\label{subsec:event}

Events in the 7 TeV and 8 TeV analyses are selected based on the above definitions of reconstructed objects and the event quality. 
All events are required to
have at least one primary vertex\footnote{The primary vertex is defined to be the reconstructed vertex with the highest $\sum \pT^2$ of the associated tracks in the event.} reconstructed from at least five tracks with $\pT>0.4$ GeV, 
and events compatible with cosmic-ray interactions are rejected.
All jets are required to pass jet quality and timing requirements and at least one lepton is required to 
match in ($\eta$, $\phi$) space with particle(s) that triggered the event.
The dilepton event sample is selected by requiring exactly two charged leptons (electrons or muons) with opposite-sign charge 
and at least two jets, including at least one that is $b$-tagged.

To suppress backgrounds from Drell-Yan and multijet processes in the $\diel$ and $\dimu$ channels in the 7 TeV analysis, 
the missing transverse momentum $\Etmiss$
is required to be greater than 60 GeV, 
and the dilepton invariant mass $\mll$ is required to be outside the $Z$ boson mass window $|\mll - 91\GeV|$ > 10 GeV. 
The dilepton invariant mass is also required to be above 15 GeV in the $\diel$ and $\dimu$ channels to reject backgrounds from bottom-quark pair and vector-meson decays. 
No $\Etmiss$ nor $\mll$ requirements are applied in the $\emu$ channel, 
but a reconstructed variable, $\HT$, defined to be the scalar sum of the $\pT$ of all selected leptons and jets in an event, 
is required to be greater than 130 GeV to suppress remaining background from $Z/\gamma^*$+jets processes at 7 TeV.
In the 8 TeV analysis the $\HT$ requirement is not applied, since the improvement is negligible due to a higher muon $\pT$ requirement than the 7 TeV analysis.

In the 7 TeV analysis, an additional requirement using the invariant mass of a jet and a lepton
is also applied to reject events where the reconstructed jet does not originate from the $\ttbar$ decay (wrong-jet events).
Exploiting the kinematics of top quark decay with the constraint from the top quark mass $\mt$,
the invariant mass of the jet with the second highest value of the $b$-tagging discriminant $j_{2}$ and either of the leptons $\ell^{+}/\ell^{-}$
is required to be less than 0.8 of $\mt$ ($m_{j_{2}\ell^{+}}/\mt<0.8$ OR $m_{j_{2}\ell^{-}}/\mt<0.8$).
This cut value was optimized to provide about 94\% selection efficiency while rejecting about 16\% of the wrong-jet events in the simulated $\ttbar$ dilepton event sample. 

Table \ref{tab:eventselection} shows a summary of the event selections for the 7 TeV and 8 TeV analyses.
The numbers of events that fulfill all selection requirements are shown in Table \ref{tab:eventtable}. 

\begin{table}[htbp]
\centering
\sisetup{round-mode=places, round-precision=0, retain-explicit-plus=true}
{
\begin{tabular}{l|c|c|c|c}
\toprule
 & \multicolumn{3}{c|}{7$\TeV$} & 8$\TeV$ \\
\midrule
Selection & \hspace{1.3cm}$\diel$\hspace{1.3cm} & \hspace{1.25cm}$\dimu$\hspace{1.25cm} & $\emu$ & $\emu$ \\
\midrule
Leptons & \multicolumn{4}{c}{Exactly 2 leptons, opposite-sign charge, isolated} \\
 & \multicolumn{4}{c}{Electrons: $\ET > 25 \GeV$, $|\eta| < 2.47$, excluding $1.37 < |\eta| < 1.52$} \\
 & \multicolumn{3}{c|}{Muons: $\pT > 20 \GeV$, $|\eta| < 2.5$} & $\pT > 25 \GeV$, $|\eta| < 2.5$ \\  
Jets    & \multicolumn{4}{c}{$\geq$ 2 jets, $\pt > 25 \GeV$, $|\eta| < 2.5$} \\ 
  & \multicolumn{4}{c}{$\geq$ 1 $b$-tagged jet at $\epsilon_b$ = 70\%} \\ 
$\mll$          & \multicolumn{2}{c|}{$|\mll-91\GeV| > 10 \GeV$, $\mll > 15 \GeV$} & None & None \\ 
$\MET$ or $\HT$ & \multicolumn{2}{c|}{$\MET>60\GeV$} & $\HT > 130\GeV$ & None \\ 
$m_{j\ell}$        & \multicolumn{3}{c|}{$m_{j_{2}\ell^{+}}/\mt<0.8$ OR $m_{j_{2}\ell^{-}}/\mt<0.8$} & None \\ 
\bottomrule
\end{tabular}
}
\caption{Summary of the event selections for the 7 TeV and 8 TeV analyses.}
\label{tab:eventselection}
\end{table}

\begin{table}[htbp]
\centering
\sisetup{round-mode=places, round-precision=0, retain-explicit-plus=true, group-integer-digits=true, group-decimal-digits=false, group-four-digits=true}
{
\begin{tabular}{lc@{ $\pm$ }cc@{ $\pm$ }cc@{ $\pm$ }cc@{ $\pm$ }c}
\toprule
 & \multicolumn{6}{c}{7$\TeV$} & \multicolumn{2}{c}{8$\TeV$} \\
\midrule
Channel & \multicolumn{2}{c}{$\diel$} & \multicolumn{2}{c}{$\dimu$} & \multicolumn{2}{c}{$\emu$}  & \multicolumn{2}{c}{$\emu$} \\
 \midrule
$\ttbar$               & \numRF{484.61}{2} & \numRF{36.24}{1} & \numRF{1421.35}{3} & \numRF{57.84}{1} & \numRF{3738.53}{3} & \numRF{169.749}{2}  & \numRF{26715}{3} &  \numRF{803}{1}  \\
$Wt$                    & \numRF{20.38}{2} & \numRF{4.34}{1}   & \numRF{57.93}{2} & \numRF{15.17}{2}  & \numRF{154.78}{3} & \numRF{23.03}{2}  & \numRF{1283}{3}  &  \numRF{108}{2}  \\
Fake leptons                      & \numRF{12.17}{2} & \numRF{6.08}{1}   & \numRF{11.44}{3} & \numRF{3.43}{2}  & \numRF{50.22}{2} & \numRF{20.09}{2}  & \numRF{232}{2}   &  \numRF{107}{2} \\
$Z(\rightarrow \tau\tau)$+jets    & \numRF{0.43}{2} & \numRF{0.33}{2}    & \numRF{2.56}{2} & \numRF{1.18}{2}      & \numRF{5.76}{2} & \numRF{1.23}{2} & \numRF{80}{2}    &  \numRF{34}{2}  \\
$Z(\rightarrow \diel/\dimu)$+jets & \numRF{2.15}{2} & \numRF{1.00}{2}    & \numRF{5.77}{1} & \numRF{4.33}{1}      & \multicolumn{2}{c}{-} & \multicolumn{2}{c}{-} \\
Diboson+jets                           & \numRF{1.03}{3} & \numRF{0.31}{2}    & \numRF{3.22}{2} & \numRF{0.96}{1}      & \numRF{8.95}{2} & \numRF{2.37}{2} & \numRF{77}{2}    &  \numRF{31}{2}   \\
\midrule
Predicted                         & \numRF{520.76}{2} & \numRF{38.8274}{1} & \numRF{1502.27}{3} & \numRF{62.3895}{1}  & \numRF{3958.25}{3} & \numRF{179.581}{2} & \numRF{28449}{3} &  \numRF{819}{1} \\
\midrule
Observed                          & \multicolumn{2}{c}{\num{532}}              & \multicolumn{2}{c}{\num{1509}}               & \multicolumn{2}{c}{\num{4038}} & \multicolumn{2}{c}{\num{28772}}  \\
\bottomrule
\end{tabular}
}
\caption{Predicted event yields and uncertainties for $\ttbar$ signal and backgrounds compared to observed event yields in the 7 TeV and 8 TeV analyses.
The uncertainties include all systematic uncertainties discussed in Section~\ref{sec:uncertainties} except $\ttbar$ modeling.}
\label{tab:eventtable}
\end{table}

\section{Reconstruction}
\label{sec:reconstruction}

To reconstruct the $\ttbar$ system the two jets identified as most likely to contain $b$-hadrons are used.
This choice improves the resolution of the $\ttbar$-system observables as the jets are more likely to have originated from top quark decay. In both the 7 TeV and 8 TeV analyses, the fractional resolution for $\mttbar$ is typically below 20\%, while for $\ptttbar$ the fractional resolution is 35\% at 100 GeV and improves as a function of $\ptttbar$. The resolution for $\absyttbar$ is on average 17\%.

An approximate four-momentum of the \ttbar{} system
is reconstructed from two leptons,
two jets, and missing transverse momentum \MET{} 
as:
\begin{eqnarray}
 E_{\rm total} &=& E(\ell_1) + E(\ell_2) + E(j_1) + E(j_2) + E_{\mathrm{T}}^{\rm miss} \nonumber \\
 p_{x} &=& p_{x}(\ell_1) + p_{x}(\ell_2) + p_{x}(j_1) + p_{x}(j_2) + E_{x}^{\rm miss} \nonumber \\
 p_{y} &=& p_{y}(\ell_1) + p_{y}(\ell_2) + p_{y}(j_1) + p_{y}(j_2) + E_{y}^{\rm miss} \nonumber \\
 p_{z} &=& p_{z}(\ell_1) + p_{z}(\ell_2) + p_{z}(j_1) + p_{z}(j_2) \nonumber
\label{eq:ttreco}
\end{eqnarray}
where $E$ indicates the energy of the corresponding objects,
the $p_{x,y,z}$ is the momentum along to $x$-, $y$-, or $z$-axis,
and the indices $\ell_1$, $\ell_2$, $j_1$, and $j_2$ indicate the two leptons and two jets, respectively. 
The $\ttbar$-system observables in consideration (invariant mass, transverse momentum, and rapidity) are obtained from this four-momentum.

Figures~\ref{fig:ttbar_system_opt2} and \ref{fig:cplots} show the distributions
of the reconstructed $\mttbar$, $\ptttbar$, and $\absyttbar$ together with the MC predictions at 7 TeV and 8 TeV, respectively.
The bottom panel shows the ratio of the data to the total prediction.
Overall there is satisfactory agreement between data and prediction. 

\begin{figure}[htbp]
\captionsetup[subfloat]{farskip=2pt,captionskip=1pt}
 \centering
 \subfloat[$\mttbar$]{
\includegraphics[width=0.325\textwidth]{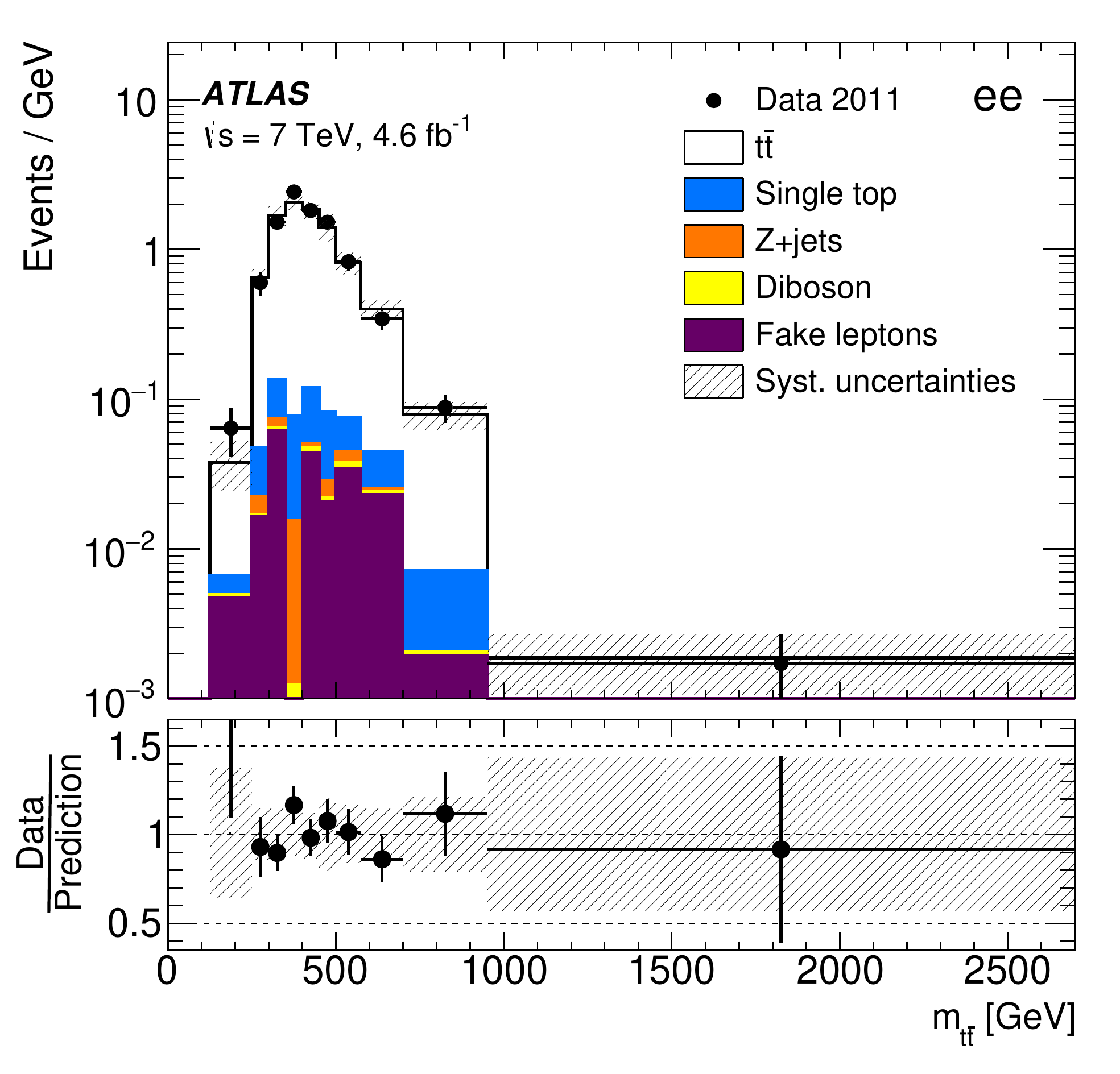} 
\includegraphics[width=0.325\textwidth]{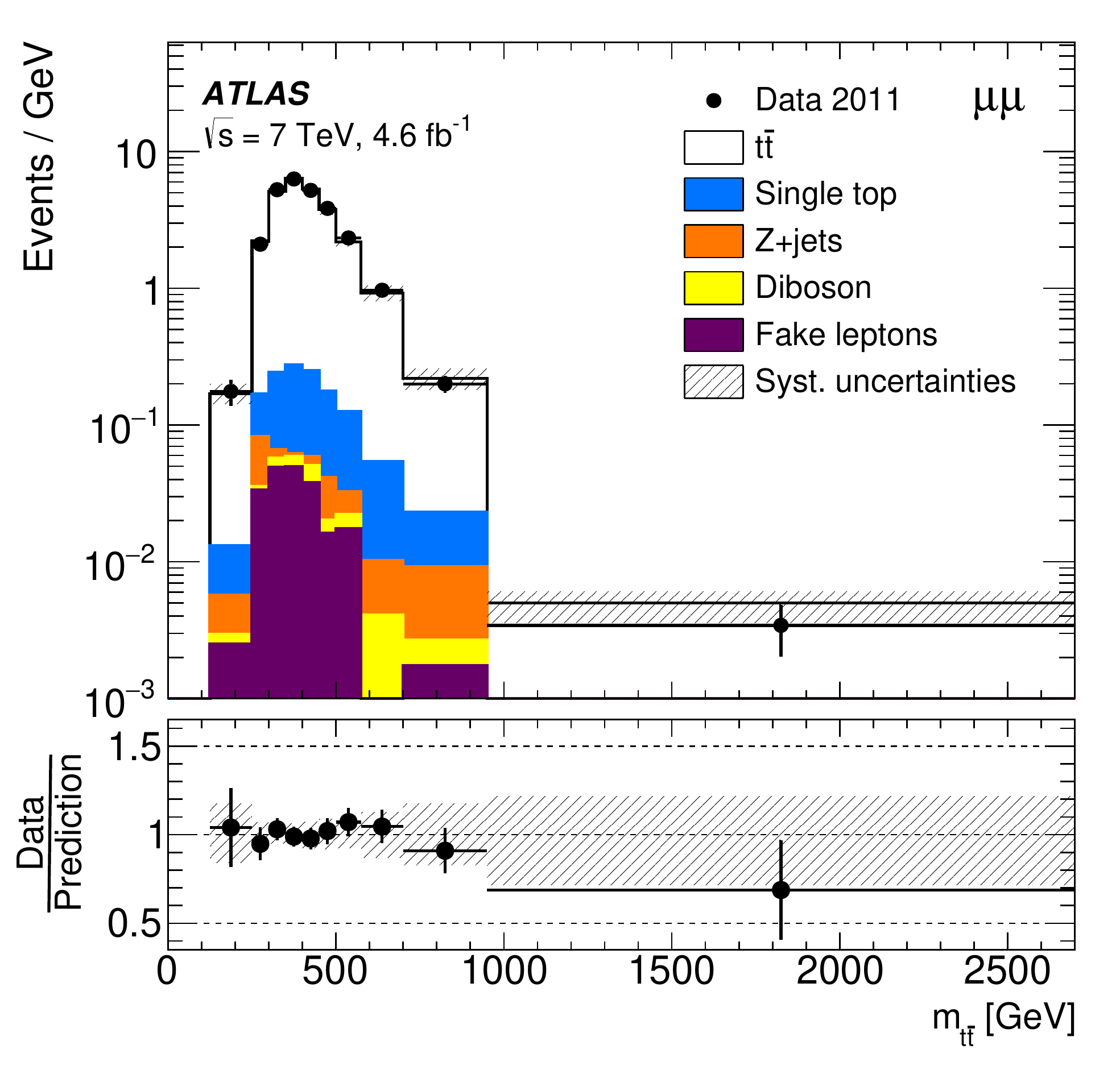} 
\includegraphics[width=0.325\textwidth]{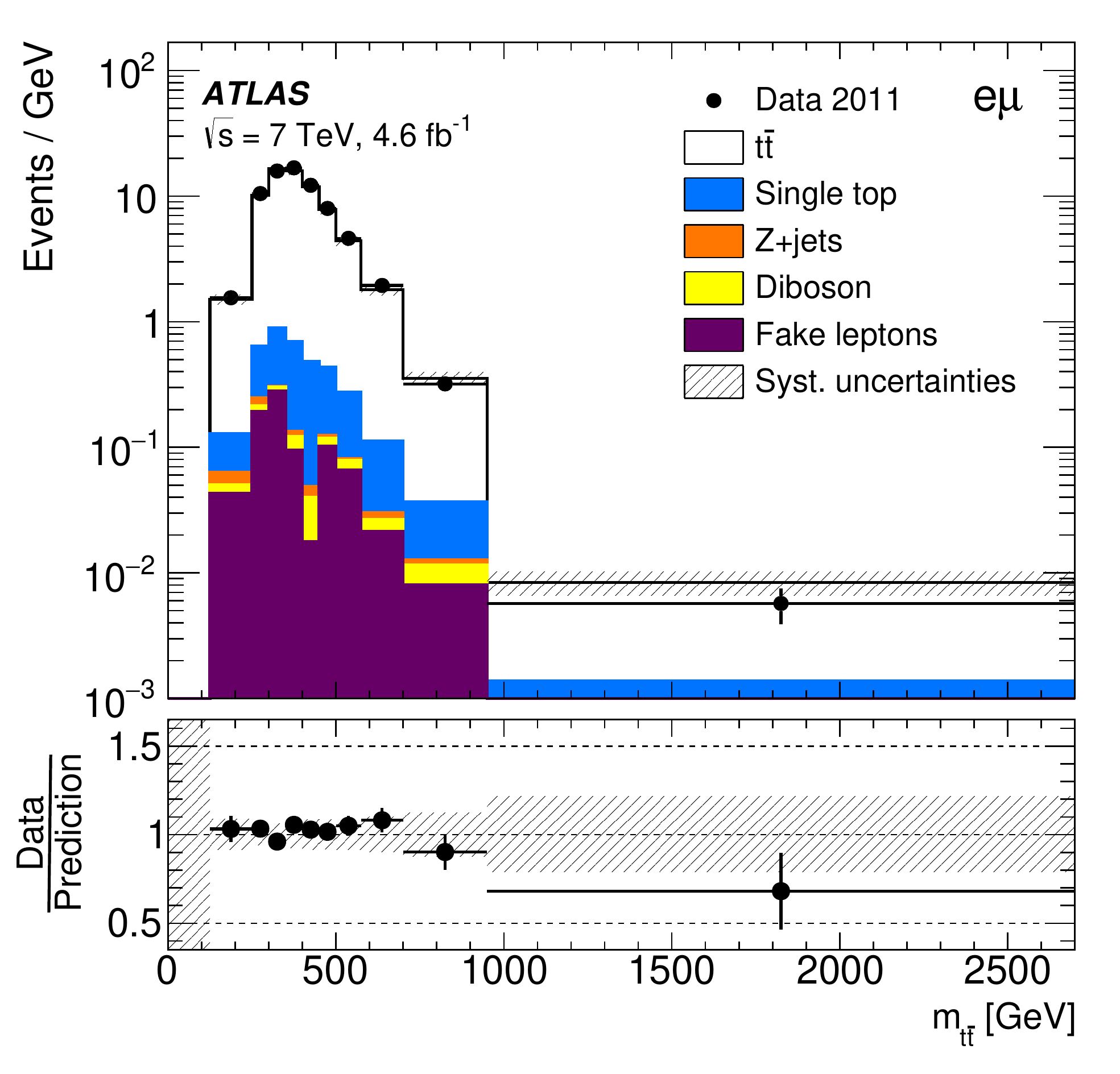} 
  \label{fig:ttbar_mass_opt2}
 }

 \subfloat[$\ptttbar$]{
\includegraphics[width=0.325\textwidth]{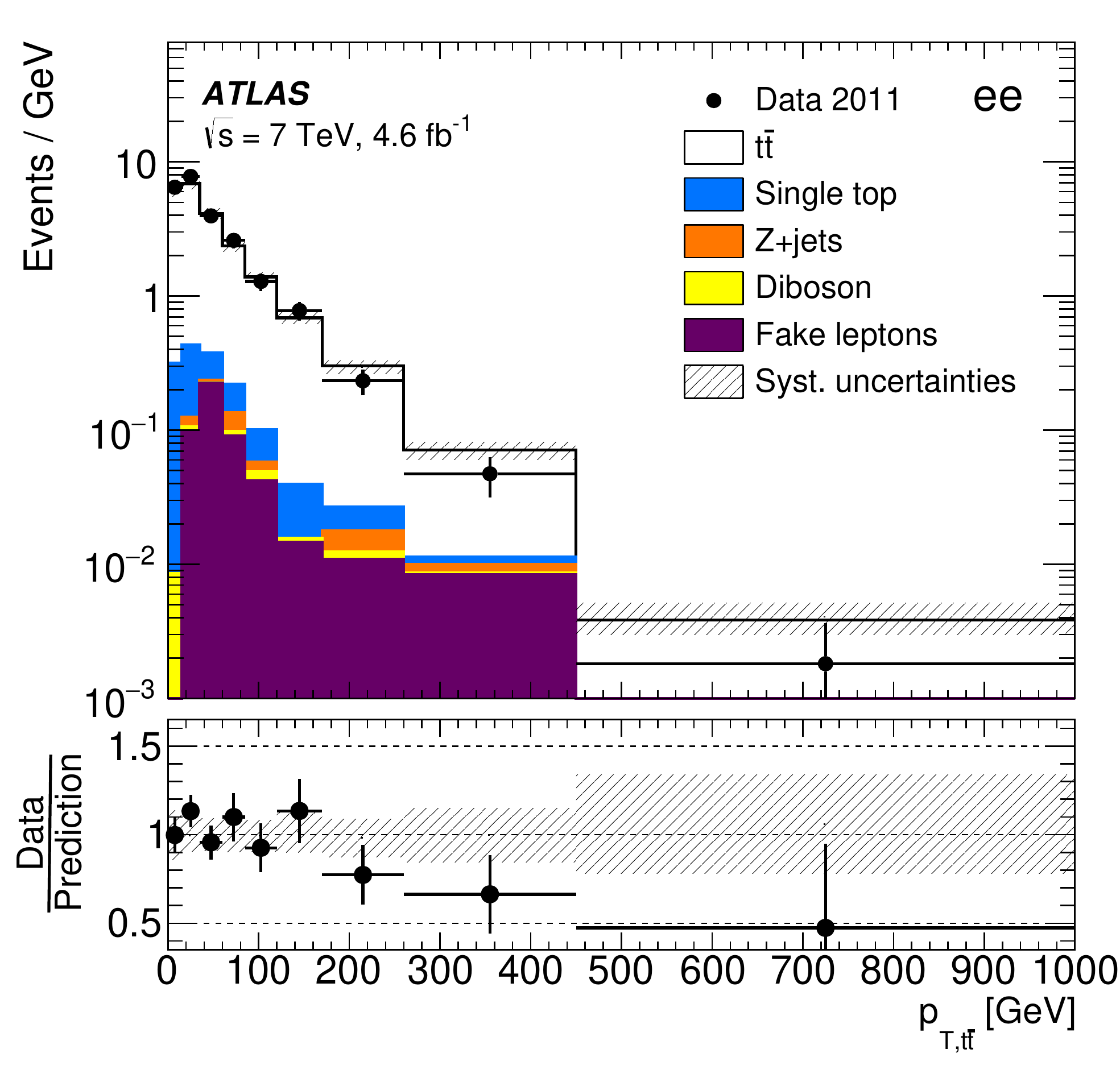} 
\includegraphics[width=0.325\textwidth]{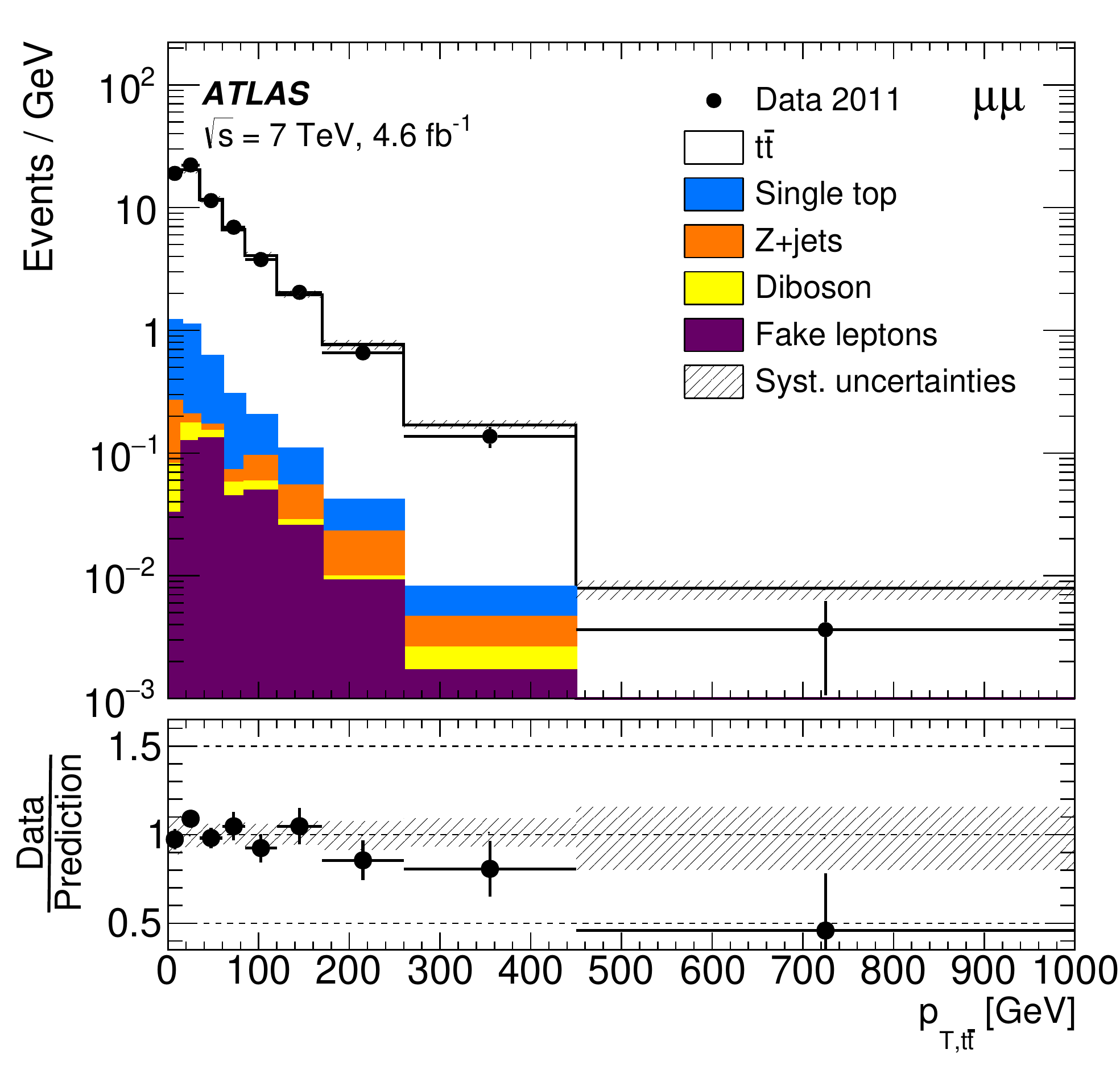} 
\includegraphics[width=0.325\textwidth]{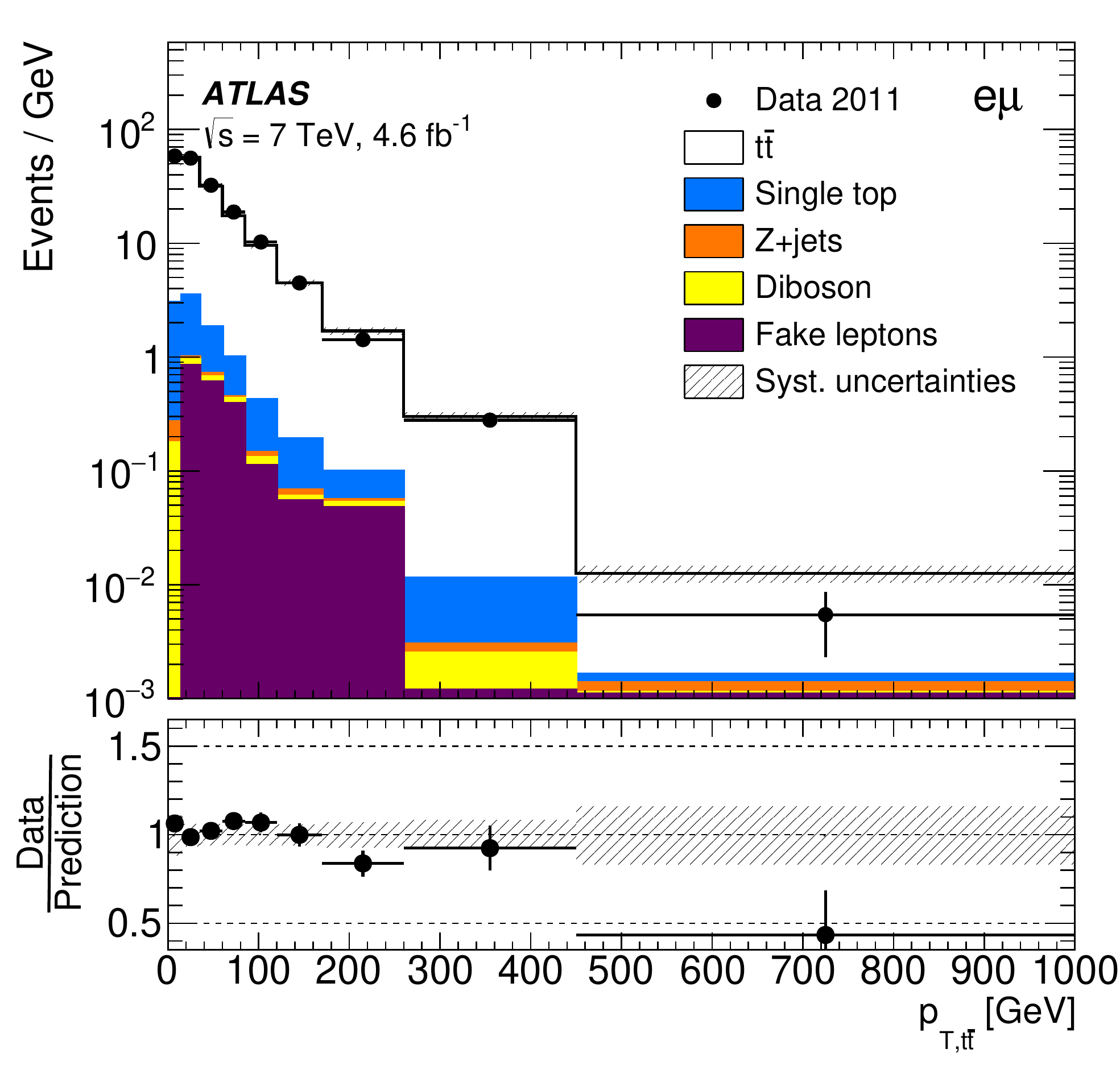} 
  \label{fig:ttbar_pt_opt2}
 }

\subfloat[$\absyttbar$]{
\includegraphics[width=0.325\textwidth]{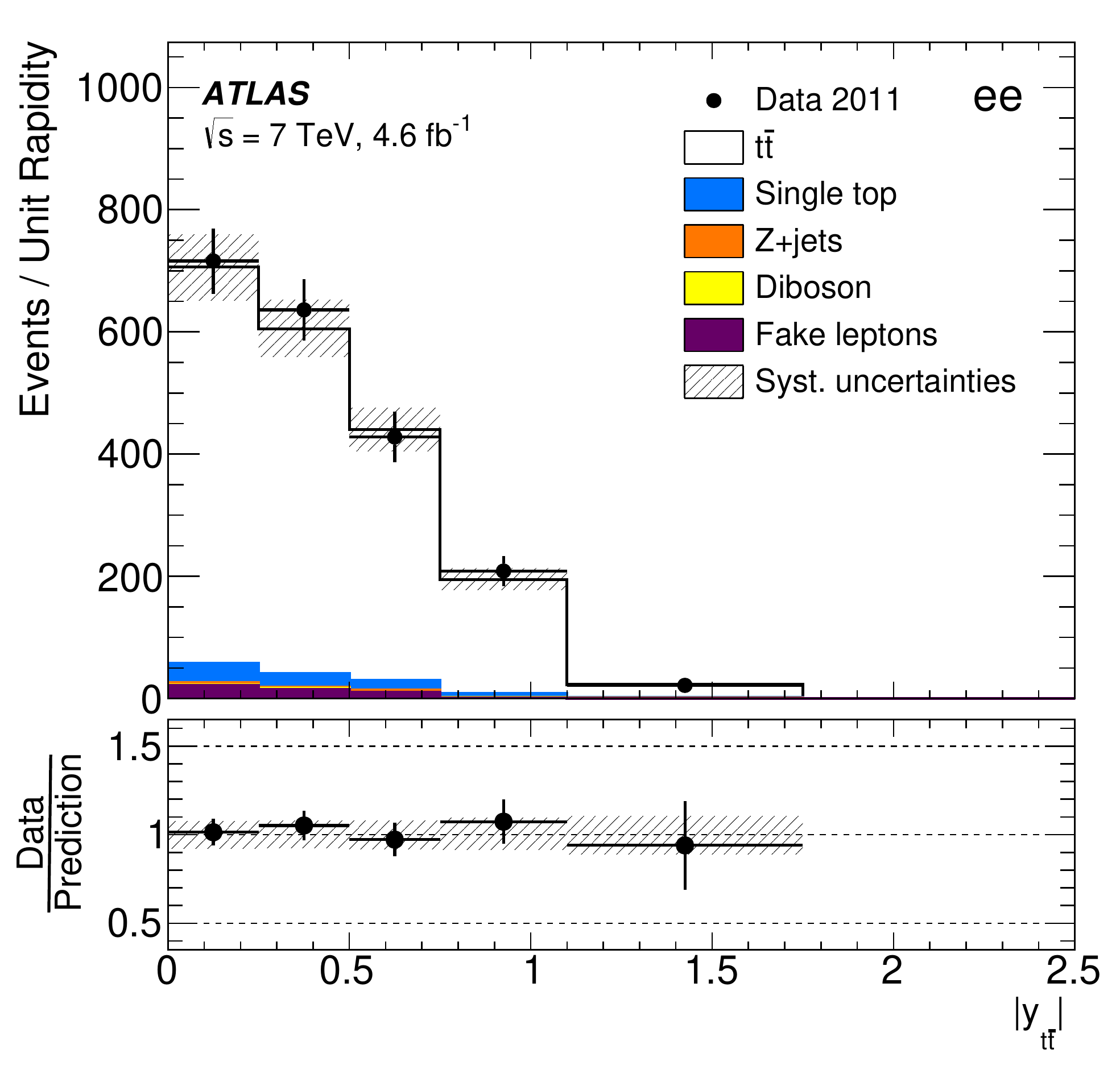} 
\includegraphics[width=0.325\textwidth]{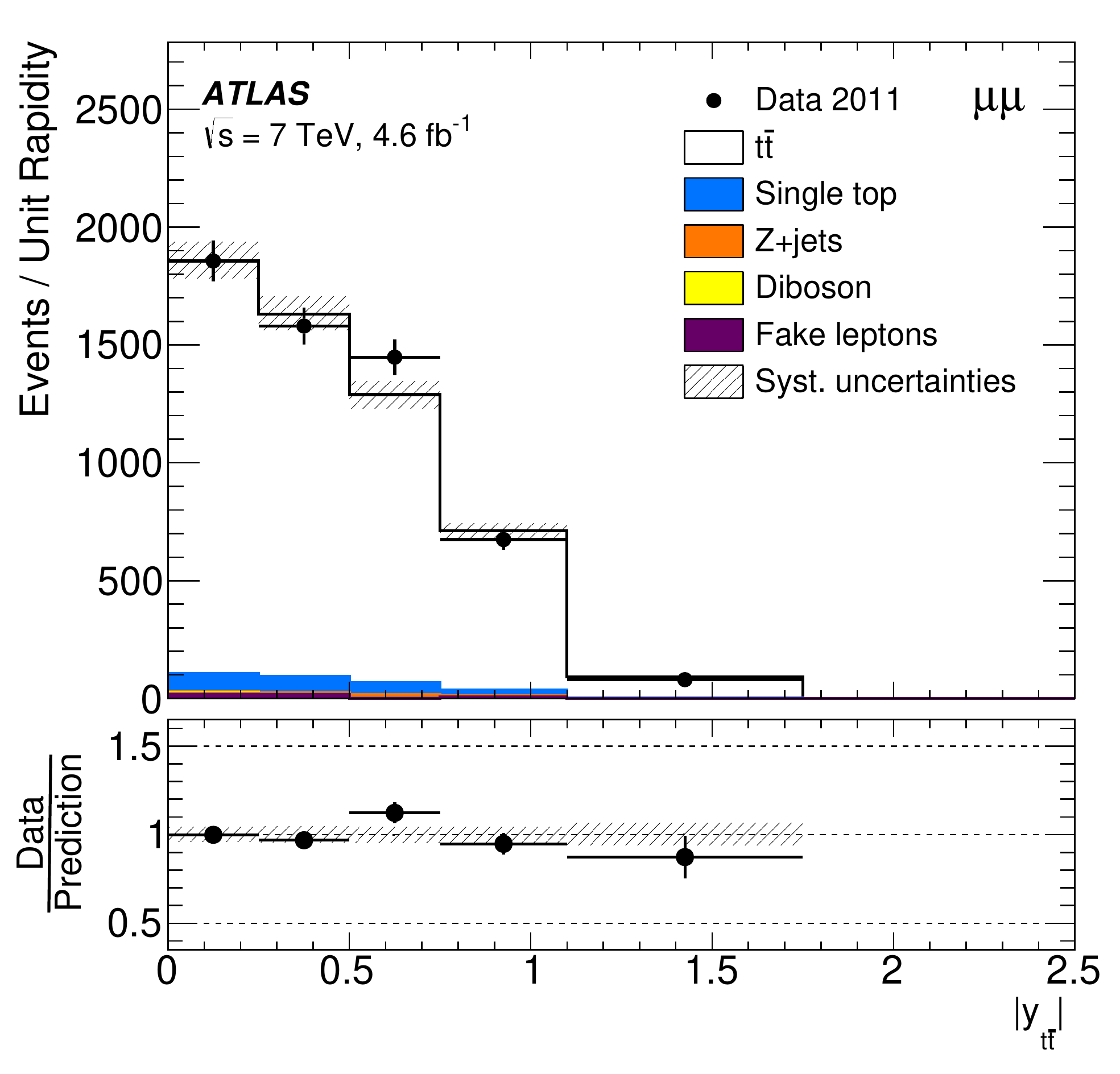} 
\includegraphics[width=0.325\textwidth]{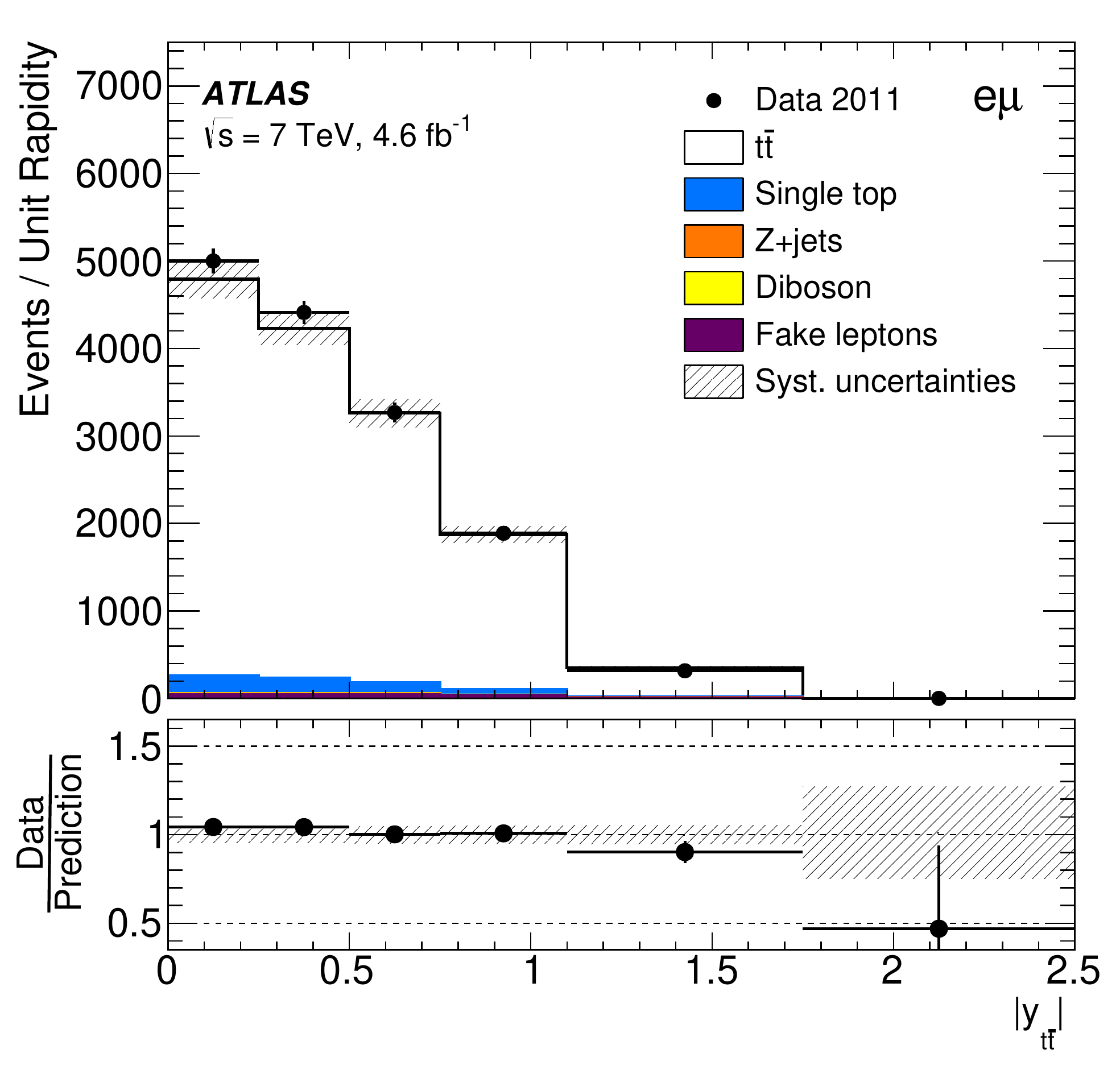} 
  \label{fig:abs_ttbar_rapidity_opt2}
}
\caption{
   Distributions of
   \protect\subref{fig:ttbar_mass_opt2} the invariant mass,
   \protect\subref{fig:ttbar_pt_opt2} the transverse momentum, 
   and 
   \protect\subref{fig:abs_ttbar_rapidity_opt2} the rapidity of the $\ttbar$ system 
   at the reconstruction level
   obtained from the $\sqrt{s}$ = 7 TeV data
   compared with the total signal and background predictions, 
   in the $\diel$ (left), $\dimu$ (center) and $\emu$ (right) channels.
   The bottom panel shows the ratio of data to prediction. 
   The error band includes all systematic uncertainties except $\ttbar$ modeling uncertainties. 
   The {\sc Powheg}+{\sc Pythia} with $h_{\rm damp}=\infty$ sample is used for the signal $\ttbar$ and is normalized to NNLO+NNLL calculations.
}
\label{fig:ttbar_system_opt2}
\end{figure}

\begin{figure}[htbp]
\captionsetup[subfloat]{farskip=2pt,captionskip=1pt}
 \centering
 \subfloat[$\mttbar$]{
 \includegraphics[width=0.325\textwidth]{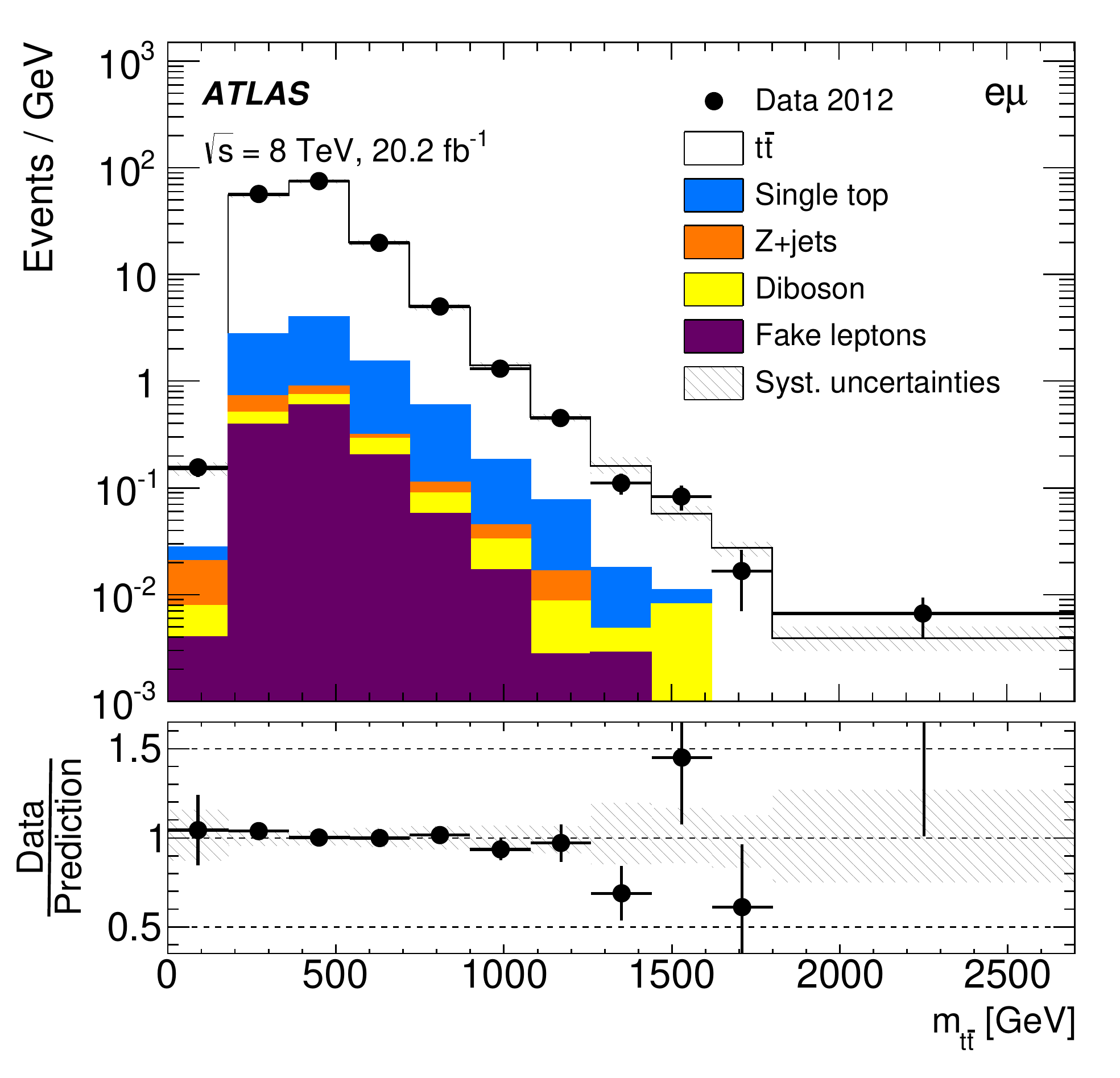}
 \label{fig:cplots_Mtt}
 }
 \subfloat[$\ptttbar$]{
 \includegraphics[width=0.325\textwidth]{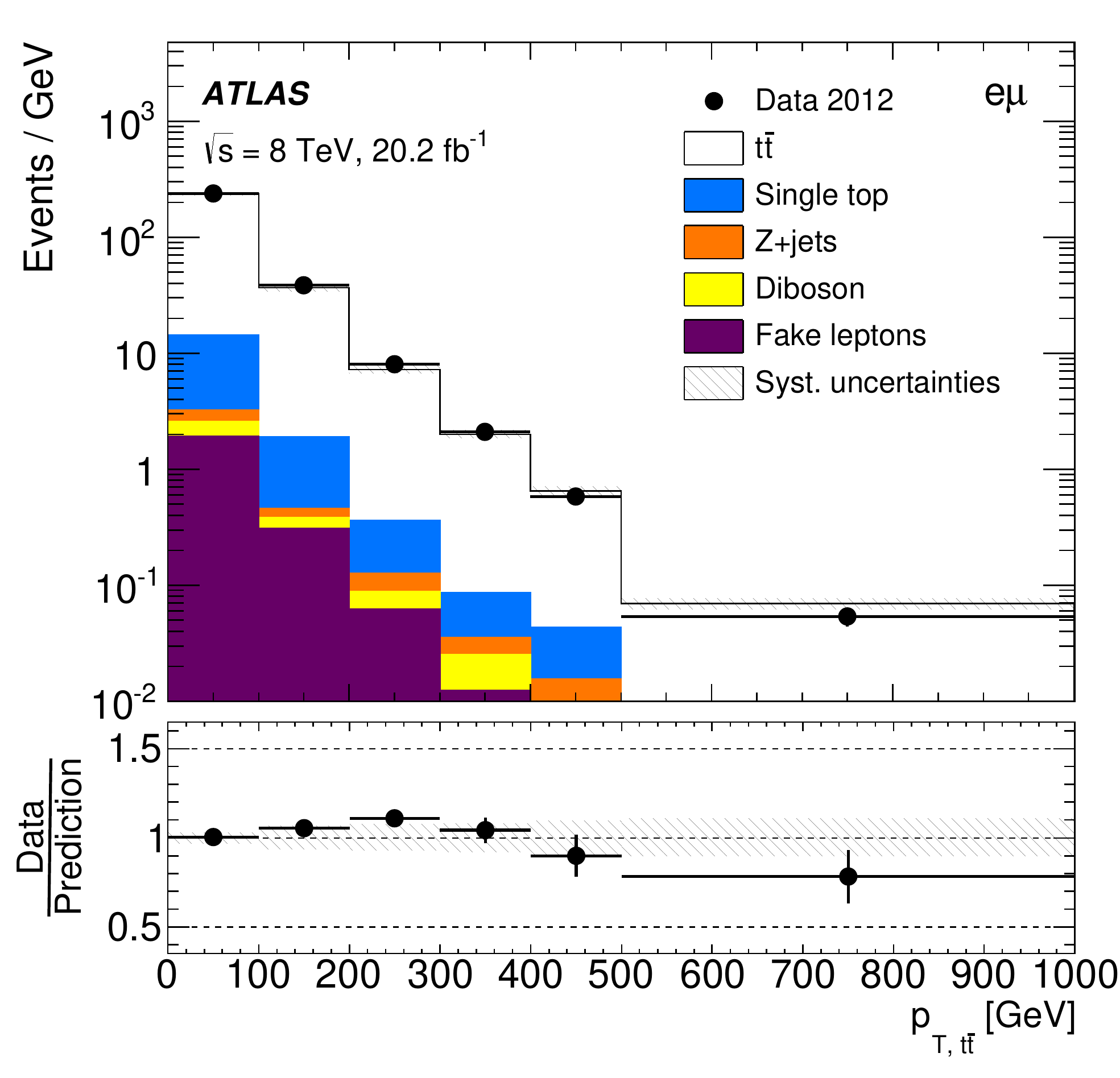}
 \label{fig:cplots_pTtt}
 }
\subfloat[$\absyttbar$]{
 \includegraphics[width=0.325\textwidth]{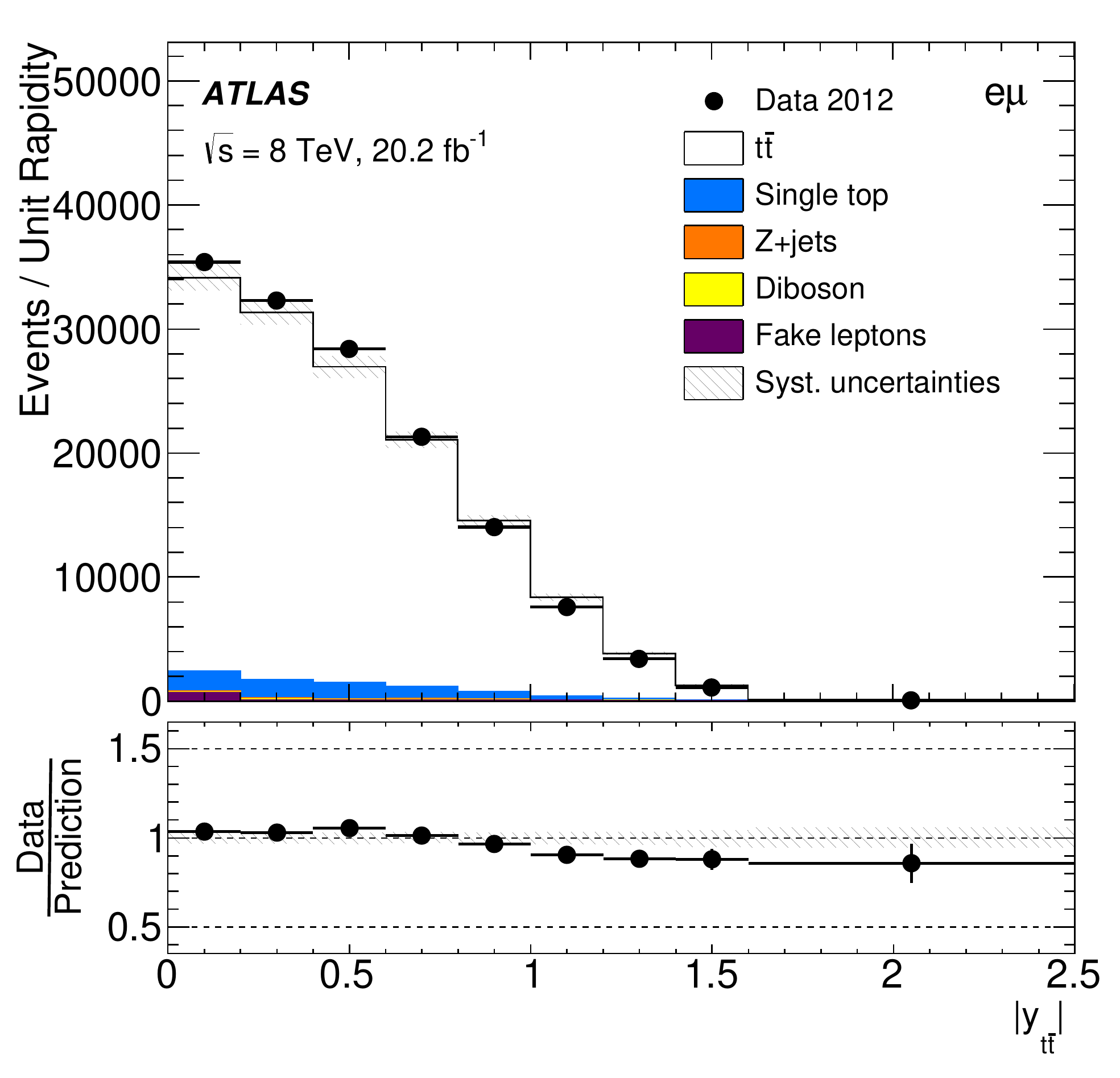}
 \label{fig:cplots_ytt}
}
\caption{
   Distributions of
   \protect\subref{fig:cplots_Mtt} the invariant mass,
   \protect\subref{fig:cplots_pTtt} the transverse momentum, 
   and 
   \protect\subref{fig:cplots_ytt} the rapidity of the $\ttbar$ system 
   at the reconstruction level
   obtained from the $\sqrt{s}$ = 8 TeV data
   compared with the total signal and background predictions.
   The bottom panel shows the ratio of data to prediction.
   The error band includes all systematic uncertainties except $\ttbar$ modeling    uncertainties. 
   The {\sc Powheg}+{\sc Pythia} with $h_{\rm damp}=\mt$ sample is used for the signal $\ttbar$ and is normalized to NNLO+NNLL calculations.
}
 \label{fig:cplots}
\end{figure}

\section{Differential cross-section determination}
\label{sec:unfolding}

The normalized differential cross-sections with respect to the $\ttbar$-system observables,
denoted as $X$, are obtained as follows.
The estimated background contributions are subtracted from the observed number of events for each bin in the distribution of the reconstructed observable.
The background-subtracted distributions are then corrected for detector acceptance and resolution effects (unfolded) 
and the efficiency to pass the event selection, 
thus extrapolated to the full phase space of $\ttbar$ production at parton level.
The differential cross-sections are finally normalized by the total $\ttbar$ cross-section,
obtained by integrating over all bins for each observable.

The differential cross-section is obtained from
\begin{equation}
  \frac{d\sigma_{\ttbar}}{dX_{i}} = \frac{1}{\Delta X_{i}\cdot {\cal L}\cdot \sum_{\alpha}({\cal B}^{\alpha} \cdot \epsilon_{i}^{\alpha})} \sum_{\alpha}\sum_{j}({\cal M}^{-1}_{ij})^{\alpha}(N^{{\rm obs}, \alpha}_{j}-N^{{\rm bkg}, \alpha}_{j}),
\label{eq:unfold}
\end{equation}
where
$i$ ($j$) indicates the bin for the observable $X$ at parton (detector) level,
$N^{\rm obs}_{j}$ is the number of observed events in data,
$N^{\rm bkg}_{j}$ is the estimated number of background events,
${\cal M}^{-1}_{ij}$ is the inverse of the migration matrix representing the correction for detector resolution effects,
$\epsilon_{i}$ is the event selection efficiency with respect to the channel,
${\cal B}$ is the branching ratio of the $\ttbar$ decays in the dilepton channel,
${\cal L}$ is the integrated luminosity, 
$\Delta X_{i}$ is the bin width, 
and $\alpha$ is the dilepton channel being considered, 
where $\alpha$ = $\diel$, $\dimu$ or $\emu$ for 7$\TeV$ 
and $\alpha$ = $\emu$ for 8$\TeV$. 
The measured cross-section at each bin $i$ represents the bin-averaged value at the bin.
The normalized differential cross-section is obtained as 
$1/\sigma_{\ttbar} \cdot d\sigma_{\ttbar}/dX_{i}$, 
where $\sigma_{\ttbar}$ is the inclusive \ttbar{} cross-section.

The unfolding from reconstruction level to parton level
is carried out using the \verb|RooUnfold| package~\cite{Adye:2011gm}
with an iterative method inspired by Bayes' theorem~\cite{bayesianUnfold}.
The number of iterations used in the unfolding procedure balances the goodness of fit and statistical uncertainties. 
The smallest number of iterations with $\chi^2$/NDF ($\chi^2$ between the unfolded and parton-level spectra over number of degrees of freedom) less than one is chosen for the distribution. 
In the 7 TeV analysis, two to four iterations are used depending on the observable; 
in the 8 TeV analysis, four iterations are used for all observables. 
The effect of varying the number of iterations by one was tested and confirmed to be negligible. 

The detector response is described using a migration matrix
that relates the generated parton-level distributions to the measured distributions.
The migration matrix ${\cal M}$ is determined using $\ttbar$ Monte Carlo samples,
where the parton-level top quark is defined as the top quark
after radiation and before decay.\footnote{
The generator status code for the top or antitop quark is required to be 3 in {\sc Pythia} and 155 in {\sc Herwig}.}
Figure~\ref{fig:resp} presents the migration matrices of $\ptttbar$ for both 7 TeV and 8 TeV in the $\emu$ channel. 
The matrix ${\cal M}_{ij}$ represents the probability for an event generated at parton level with $X$ in bin $i$ to have a reconstructed $X$ in bin $j$,
so the elements of each row add up to unity (within rounding uncertainties).
The probability for the parton-level events to remain in the same bin in the measured distribution is shown in the diagonal, and the off-diagonal elements represent the fraction of parton-level events that migrate into other bins. 
The fraction of events in the diagonal bins are the highest for $\ptttbar$, 
while for other observables more significant migrations are present due to the effect of $p_{z}$ of the undetected neutrinos in the reconstruction. In the 7 TeV analysis, the effect of bin migrations in the $\diel$ and $\dimu$ channels are similar to those in the $\emu$ channel. 
In the 8 TeV analysis, the bin boundaries for $\mttbar$ and $\absyttbar$ are determined separately for the parton-level and reconstruction-level observables, based on the migrations between them.

The event selection efficiency $\epsilon_i$ for each bin $i$ is evaluated as the ratio of the parton-level spectra before and after implementing the event selection at the reconstruction level. 
In both the 7 TeV and 8 TeV analyses, 
the efficiencies generally increase towards higher $\mttbar$ and $\ptttbar$,
while at high values of $\absyttbar$
the efficiency decreases
due to leptons and jets falling outside the required pseudorapidity range for reconstructed leptons and jets. 
The efficiencies are 
typically in the range of 15--20\%
for the $\emu$ channel at both 7 and 8 TeV,
and 
3--5\% and 8--13\% 
for the $\diel$ and $\dimu$ channels, respectively, in the 7 TeV analysis.
The lower values in the same-flavor channels are due to the rejection cuts for Drell-Yan and $Z\rightarrow\ell\ell$ events in these channels, 
while isolation requirements that are more restrictive for electrons than for muons in 7 TeV analyses 
result in further lowered efficiencies in the $\diel$ channel. 

The bin width for each observable is determined
by considering the resolution of the observable and the statistical precision in each bin.
In the 7 TeV analysis,  
the bin widths are set to be the same as the ones used in the previous 7$\TeV$ ATLAS measurement in the $\ell$+jets channel~\cite{TOPQ-2012-08} due to comparable resolutions for each observable, 
and to enable a direct comparison of the results between the two channels. 
For the 8 TeV analysis,
the determined bin widths are generally finer than the bin widths for the 7 TeV analysis
due to the larger dataset available.

Possible biases due to the use of the MC generator in the unfolding procedure are assessed by altering the shape of the parton-level spectra in simulation using continuous functions. 
The altered shapes studied cover the difference observed between the default MC and data for each observable. 
These studies verify that the altered shapes are recovered by the unfolding based on the nominal migration matrices within statistical uncertainties.

A multichannel combination is performed in the 7$\TeV$ analysis
by summing the background-subtracted observed events 
corrected by the migration matrix and the event selection efficiency over channels.
The results obtained from the combined dilepton channel are consistent with those from the individual channels. 

\begin{figure}[htbp]
\captionsetup[subfloat]{farskip=2pt,captionskip=1pt}
 \centering
 \subfloat[$\ptttbar$ for 7$\TeV$ analysis in $\emu$ channel]{
  \includegraphics[width=0.33\textwidth]{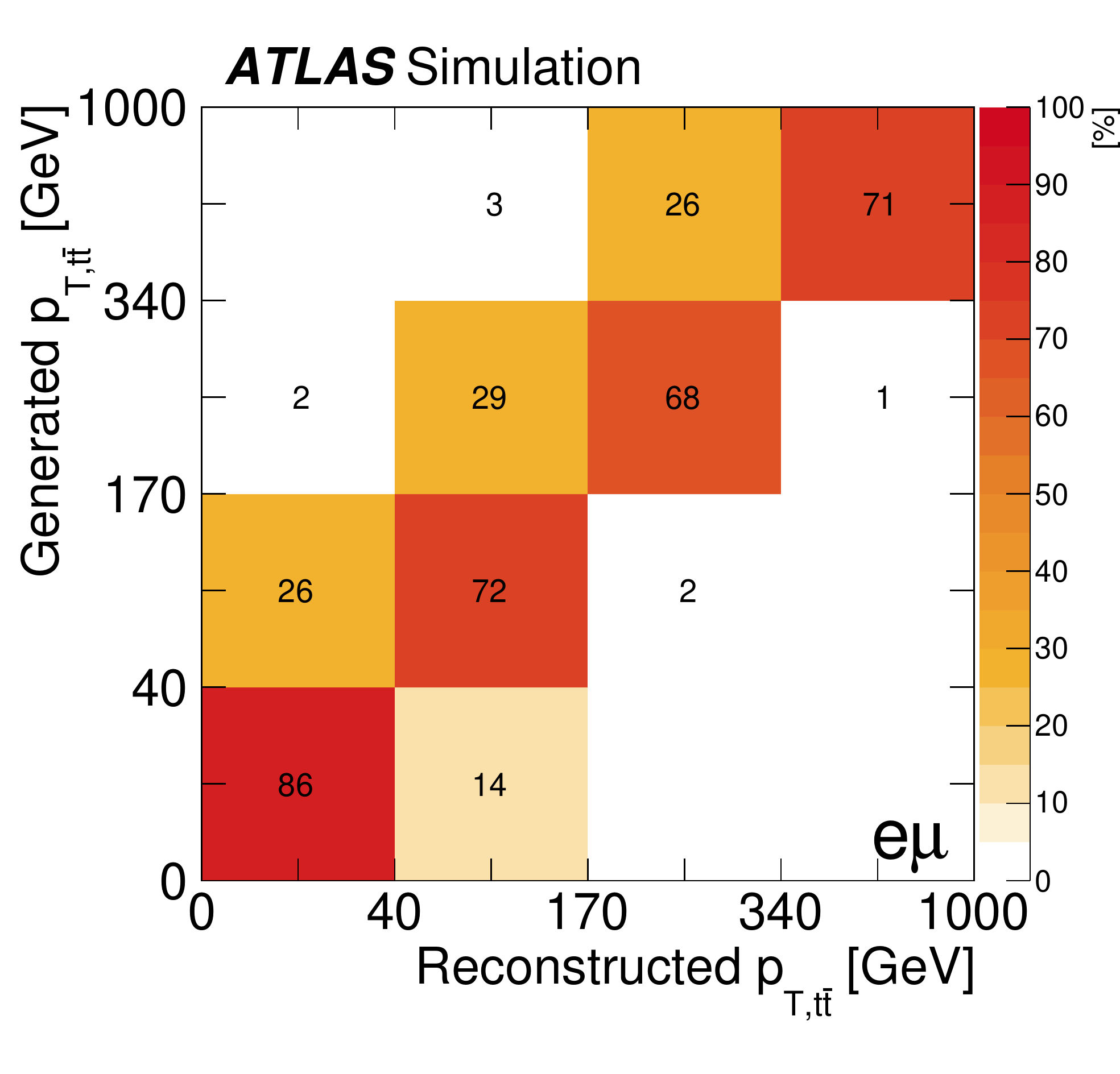} 
  \label{fig:resp_pttt_emu_7TeV}
 }
 \subfloat[$\ptttbar$ for 8$\TeV$ analysis  in $\emu$ channel]{
  \includegraphics[width=0.33\textwidth]{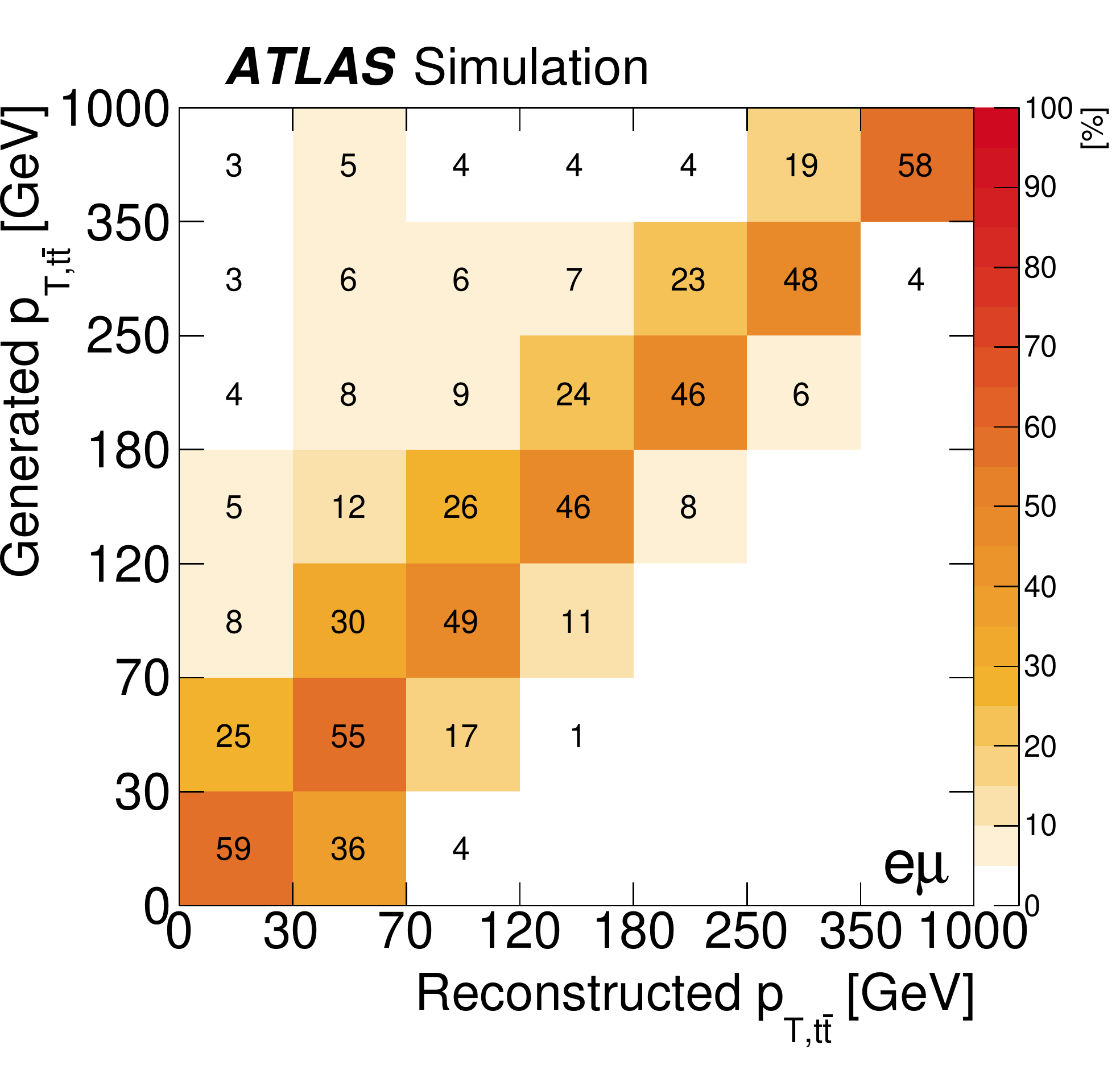} 
  \label{fig:resp_pttt_emu_8TeV}
 }
 \caption{The migration matrix of 
$\ptttbar$ represented in probability for (a) 7$\TeV$ and (b) 8$\TeV$ in the $\emu$ channel, 
obtained from $\ttbar$ simulation with the {\sc Powheg+Pythia} generator.
Different $h_{\rm damp}$ parameters are used at 7 TeV ($h_{\rm damp}=\infty$) and 8 TeV ($h_{\rm damp}=\mt$) in the {\sc Powheg+Pythia} sample, 
where the effect of the different $h_{\rm damp}$ in the migration matrix is negligible. 
Elements in each row add up to unity.
Empty elements indicate
either a probability of lower than 0.5\%
or no events are present.
}
 \label{fig:resp}
\end{figure}

\section{Uncertainties}
\label{sec:uncertainties}

Various sources of systematic uncertainty affect the measurement and are discussed below.
The systematic uncertainties due to signal modeling and detector modeling 
affect the estimation of the detector response and the signal reconstruction efficiency.
The systematic uncertainties due to the background estimation and the detector modeling
affect the background subtraction.

The covariance matrix due to the statistical and systematic uncertainties for each normalized unfolded spectrum
is obtained by evaluating the correlations between the bins for each uncertainty contribution. 
In particular, 
the correlations due to statistical fluctuations are evaluated from an ensemble of pseudoexperiments, each by
varying the data event counts independently in each bin and propagating the variations through the unfolding procedure.

\subsection{Signal modeling uncertainties}

The signal modeling uncertainties are estimated by repeating the full analysis procedure, 
using an alternative MC sample to derive the migration matrix
and the corrections for selection efficiency.
The differences between the results obtained using the alternative and nominal MC samples
are taken as systematic uncertainties.

At $\sqrt{s}$ = 7 TeV, the uncertainties due to the choice of generator are estimated by comparing 
{\sc Powheg+Pythia} and {\sc MC@NLO+Herwig} signal MC samples.
The uncertainty is found to be up to 2\% in $\mttbar$ and $\absyttbar$, and in the range of 2--19\% in $\ptttbar$ with larger values with increasing $\ptttbar$, due to the difference at the parton level between the two MC $\ttbar$ samples in the high $\ptttbar$ region. 
At \rts = 8 TeV, the uncertainties related to the generator
are estimated using {\sc Powheg+Herwig} and {\sc MC@NLO+Herwig} signal MC samples, 
and the uncertainties due to parton shower and hadronization
are estimated using {\sc Powheg+Pythia} and {\sc Powheg+Herwig} signal MC samples.
These uncertainties are typically less than 10\% (3\%) in $\mttbar$ and $\ptttbar$ ($\absyttbar$),
and increase to 20\% at large $\mttbar$ in the case of generator uncertainty.
 
The effects due to modeling of extra radiation in $\ttbar$ events
are assessed at both the matrix element and parton shower levels.
At $\sqrt{s}$ = 7 TeV, the uncertainty due to matrix element renormalization and factorization scales 
is evaluated using {\sc MC@NLO+Herwig} samples with varied renormalization/factorization scales, 
and the uncertainty due to parton showering in different initial-state and final-state radiation (ISR/FSR) conditions 
is estimated using two different {\sc Alpgen+Pythia} samples
with varied radiation settings.
The overall effects in both cases are less than 1\% in $\absyttbar$ and up to 6\% for $\mttbar$ and $\ptttbar$ with the larger values towards higher values of $\mttbar$ and $\ptttbar$.
At $\sqrt{s}$ = 8 TeV, the treatment of these uncertainties was improved by using {\sc Powheg+Pythia} samples with 
tuned parameters to span the variations in radiation compatible with the ATLAS $\ttbar$ gap fraction measurements 
at $\sqrt{s}$ = 7 TeV~\cite{TOPQ-2011-21} as discussed in detail in Ref.~\cite{ATL-PHYS-PUB-2014-005}. 
The samples have varied renormalization/factorization scales and $h_{\rm damp}$ parameter values, 
resulting in either more or less radiation than the nominal signal sample.
The overall impact is typically less than 2\% for all observables, and up to 4\% towards higher values of $\ptttbar$.

The uncertainties due to the choice of PDFs,
which affect most significantly the signal selection efficiency,
are estimated
based on the PDF4LHC recommendations~\cite{Botje:2011sn}
using the {\sc MC@NLO+Herwig} sample with three different NLO PDF sets: 
CT10~\cite{CT10}, MSTW2008nlo68cl~\cite{MSTW}, and NNPDF2.3~\cite{NNPDF}.
An intra-PDF uncertainty is obtained for each PDF set
by following its respective prescription while an inter-PDF
uncertainty is computed as the envelope of the three intra-PDF uncertainties.
The overall effect is less than 2\% for all observables in both the 7 TeV and 8 TeV measurements 
(except for the highest $\absyttbar$ bin at 8 TeV where the effect is up to 8\%). 

The dependence of the $\ttbar$-system observables on the top quark mass $\mt$ is evaluated at $\sqrt{s}=$ 7 TeV using $\ttbar$ samples with different mass points at 170 GeV and 175 GeV to unfold the data, 
then the difference of the results at the two mass points is taken and divided by the difference $\Delta \mt$
to extract the difference of the differential cross-section per GeV change of $\Delta \mt$. 
These studies show that the dependence of the differential cross-sections on the $\mt$ is 
no more than 1\% per GeV for all kinematic observables.
These variations are not included in the total uncertainty. 

\subsection{Background modeling uncertainties}

Uncertainties arising from the background estimates are evaluated by repeating 
the full analysis procedure, varying the background contributions by $\pm 1\sigma$
from the nominal values.
The differences between the results obtained using the nominal and the varied background estimations
are taken as systematic uncertainties.

The uncertainties due to the $Wt$ background modeling are estimated 
by comparing the inclusive ``diagram removal'' and inclusive ``diagram subtraction'' samples.
The uncertainty is typically below 1\%, 
except for high $\mttbar$ and $\ptttbar$ bins where the uncertainty is up to about 5\% and 2\%, respectively. 

The relative uncertainties of 7.7\% (7 TeV) and 6.8\% (8 TeV) in the predicted cross-section of $Wt$ production are applied in all bins of the differential cross-sections. 
An uncertainty of 5\% is assigned to the predicted diboson cross-section, with an additional uncertainty of 24\% per additional selected jet added in quadrature to account for the assumption that the ($W$ + $n$ + 1 jets)/($W$ + $n$ jets) ratio is constant~\cite{Alwall:2007fs,TOPQ-2010-01}. 
The overall impact of these uncertainties is less than 1\%.

For the $Z$+jets background, in the $\emu$ channel only the $Z(\rightarrow\tau\tau)$+jets process contributes,
while the $Z(\rightarrow\diel)$+jets ($Z(\rightarrow\dimu)$+jets) process contributes only to the $\diel$ ($\dimu$) channel. 
An inclusive uncertainty of 4\% is assigned to the predicted cross-section of $Z(\rightarrow\tau\tau)$+jets, 
with an additional uncertainty of 24\% per additional selected jet added in quadrature.
The $Z(\rightarrow\diel/\dimu)$+jets background is estimated by a data-driven method~\cite{TOPQ-2010-01,TOPQ-2011-01} that uses a control region populated with $Z$ events. The uncertainty is evaluated by varying the control region (defined by $|m_{\ell\ell}-m_Z|<10\GeV$ and $\Etmiss>30\GeV$) by $\pm$5 GeV in $\Etmiss$.
The overall impact of these uncertainties is less than 1\% in both the 7 TeV and 8 TeV measurements.

The fake-lepton contribution is estimated directly from data, 
using a matrix method~\cite{TOPQ-2010-01} in 7 TeV data
and the same-sign dilepton events in the 8 TeV data sample~\cite{TOPQ-2013-04}.
In the 7 TeV analysis, 
the uncertainty of the fake-lepton background is evaluated by 
considering the uncertainties in the real- and fake-lepton efficiency measurements
and by comparing results obtained from different matrix methods. 
In the 8 TeV analysis a conservative uncertainty of 50\% is assigned to the fake-lepton background~\cite{TOPQ-2013-04}. 
The impact of the uncertainty is typically less than 1\% in all observables, except in high-$\mttbar$ and high-$\ptttbar$ bins where it is up to 
5\%. 

\subsection{Detector modeling uncertainties}

The uncertainties due to the detector modeling
are estimated for each bin based on the methods described in Ref.~\cite{TOPQ-2013-04}.
They affect the detector response including signal reconstruction efficiency
and the estimation of background events that passed all event selections and their kinematic distribution.
The full analysis procedure is repeated with the varied detector modeling, and the difference between the results using the nominal and the varied modeling is taken as a systematic uncertainty.

The lepton reconstruction efficiency in simulation
is calibrated by correction factors derived from measurements
of these efficiencies in data using control regions enriched in $Z\rightarrow\ell\ell$ events.
The lepton trigger and reconstruction efficiency correction factors,
energy scale, and resolution 
are varied within the uncertainties in the $Z\to \ell\ell$ measurements~\cite{PERF-2010-04,ATLAS-CONF-2013-088}.

The jet energy scale (JES) uncertainty is derived using a combination of simulations, test beam data and
{\it in situ} measurements~\cite{PERF-2012-01,PERF-2011-03,PERF-2011-05}.
Additional contributions from the jet flavor composition, calorimeter response
to different jet flavors, and pileup are taken into account.
Uncertainties in the jet energy resolution are
obtained with an {\it in situ} measurement of the jet response balance
in dijet events~\cite{PERF-2011-04}.

The difference in $b$-tagging efficiency between data and MC simulation is estimated
in lepton+jets $\ttbar$ events
with the selected jet containing a $b$-hadron on the leptonic side~\cite{ATLAS-CONF-2012-097}.
Correction factors are also applied for jets originating from light hadrons that are
misidentified as jets containing $b$-hadrons.
The associated systematic uncertainties are computed by varying the correction factors
within their uncertainties.

The uncertainty associated with \MET{}
is calculated by propagating the energy scale and resolution
systematic uncertainties to all jets and leptons in the \MET{} calculation.
Additional \MET{} uncertainties arising from energy deposits not associated
with any reconstructed objects are also included~\cite{PERF-2011-07}.

The uncertainty due to the finite size of the MC simulated samples are evaluated
by varying the content of the migration matrix with a Poisson distribution.
The standard deviation of the ensemble of results unfolded with the varied matrices is taken as the uncertainty.
The effect is more significant in the 7 TeV analysis (up to 3\% in high-$\mttbar$ and high-$\ptttbar$ bins),
due to the smaller size of the MC simulation sample available at 7 TeV.
In the 8 TeV analysis, while the MC statistical uncertainty is less significant (sub-percent overall),
an additional uncertainty is included to account for the bias introduced by the unfolding procedure
due to the observed deviation between data and the predicted $\ttbar$ events.
The typical size of the bias is less than 1\%, and increases towards higher $\mttbar$, $\ptttbar$, and $\absyttbar$ up to about 4\%.
The bias in the 7 TeV analysis is taken into account
by choosing an unfolding parameter based on the level of bias for an observable, 
which is reflected in the data statistical uncertainty and thus not included as a systematic uncertainty.

The uncertainty in the integrated luminosity
is estimated to be 
1.8\% for \rts = 7\,\TeV{}~\cite{DAPR-2011-01}
and 1.9\% for \rts = 8\,\TeV{}~\cite{DAPR-2013-01}.
The effect of the uncertainty is substantially reduced in the normalized differential cross-sections due to large bin-to-bin correlations.

\subsection{Summary of the main sources of systematic uncertainty}

For $\mttbar$, the largest systematic uncertainties come from 
signal modeling (including generator choice, 
parton showering and hadronization, 
and extra radiation),
JES, 
and $Wt$ background modeling (at large $\mttbar$). 
The uncertainty due to signal modeling in $\mttbar$ is generally smaller at 7 TeV because of the requirement on the jet-lepton invariant mass, which reduces the fraction of wrong-jet events used to reconstruct the $\ttbar$-system, is applied in the 7 TeV analysis but not in the 8 TeV analysis. 
For $\ptttbar$, 
the uncertainty from signal modeling (including generator choice, 
parton showering and hadronization, 
and extra radiation)
is the largest, followed by JES.
The main uncertainties for $\absyttbar$ come from PDF and signal generator choice. 

\section{Results}
\label{sec:result}

The unfolded parton-level normalized differential cross-sections for $\sqrt{s}=7\TeV$ and $\sqrt{s}=8\TeV$ are shown
in Table~\ref{tab:summary_dxs_norm_ll} 
and Table~\ref{tab:xsec_summary_norm}, respectively.
The total inclusive $\ttbar$ cross-sections, evaluated by integrating the spectra before the normalization, agree with the theoretical calculations and other inclusive measurements within uncertainties at both energies. 
The estimated uncertainties include all sources discussed in Section~\ref{sec:uncertainties}. 

Comparisons of the data distributions with different SM predictions 
are quantified by computing $\chi^2$ values 
and inferring $p$-values (probability of obtaining a $\chi^2$ is larger than or equal to the observed value) 
from the $\chi^2$ values and the number of degrees of freedom (NDF).
The $\chi^2$ is defined as
\begin{equation}
 \chi^{2}=V^{\rm T}\cdot {\rm Cov}^{-1}\cdot V
\end{equation}
where $V$ is the vector of the differences between the data and the theoretical predictions,
and ${\rm Cov}^{-1}$ is the inverse of the full bin-to-bin covariance matrix.
Due to the normalization constraint in the derivation of normalized differential cross-sections,
the NDF and the rank of the covariance matrix
is reduced by one unit to $N_{\rm b}-1$, 
where $N_{\rm b}$ is the number of bins in the spectrum being considered.
Consequently, one of the $N_{\rm b}$ elements in $V$ 
and the corresponding row and column in the $N_{\rm b} \times N_{\rm b}$ full covariance matrix ${\rm Cov}$ 
is discarded, 
and the $N_{\rm b}-1 \times N_{\rm b}-1$ submatrix obtained in this way is invertible, 
allowing the $\chi^2$ to be computed. 
The $\chi^2$ value does not depend on which element is discarded from the vector $V_{N_{\rm b}-1}$
and the corresponding sub-matrix ${\rm Cov}_{N_{\rm b}-1}$.
The evaluation of $\chi^2$ under the normalization constraint 
follows the same procedure as described in Refs.~\cite{TOPQ-2012-08,TOPQ-2015-06}.

The comparison of the measured normalized distributions to 
predictions from different MC generators of $\ttbar$ production are shown graphically in 
Figure~\ref{fig:dxs_norm_mc_ll} for $\sqrt{s}=7\TeV$
and 
Figure~\ref{fig:Normalized} for $\sqrt{s}=8\TeV$, with the corresponding $p$-values comparing the measured spectra to the predictions from the MC generators in Table~\ref{tab:chi2_dxs_norm_ll} and Table~\ref{tab:chi2_norm}.
Predictions from 
{\sc Powheg+Pythia} with $h_{\rm damp} = \mt$,
{\sc MC@NLO+Herwig},
{\sc Powheg+Pythia} with $h_{\rm damp} = \infty$,
and {\sc Powheg+Herwig} 
are used for comparison with data. 
In the 7 TeV analysis, {\sc Alpgen+Herwig} is also used for the comparison, 
as it was the default sample used in the differential measurement in the $\ell$+jets channel by ATLAS~\cite{TOPQ-2012-08}.
Both NLO generators ({\sc Powheg} and {\sc MC@NLO}) use the NLO CT10~\cite{CT10} PDF set, while {\sc Alpgen+Herwig} uses the LO CTEQ6L1~\cite{cteq} PDF set. 

Most of the generators agree with data in a wide kinematic range of the distributions.
The $\mttbar$ spectrum is well described by most of the generators
at both 7 TeV and 8 TeV, 
except for {\sc Powheg+Pythia} 
in the highest $\mttbar$ bin in the 7 TeV analysis.
For $\ptttbar$, agreement with {\sc Powheg+Pythia} with $h_{\rm damp} = \infty$ is particularly bad
due to a harder $\ptttbar$ spectrum than data at both 7 TeV and 8 TeV. 
Better agreement with data is obtained from {\sc Powheg+Pythia} with $h_{\rm damp} = \mt$. 
This is consistent with the studies in Refs.~\cite{ATL-PHYS-PUB-2015-002,ATL-PHYS-PUB-2015-011} 
using data from the $\sqrt{s}$ = 7 TeV ATLAS parton-level measurement in the $\ell$+jets channel~\cite{TOPQ-2012-08}.
In both the 7 TeV and 8 TeV analyses, {\sc MC@NLO+Herwig} describes the $\ptttbar$ spectrum well also.
Similar good agreement is also observed in 7 TeV and 8 TeV parton-level measurements by ATLAS in the $\ell$+jets channel~\cite{TOPQ-2012-08,TOPQ-2015-06}. 
For $\absyttbar$, all the generators show fair agreement with data in the 7 TeV analysis,
while at 8 TeV, none of the generators provides an adequate description of $\absyttbar$. 
This difference in the level of agreement is due to the improved statistical precision and finer binning in $\absyttbar$ for the 8 TeV analysis. 
The increasing discrepancy between data and MC prediction with increasing $\absyttbar$ is also observed at the reconstructed level for both energies, as shown in Figure~\ref{fig:ttbar_system_opt2} and Figure~\ref{fig:cplots}. 
This observation is also consistent with the results of the ATLAS differential cross-section measurements in the $\ell$+jets channel, at both 7 and 8 TeV\cite{TOPQ-2012-08,TOPQ-2015-06}.

Figure~\ref{fig:Normalized4PDF} shows the normalized differential cross-sections at $\sqrt{s}=8\TeV$ 
compared with the predictions of {\sc MC@NLO+Herwig} reweighted with different PDF sets:
CT10, MSTW2008nlo68cl, NNPDF2.3, and HERAPDF15NLO.
The hatched bands show the uncertainty of each PDF set.
All predictions are compatible with the measured cross-sections within the uncertainties in the
cases of $\mttbar$ and $\ptttbar$.
However, for $\absyttbar$,
the {\sc MC@NLO+Herwig} sample with the CT10 PDF set does not agree with the measured cross-sections
at $\absyttbar \sim 1.6$.
Using NNPDF or HERAPDF significantly improves the agreement.
The corresponding $p$-values are shown in Table~\ref{tab:chi2_norm_pdf}. 

Figure~\ref{fig:Normalized3IFSR} and Table~\ref{tab:chi2_norm_ifsr} show the comparison of the measured normalized differential cross-sections at $\sqrt{s}=8\TeV$ 
to {\sc Powheg+Pythia} with different levels of radiation. 
The nominal sample (with $h_{\rm damp}=\mt$) and two other samples, one with lower radiation ($h_{\rm damp}=\mt$ and $\mu=2.0$) and one with higher radiation ($h_{\rm damp}=2.0\mt$ and $\mu=0.5$) than the nominal one, are used in the comparison.
The $\ptttbar$ spectrum, particularly sensitive to radiation activity, shows that
the nominal sample has better agreement with data.
This observation is also consistent with the studies in Refs.~\cite{ATL-PHYS-PUB-2015-002,ATL-PHYS-PUB-2015-011}. 

The parton-level measured distributions are also compared to fixed-order QCD calculations. 
Figure~\ref{fig:thll_dxs_norm_mtt_pttt} and Figure~\ref{fig:Normalized_NLO} show 
the comparison with theoretical QCD NLO+NNLL predictions
for $\mttbar$~\cite{NLO_NNLL_mttbar} 
and $\ptttbar$~\cite{NLO_NNLL_ptttbar1,NLO_NNLL_ptttbar2}
distributions at $\sqrt{s}=7\TeV$ and $\sqrt{s}=8\TeV$, respectively,
and the corresponding $p$-values are given in Table~\ref{tab:chi2_dxs_norm_ll_th}.
The predictions are calculated using the mass of the $\ttbar$ system as the dynamic scale of the process
and the MSTW2008nnlo PDF~\cite{MSTW} set.
The NLO+NNLL calculation shows a good agreement in the $\mttbar$ spectrum
and a large discrepancy for high values of $\ptttbar$ in measurements at both $\sqrt{s}=7\TeV$ and $\sqrt{s}=8\TeV$.
Figure~\ref{fig:Normalized_NNLO} shows
the comparison of a full NNLO calculation~\cite{Czakon:2015owf} to the $\mttbar$ and $|\yttbar|$ measurements at $\sqrt{s}$ = 8 TeV.
The full NNLO calculation is evaluated using 
the fixed 
scale $\mu=\mt$
and the MSTW2008nnlo PDF~\cite{MSTW}.
The range of the NNLO prediction does not fully cover the highest bins in $\mttbar$ and $|\yttbar|$ and thus no prediction is shown in those bins.

The $\sqrt{s}=7\TeV$ results, 
together with previous results reported in $\ell$+jets channel by ATLAS~\cite{TOPQ-2012-08}, 
are summarized with the SM predictions in Figure~\ref{fig:summary_all_dxs_norm}.
This direct comparison can be performed due to the same bin widths of the $\ttbar$-system observables used in both analyses. 
All distributions are plotted as ratios with respect to dilepton channel results. 
The normalized results from both the dilepton and $\ell$+jets channels are consistent with each other in all $\ttbar$-system variables within the uncertainties of the measurements.

\begin{table}[htbp]
\centering
\sisetup{round-mode=figures, round-precision=2, retain-explicit-plus=true, separate-uncertainty, group-integer-digits=true, group-decimal-digits=false, group-four-digits=true}
\begin{tabular}{cccc}
  \toprule
{$\mttbar$ [GeV]} & {$\frac{1}{\sigma} \frac{d\sigma}{d\mttbar}$ [$10^{-3}\GeV^{-1}$]} & {Stat. [\%]} & {Syst. [\%]}  \\
  \midrule
250--450 & \numRP{2.41426}{2}$\pm$\numRP{0.0802242}{2} & $\pm$\num{1.59179} & $\pm$\num{2.91686}  \\
450--550 & \numRP{2.78669}{2}$\pm$\numRP{0.0485375}{2} & $\pm$\num{1.40859} & $\pm$\num{1.02449}  \\
550--700 & \numRP{1.09332}{2}$\pm$\numRP{0.0604522}{2} & $\pm$\num{3.09392} & $\pm$\num{4.58259}  \\
700--950 & \numRP{0.251502}{3}$\pm$\numRP{0.0231462}{3} & $\pm$\num{5.73509} & $\pm$\num{7.19774}  \\
950--2700 & \numRP{0.00663185}{4}$\pm$\numRP{0.00141049}{4} & $\pm$\num{16.2267} & $\pm$\num{13.7493}  \\
& & & \\
  \toprule
{$\ptttbar$ [GeV]} & {$\frac{1}{\sigma} \frac{d\sigma}{d\ptttbar}$ [$10^{-3}\GeV^{-1}$]} & {Stat. [\%]} & {Syst. [\%]}  \\
  \midrule
0--40 & \numRP{13.492}{1}$\pm$\numRP{0.653202}{1} & $\pm$\num{1.18917} & $\pm$\num{4.6931}  \\
40--170 & \numRP{3.14436}{2}$\pm$\numRP{0.167031}{2} & $\pm$\num{1.48227} & $\pm$\num{5.10108}  \\
170--340 & \numRP{0.268961}{3}$\pm$\numRP{0.0327834}{3} & $\pm$\num{6.14172} & $\pm$\num{10.5284}  \\
340--1000 & \numRP{0.00883629}{4}$\pm$\numRP{0.00256648}{4} & $\pm$\num{18.6101} & $\pm$\num{22.2994}  \\
& & & \\
  \toprule
{$|\yttbar|$ } & {$\frac{1}{\sigma} \frac{d\sigma}{d\absyttbar}$ } & {Stat. [\%]} & {Syst. [\%]}  \\
  \midrule
0--0.5 & \numRP{0.825575}{3}$\pm$\numRP{0.0194582}{3} & $\pm$\num{1.89434} & $\pm$\num{1.40235}  \\
0.5--1 & \numRP{0.64306}{3}$\pm$\numRP{0.0179726}{3} & $\pm$\num{1.84211} & $\pm$\num{2.10187}  \\
1--2.5 & \numRP{0.177121}{3}$\pm$\numRP{0.00729457}{3} & $\pm$\num{2.8409} & $\pm$\num{2.9817}  \\
  \bottomrule
\end{tabular}
\label{tab:summary_dxs_norm_abyttbar_ll}
\caption{Normalized $\ttbar$ differential cross-sections for the different $\ttbar$ kinematic variables at $\sqrt{s}$ = 7 TeV. 
The cross-sections in the last bins include events (if any) beyond of the bin edges. 
The uncertainties quoted in the second column represent the statistical and systematic uncertainties added in quadrature.
}
\label{tab:summary_dxs_norm_ll}
\end{table}

\begin{table}[htbp]
 \begin{center}
 \sisetup{round-mode=figures, round-precision=2, retain-explicit-plus=true, separate-uncertainty, group-integer-digits=true, group-decimal-digits=false, group-four-digits=true}
 \begin{tabular}{cccc}
 \toprule
 {$\mttbar$ [GeV]} & {$\frac{1}{\sigma} \frac{d\sigma}{d\mttbar}$ [$10^{-3}\GeV^{-1}$]} & {Stat. [\%]} & {Syst. [\%]}  \\
 \midrule
 250--450 & \numRP{2.4074325}{2}$\pm$\numRP{0.06527}{2} & $\pm$\num{1.13651} & $\pm$\num{6.0409} \\
 450--570 & \numRP{2.55699}{2}$\pm$\numRP{0.05376}{2} & $\pm$\num{1.0528} & $\pm$\num{1.8587} \\
 570--700 & \numRP{0.967977}{2}$\pm$\numRP{0.08342}{2} & $\pm$\num{1.6335} & $\pm$\num{8.4166} \\
 700--850 & \numRP{0.345951}{2}$\pm$\numRP{0.04515}{2} & $\pm$\num{2.53026} & $\pm$\num{12.6573} \\
 850--1000 & \numRP{0.128619}{3}$\pm$\numRP{0.02249}{3} & $\pm$\num{3.55797} & $\pm$\num{16.9644} \\
 1000--2700 & \numRP{0.00861843}{4}$\pm$\numRP{0.00243}{4} & $\pm$\num{6.58544} & $\pm$\num{23.4147} \\
& & & \\
 \toprule
 {$\ptttbar$ [GeV]} & {$\frac{1}{\sigma} \frac{d\sigma}{d\ptttbar}$ [$10^{-3}\GeV^{-1}$]} & {Stat. [\%]} & {Syst. [\%]}  \\
 \midrule
 0--30 & \numRP{14.3409}{1}$\pm$\numRP{1.0038}{1} & $\pm$\num{1.15801} & $\pm$\num{6.8662} \\
 30--70 & \numRP{7.59835}{2}$\pm$\numRP{0.1596}{2} & $\pm$\num{1.08914} & $\pm$\num{1.8510} \\
 70--120 & \numRP{2.94231}{2}$\pm$\numRP{0.2793}{2} & $\pm$\num{1.75359} & $\pm$\num{9.2875} \\
 120--180 & \numRP{1.13949}{2}$\pm$\numRP{0.11514}{2} & $\pm$\num{2.65273} & $\pm$\num{9.4565} \\
 180--250 & \numRP{0.423467}{2}$\pm$\numRP{0.0441}{2} & $\pm$\num{4.01751} & $\pm$\num{9.7130} \\
 250--350 & \numRP{0.142997}{3}$\pm$\numRP{0.01806}{3} & $\pm$\num{6.0417} & $\pm$\num{11.4005} \\
 350--1000 & \numRP{0.00986377}{4}$\pm$\numRP{0.00147}{4} & $\pm$\num{8.93817} & $\pm$\num{11.6486} \\
& & & \\
 \toprule
 {$\absyttbar$} & {$\frac{1}{\sigma} \frac{d\sigma}{d\yttbar}$} & {Stat. [\%]} & {Syst. [\%]}  \\
 \midrule
 0.0--0.4 & \numRP{0.821127}{3}$\pm$\numRP{0.02134938}{3} & $\pm$\num{1.29998} & $\pm$\num{2.2403} \\
 0.4--0.8 & \numRP{0.721364}{3}$\pm$\numRP{0.018034}{3} & $\pm$\num{1.34046} & $\pm$\num{2.1394} \\
 0.8--1.2 & \numRP{0.499483}{3}$\pm$\numRP{0.01298648}{3} & $\pm$\num{1.64534} & $\pm$\num{2.0287} \\
 1.2--2.0 & \numRP{0.206433}{3}$\pm$\numRP{0.0061929}{3} & $\pm$\num{2.38768} & $\pm$\num{1.8787} \\
 2.0--2.8 & \numRP{0.0225804}{4}$\pm$\numRP{0.00230316}{4} & $\pm$\num{8.26282} & $\pm$\num{9.878} \\
 \bottomrule
 \end{tabular}
 \caption{Normalized $\ttbar$ differential cross-sections for the different $\ttbar$ kinematic variables at $\sqrt{s}$ = 8 TeV.
The uncertainties quoted in the second column represent the statistical and systematic uncertainties added in quadrature.
}
 \label{tab:xsec_summary_norm}
 \end{center}
\end{table}

\clearpage 

\begin{figure}[htbp]
  \centering
  \subfloat[$\mttbar$]{
    \includegraphics[width=0.45\textwidth]{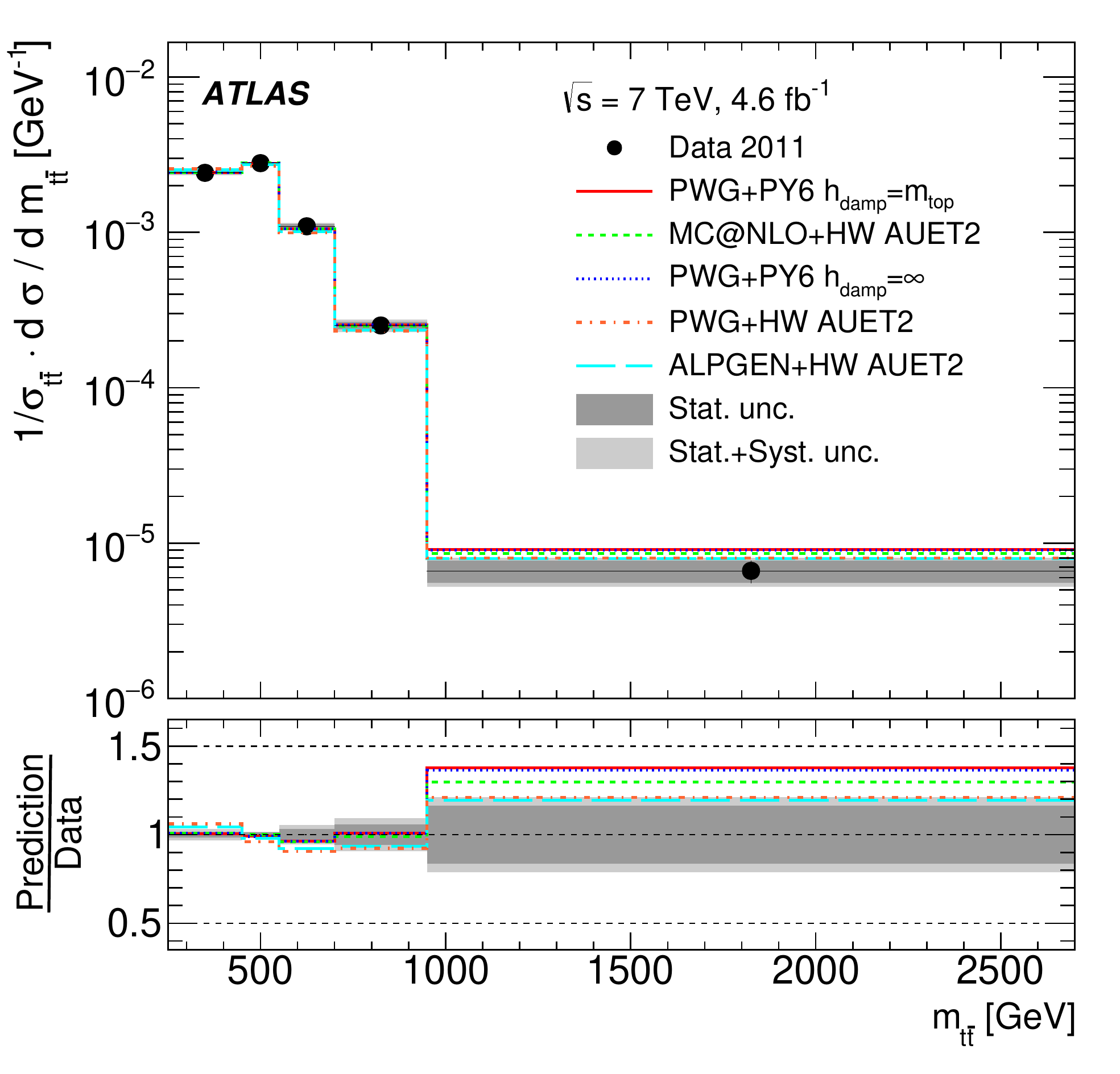} 
    \label{fig:dxs_mc_norm_ll_mtt}
  }
  \hspace{8pt}
  \subfloat[$\ptttbar$]{
    \includegraphics[width=0.45\textwidth]{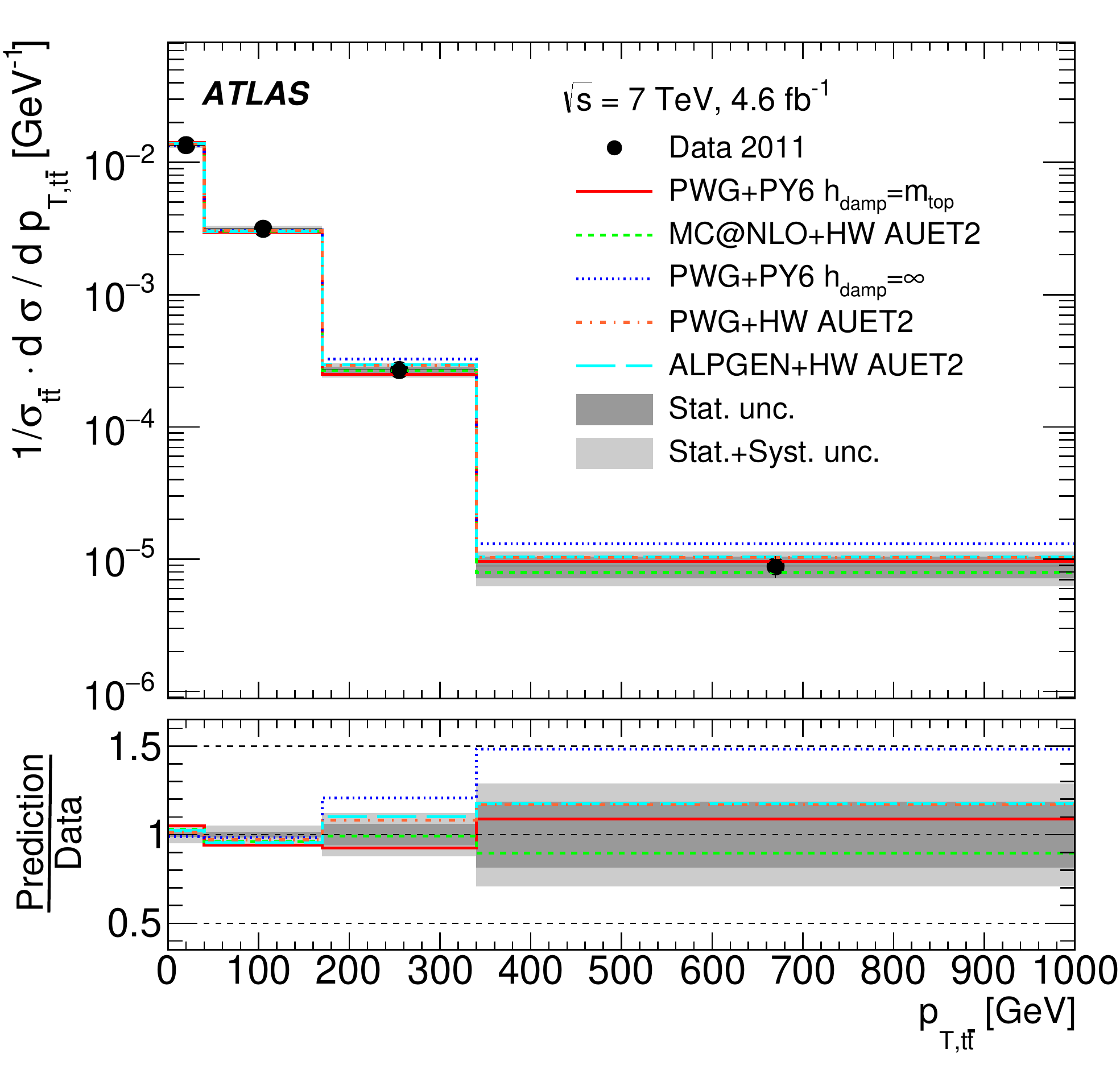} 
    \label{fig:dxs_mc_norm_ll_pttt}
  }
  \hspace{8pt}
  \subfloat[$\absyttbar$]{
    \includegraphics[width=0.45\textwidth]{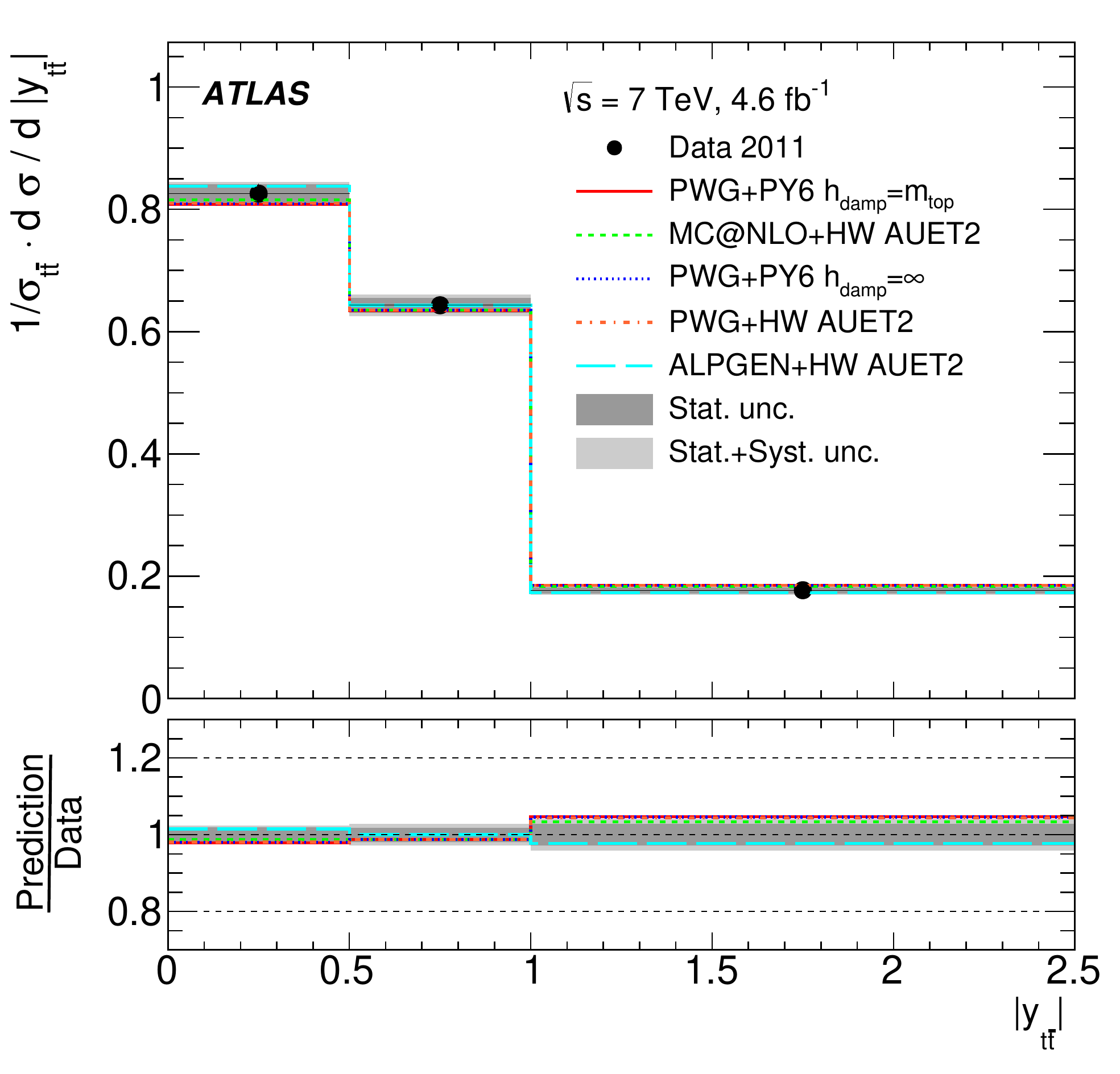} 
    \label{fig:dxs_mc_norm_ll_absytt}
  }
  \caption{
  Normalized $\ttbar$ differential cross-sections as a function of the 
  \protect\subref{fig:dxs_mc_norm_ll_mtt} invariant mass ($\mttbar$)
  \protect\subref{fig:dxs_mc_norm_ll_pttt} transverse momentum ($\ptttbar$)
  and \protect\subref{fig:dxs_mc_norm_ll_absytt} absolute value of the rapidity ($\absyttbar$)
  of the $\ttbar$ system at $\sqrt{s}=7$ TeV
  measured in the dilepton channel
  compared to theoretical predictions from MC generators.
  All generators use the NLO CT10~\cite{CT10} PDF, except for {\sc Alpgen+Herwig} using the LO CTEQ6L1 PDF. 
  The bottom panel shows the ratio of prediction to data. 
  The light (dark) gray band includes the total (data statistical) uncertainty in the data in each bin.
  }
  \label{fig:dxs_norm_mc_ll}
\end{figure}

\begin{figure}[htbp]
  \captionsetup[subfloat]{farskip=2pt,captionskip=1pt}
  \centering
  \subfloat[$\mttbar$]{
    \includegraphics[width=0.45\textwidth]{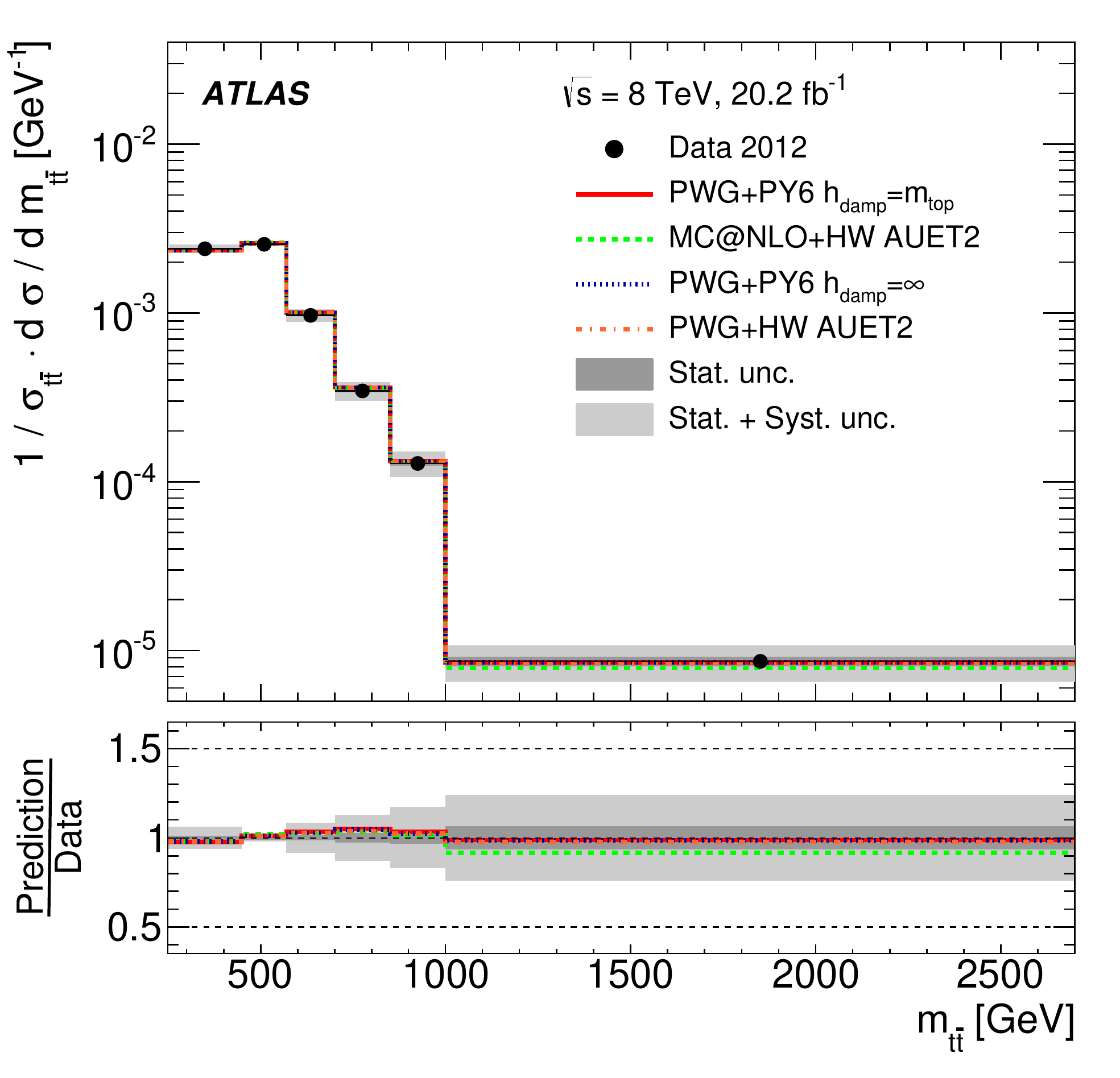}
    \label{fig:Normalized_Mtt}
  }
  \hspace{8pt}
  \subfloat[$\ptttbar$]{
    \includegraphics[width=0.45\textwidth]{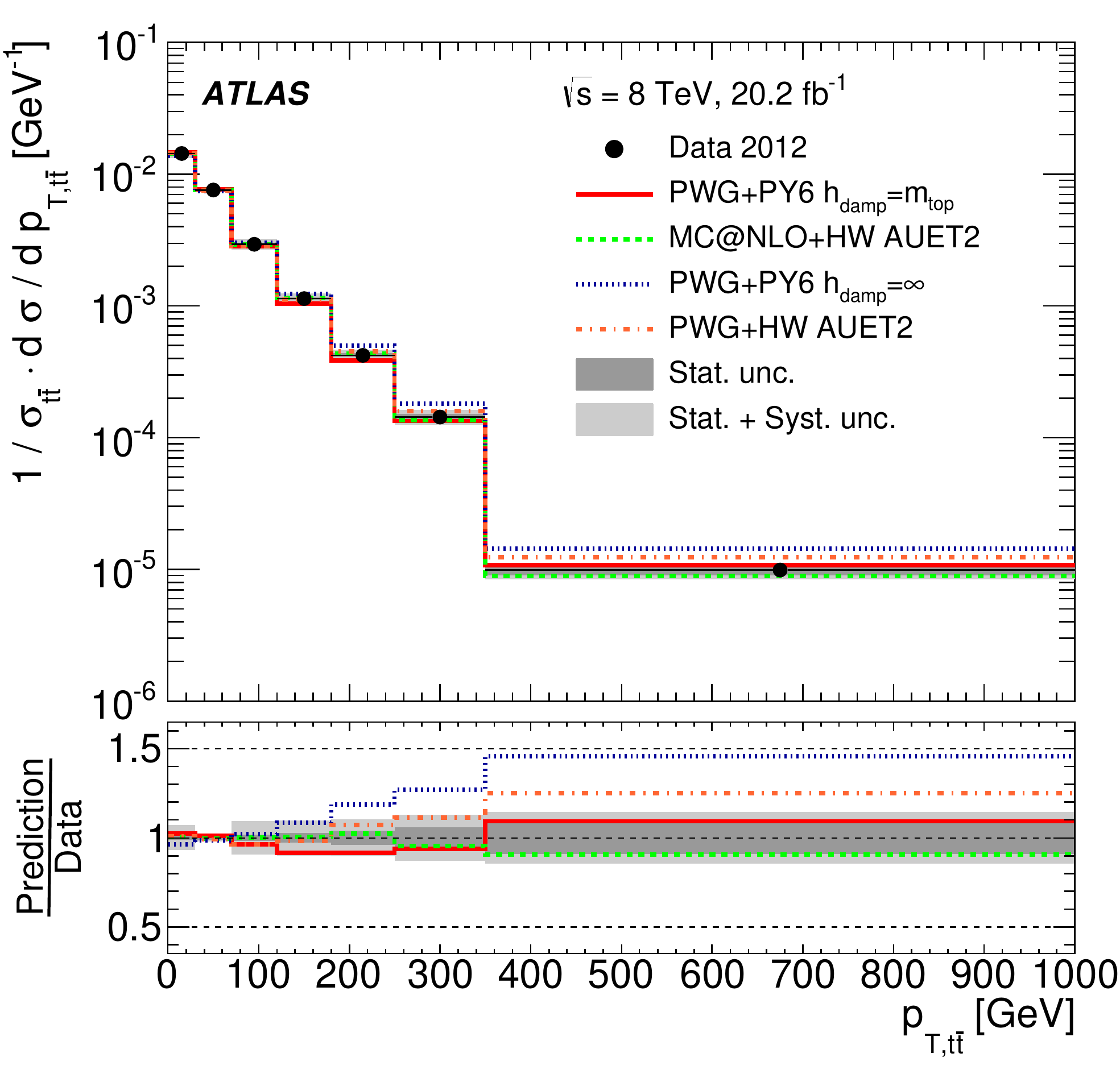}
    \label{fig:Normalized_pTtt}
  }
  \hspace{8pt}
  \subfloat[$\absyttbar$]{
    \includegraphics[width=0.45\textwidth]{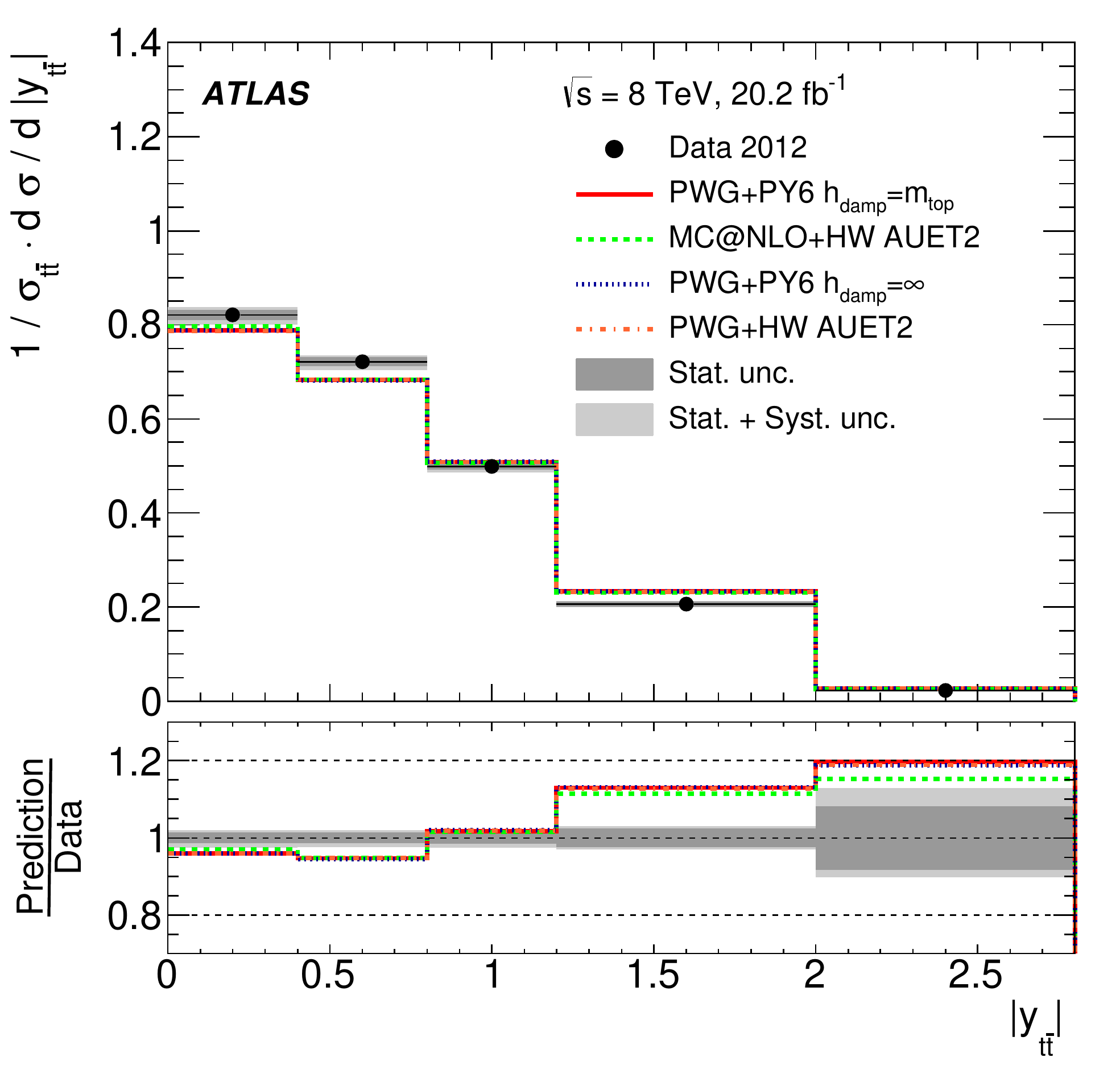}
    \label{fig:Normalized_ytt}
  }
  \caption{
  Normalized $\ttbar$ differential cross-sections as a function of the 
  \protect\subref{fig:Normalized_Mtt} invariant mass ($\mttbar$)
  \protect\subref{fig:Normalized_pTtt} transverse momentum ($\ptttbar$)
  and \protect\subref{fig:Normalized_ytt} absolute value of the rapidity ($\absyttbar$)
  of the $\ttbar$ system at $\sqrt{s}=8$ TeV
  measured in the dilepton $\emu$ channel
  compared to theoretical predictions from MC generators.
  All generators use the NLO CT10~\cite{CT10} PDF. 
  The bottom panel shows the ratio of prediction to data. 
  The light (dark) gray band includes the total (data statistical) uncertainty in the data in each bin.
 }
  \label{fig:Normalized}
\end{figure}

\begin{figure}[htbp]
  \captionsetup[subfloat]{farskip=2pt,captionskip=1pt}
  \centering
  \subfloat[$\mttbar$]{
    \includegraphics[width=0.45\textwidth]{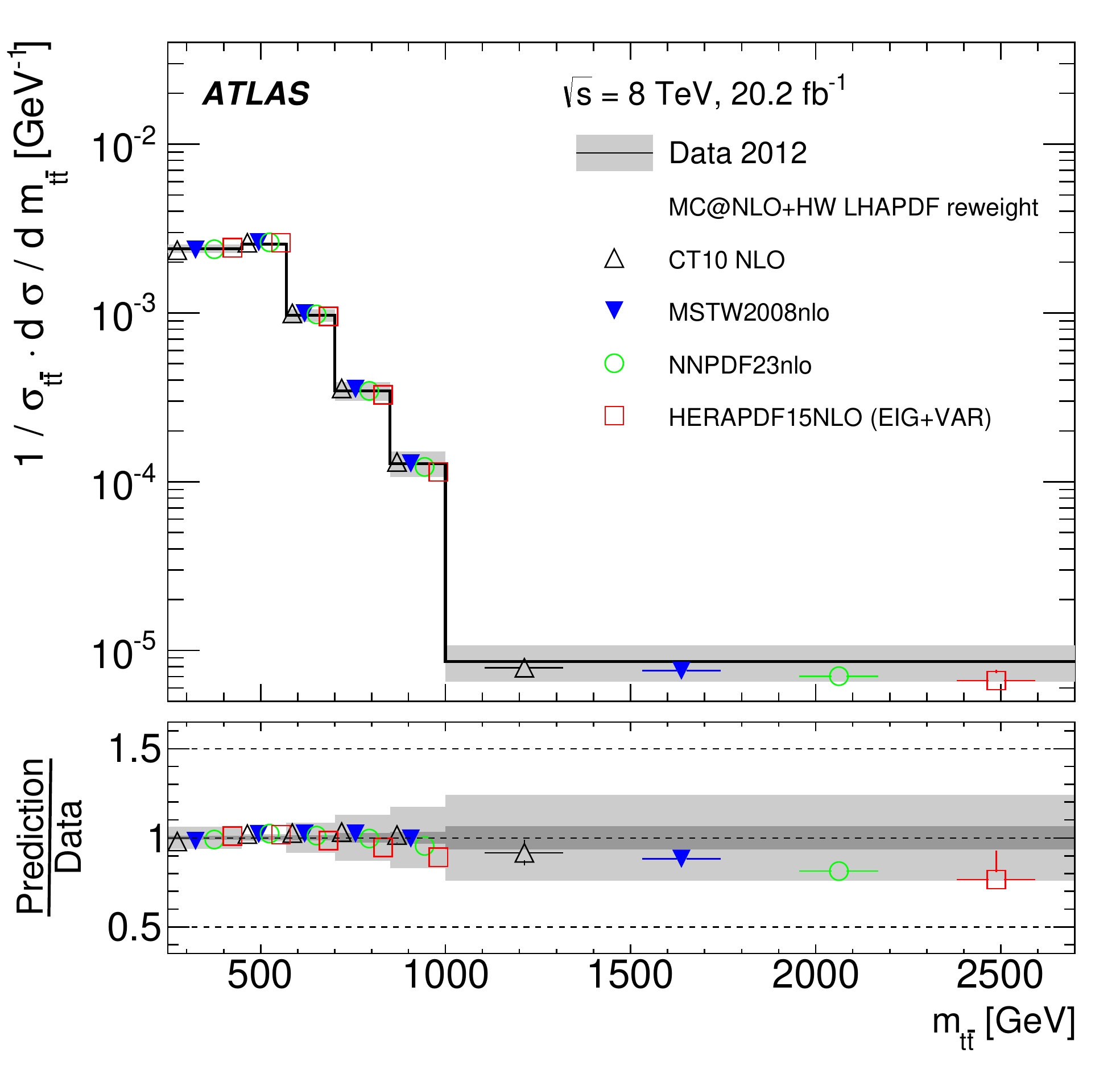}
    \label{fig:Normalized_PDF_Mtt}
  }
  \hspace{8pt}
  \subfloat[$\ptttbar$]{
    \includegraphics[width=0.45\textwidth]{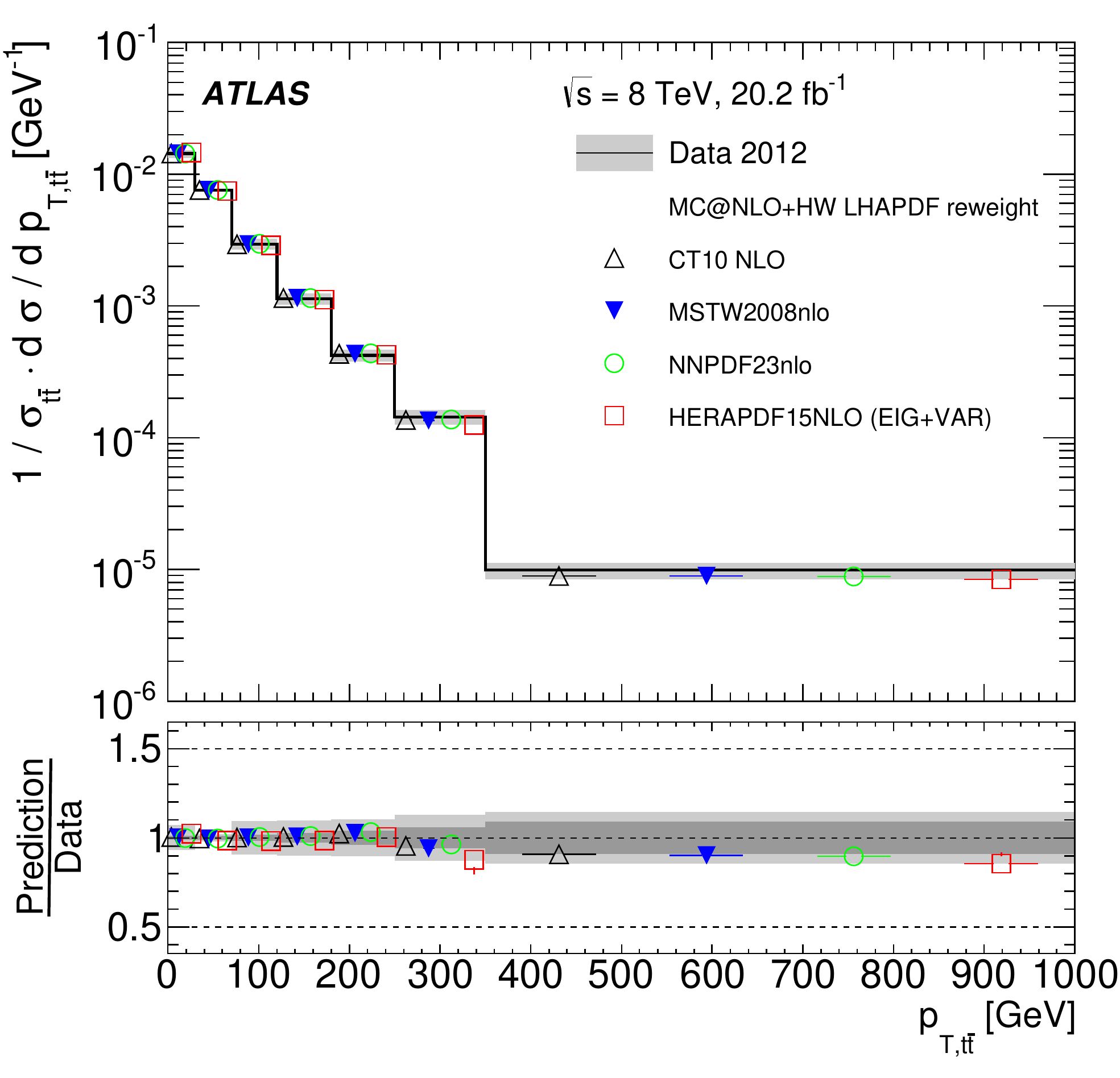}
    \label{fig:Normalized_PDF_pTtt}
  }
  \hspace{8pt}
  \subfloat[$\absyttbar$]{
    \includegraphics[width=0.45\textwidth]{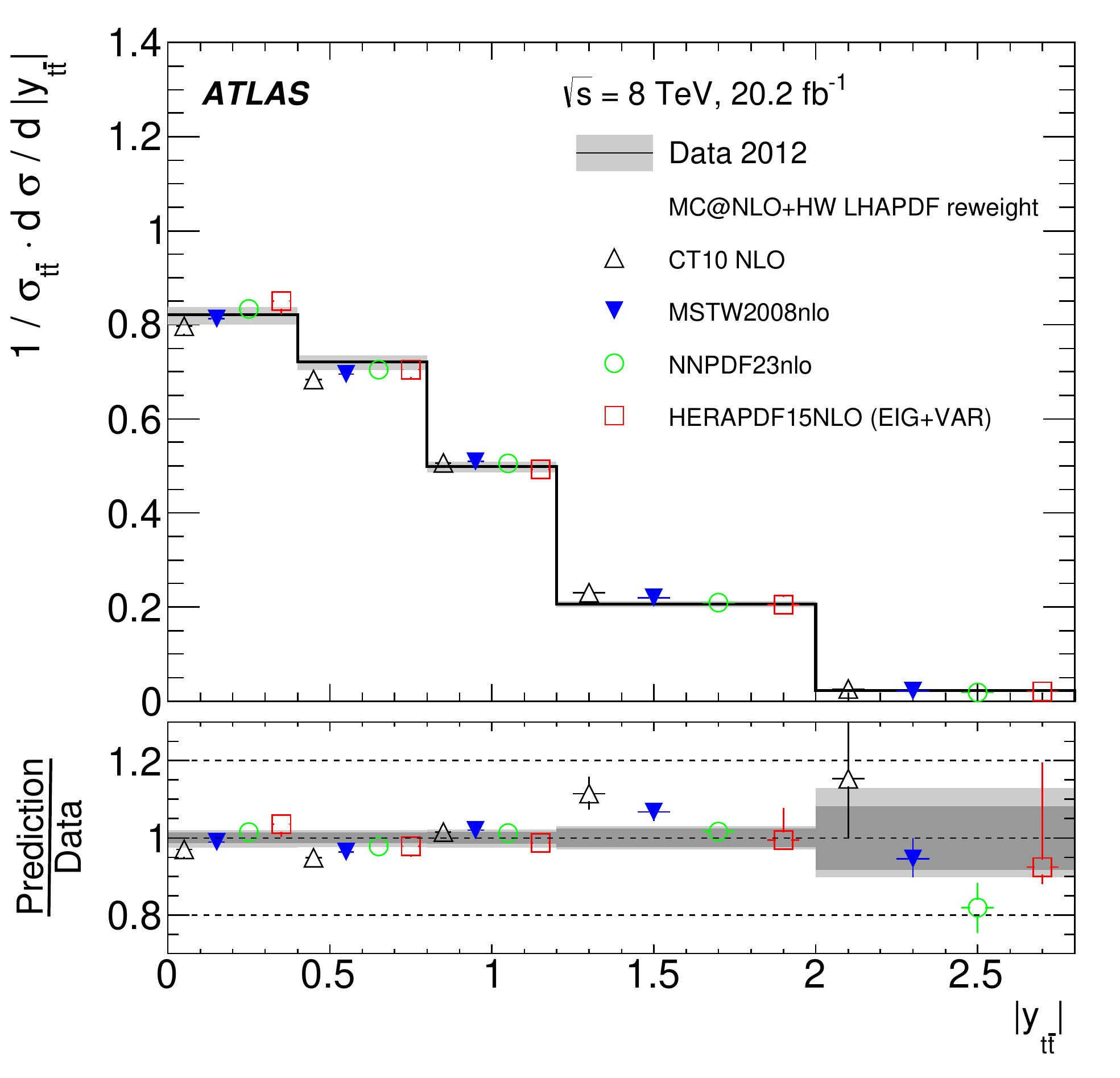}
    \label{fig:Normalized_PDF_ytt}
  }
  \caption{
  Normalized $\ttbar$ differential cross-sections as a function of the 
  \protect\subref{fig:Normalized_PDF_Mtt} invariant mass ($\mttbar$)
  \protect\subref{fig:Normalized_PDF_pTtt} transverse momentum ($\ptttbar$)
  and \protect\subref{fig:Normalized_PDF_ytt} absolute value of the rapidity ($\absyttbar$)
  of the $\ttbar$ system at $\sqrt{s}=8$ TeV
  measured in the dilepton $\emu$ channel
  compared to different PDF sets.
  The {\sc MC@NLO+Herwig} generator is reweighted using the PDF sets to produce the different predictions.
  The bottom panel shows the ratio of prediction to data. 
  The light (dark) gray band includes the total (data statistical) uncertainty in the data in each bin.
 }
  \label{fig:Normalized4PDF}
\end{figure}

\begin{figure}[htbp]
  \captionsetup[subfloat]{farskip=2pt,captionskip=1pt}
  \centering
  \subfloat[$\mttbar$]{
    \includegraphics[width=0.45\textwidth]{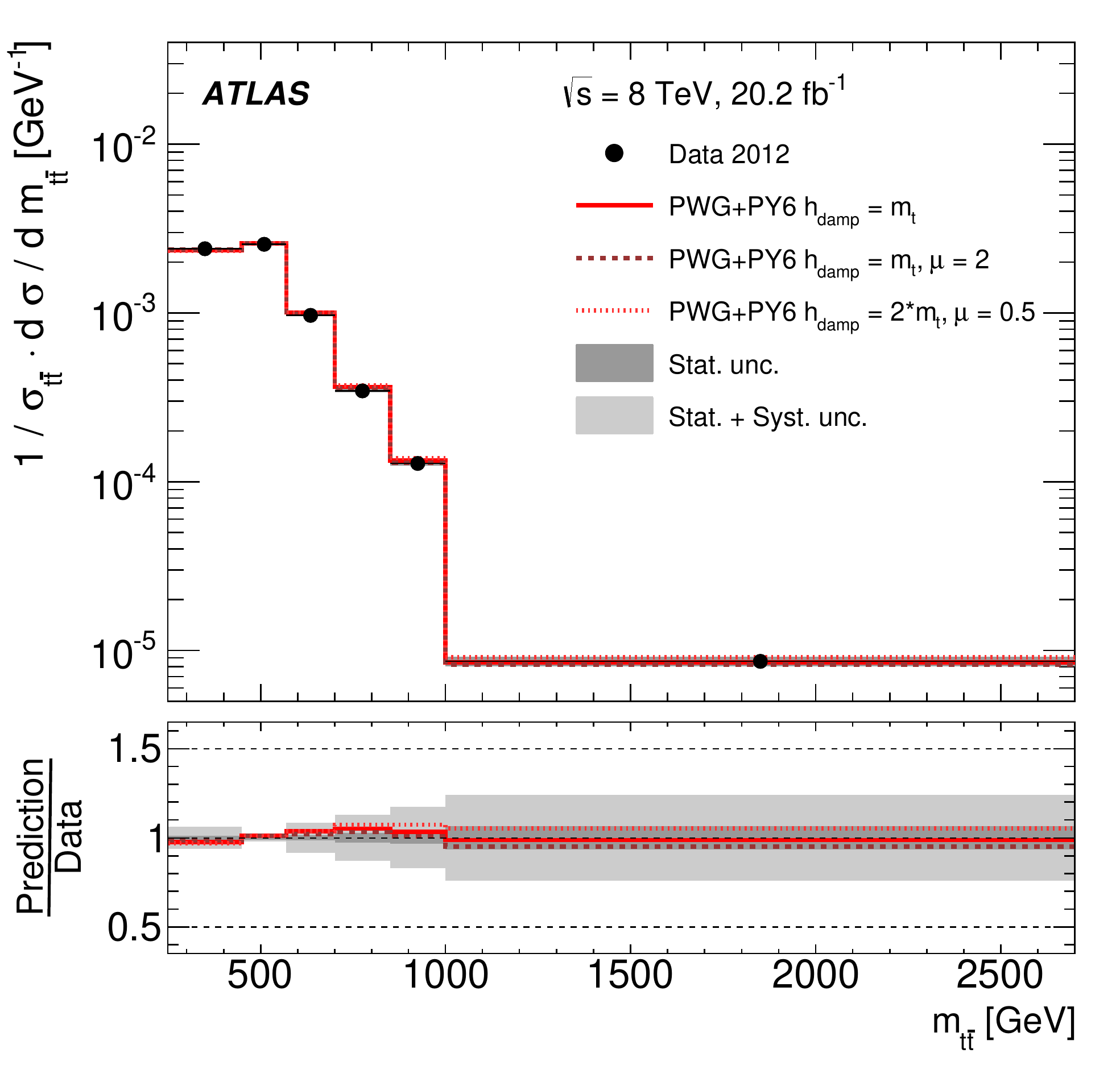}
    \label{fig:Normalized_IFSR_Mtt}
  }
  \hspace{8pt}
  \subfloat[$\ptttbar$]{
    \includegraphics[width=0.45\textwidth]{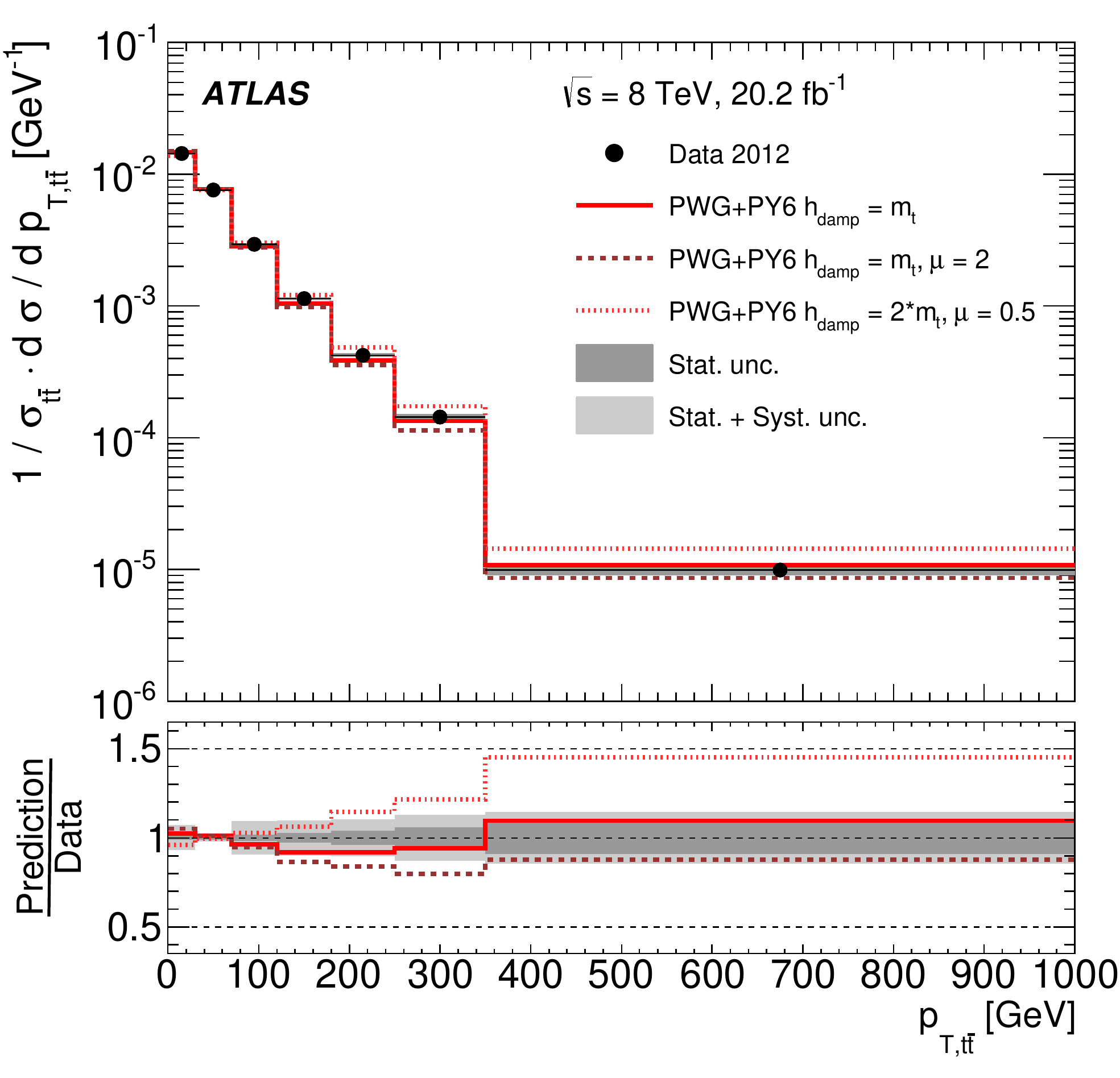}
    \label{fig:Normalized_IFSR_pTtt}
  }
  \hspace{8pt}
  \subfloat[$\absyttbar$]{
    \includegraphics[width=0.45\textwidth]{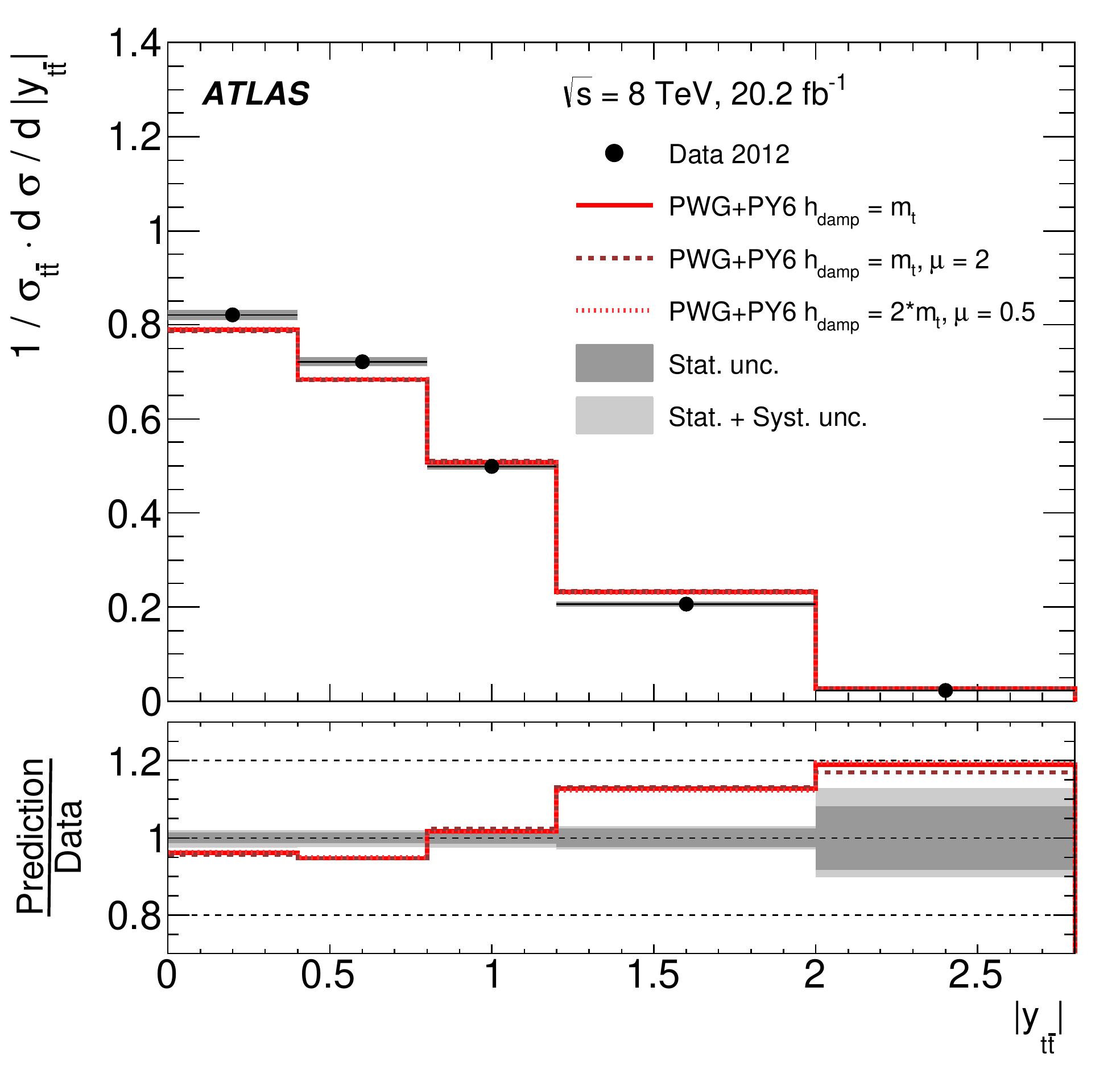}
    \label{fig:Normalized_IFSR_ytt}
  }
  \caption{
  Normalized $\ttbar$ differential cross-sections as a function of the 
  \protect\subref{fig:Normalized_IFSR_Mtt} invariant mass ($\mttbar$),
  \protect\subref{fig:Normalized_IFSR_pTtt} transverse momentum ($\ptttbar$),
  and \protect\subref{fig:Normalized_IFSR_ytt} absolute value of the rapidity ($\absyttbar$),
  of the $\ttbar$ system at $\sqrt{s}=8$ TeV
  measured in the dilepton $\emu$ channel
  compared to theoretical predictions from MC generators.
  The {\sc Powheg+Pythia} generator with different levels of radiation 
  are used for the predictions.
  All generators use the NLO CT10~\cite{CT10} PDF. 
  The bottom panel shows the ratio of prediction to data. 
  The light (dark) gray band includes the total (data statistical) uncertainty in the data in each bin.
    }
  \label{fig:Normalized3IFSR}
\end{figure}

\begin{figure}[htbp]
\centering
\subfloat[$\mttbar$]{
\includegraphics[width=0.45\textwidth]{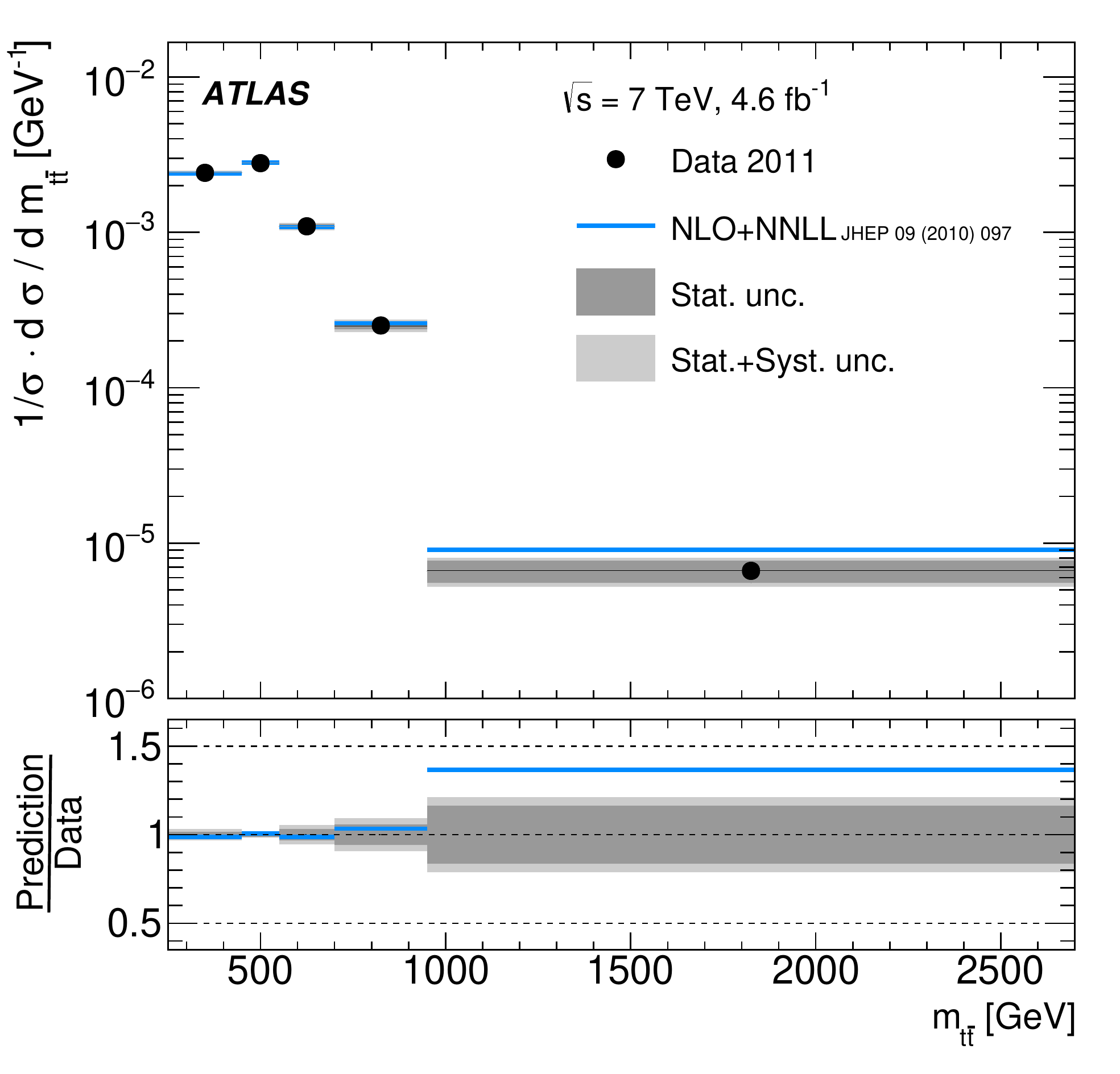} 
\label{fig:thll_dxs_norm_mtt}
}
\hspace{8pt}
\subfloat[$\ptttbar$]{
\includegraphics[width=0.45\textwidth]{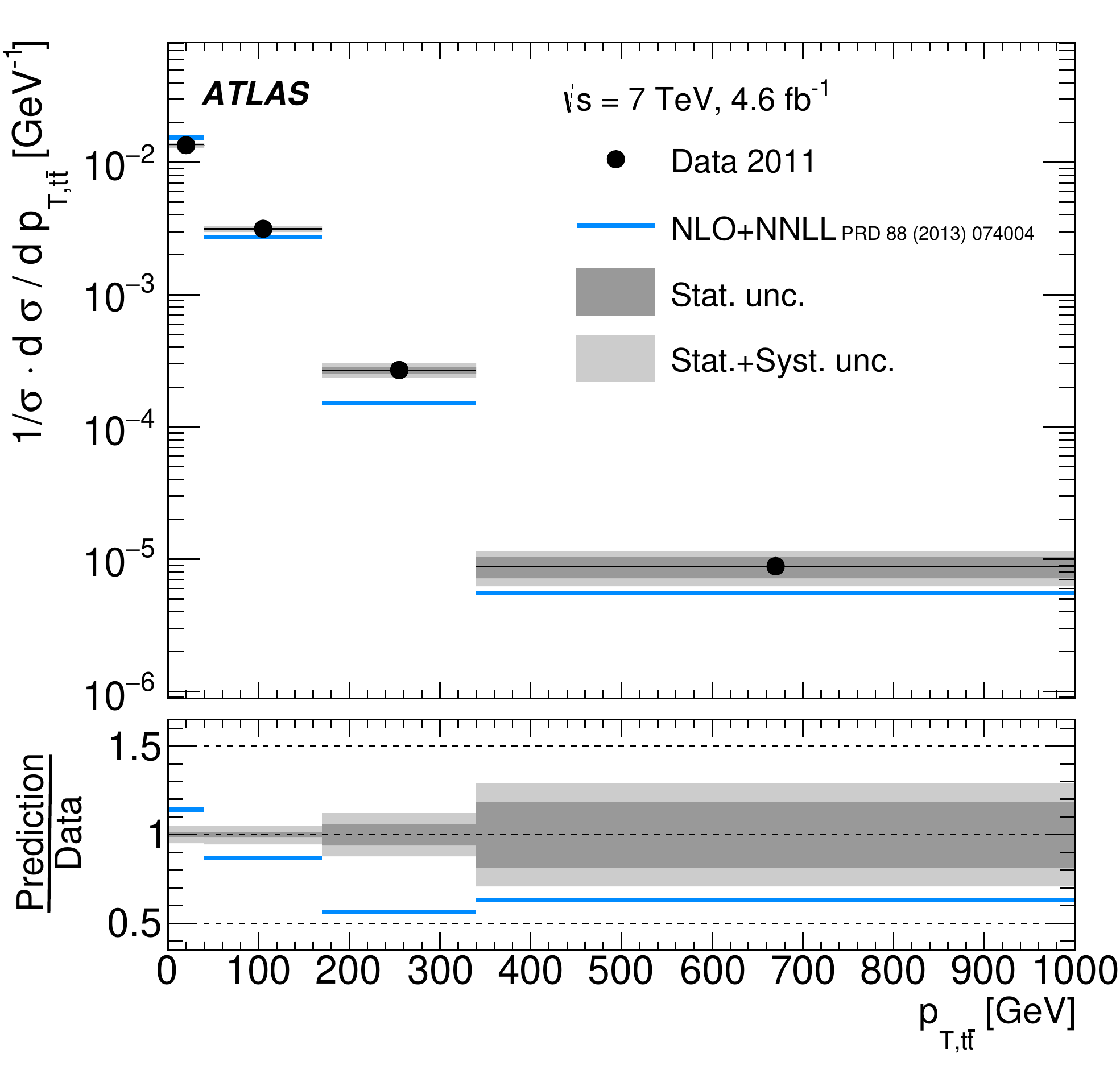} 
\label{fig:thll_dxs_norm_pttt}
}
\caption{
  Normalized $\ttbar$ differential cross-sections as a function of the 
  \protect\subref{fig:thll_dxs_norm_mtt} invariant mass ($\mttbar$)
  and \protect\subref{fig:thll_dxs_norm_pttt} transverse momentum ($\ptttbar$)
  of the $\ttbar$ system at $\sqrt{s}=7$ TeV 
  measured in the dilepton channel 
compared with theoretical QCD calculations at NLO+NNLL level. 
  The predictions are calculated using the MSTW2008nnlo PDF.
  The bottom panel shows the ratio of prediction to data. 
  The light (dark) gray band includes the total (data statistical) uncertainty in the data in each bin.
}
\label{fig:thll_dxs_norm_mtt_pttt}
\end{figure}

\begin{figure}[htbp]
\centering
\subfloat[$\mttbar$]{
\includegraphics[width=0.45\textwidth]{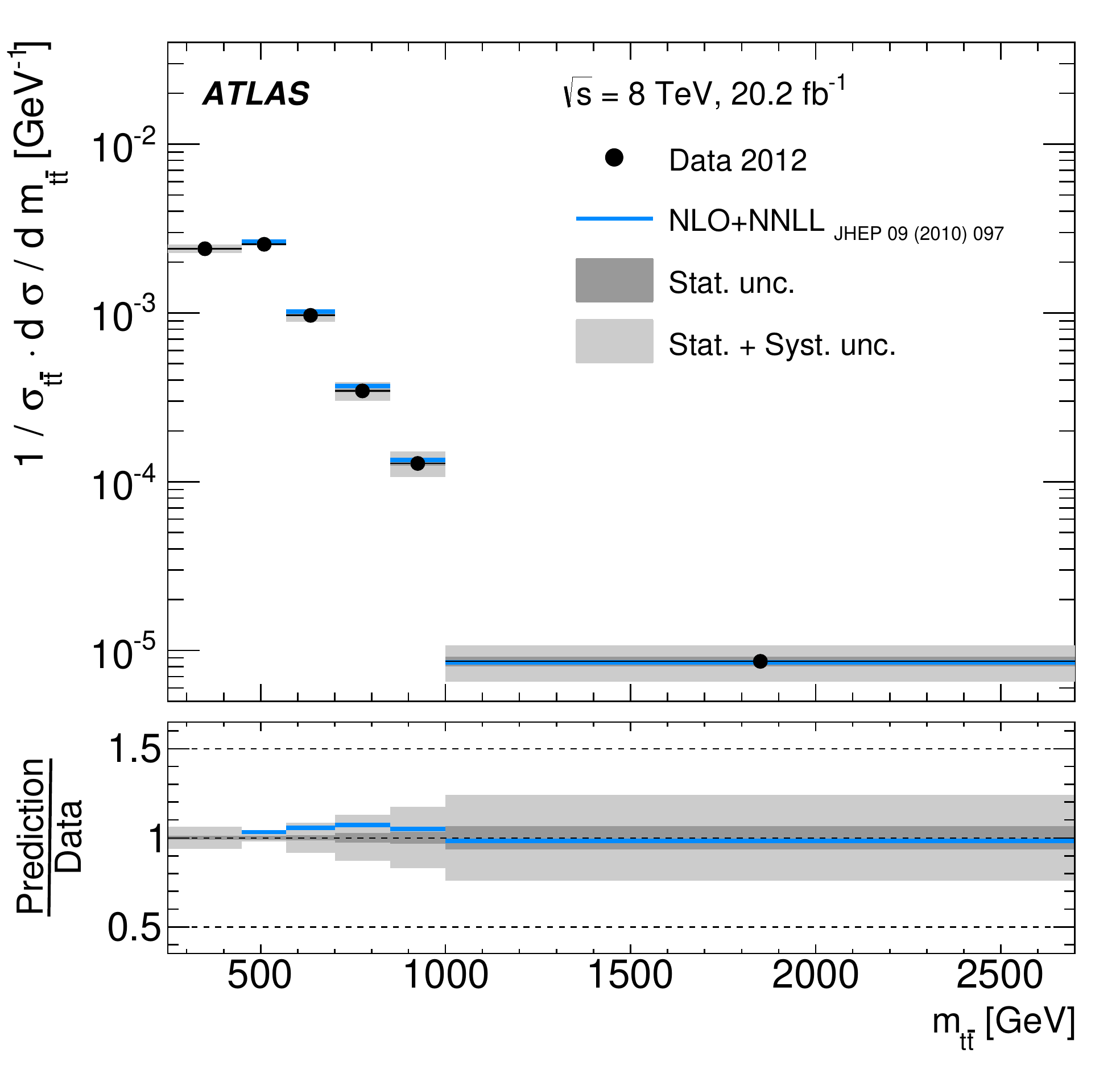} 
\label{fig:Normalized_NLO_Mtt}
}
\hspace{8pt}
\subfloat[$\ptttbar$]{
\includegraphics[width=0.45\textwidth]{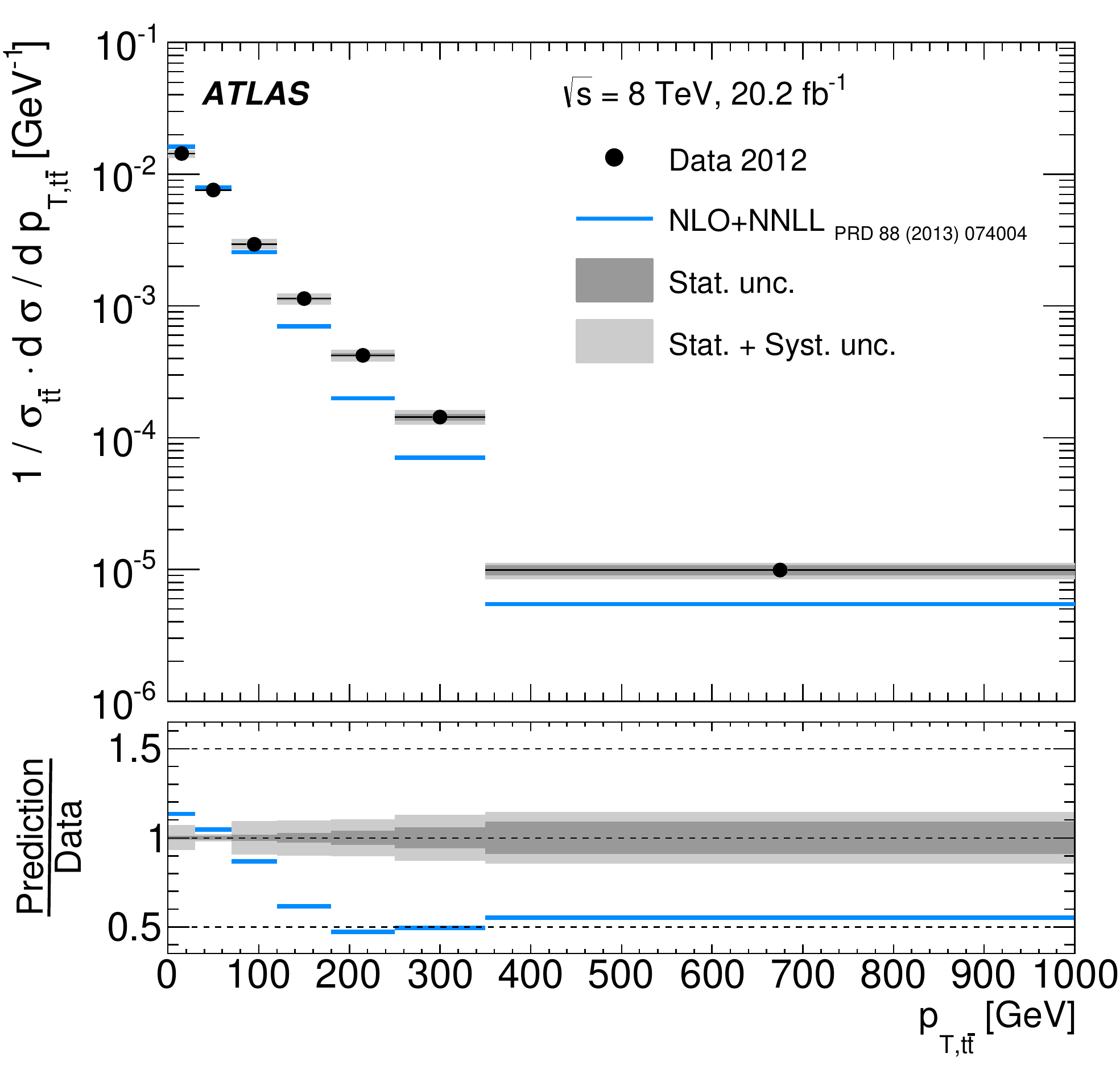} 
\label{fig:Normalized_NLO_pTtt} 
}
\caption{
  Normalized $\ttbar$ differential cross-sections as a function of the 
  \protect\subref{fig:Normalized_NLO_Mtt} invariant mass ($\mttbar$)
  and \protect\subref{fig:Normalized_NLO_pTtt} transverse momentum ($\ptttbar$)
  of the $\ttbar$ system at $\sqrt{s}=8$ TeV 
  measured in the dilepton $\emu$ channel 
  compared with theoretical QCD calculations at NLO+NNLL level.
  The predictions are calculated using the MSTW2008nnlo PDF.
  The bottom panel shows the ratio of prediction to data. 
  The light (dark) gray band includes the total (data statistical) uncertainty in the data in each bin.
}
\label{fig:Normalized_NLO} 
\end{figure}

\begin{figure}[htbp]
\centering
\subfloat[$\mttbar$]{
\includegraphics[width=0.45\textwidth]{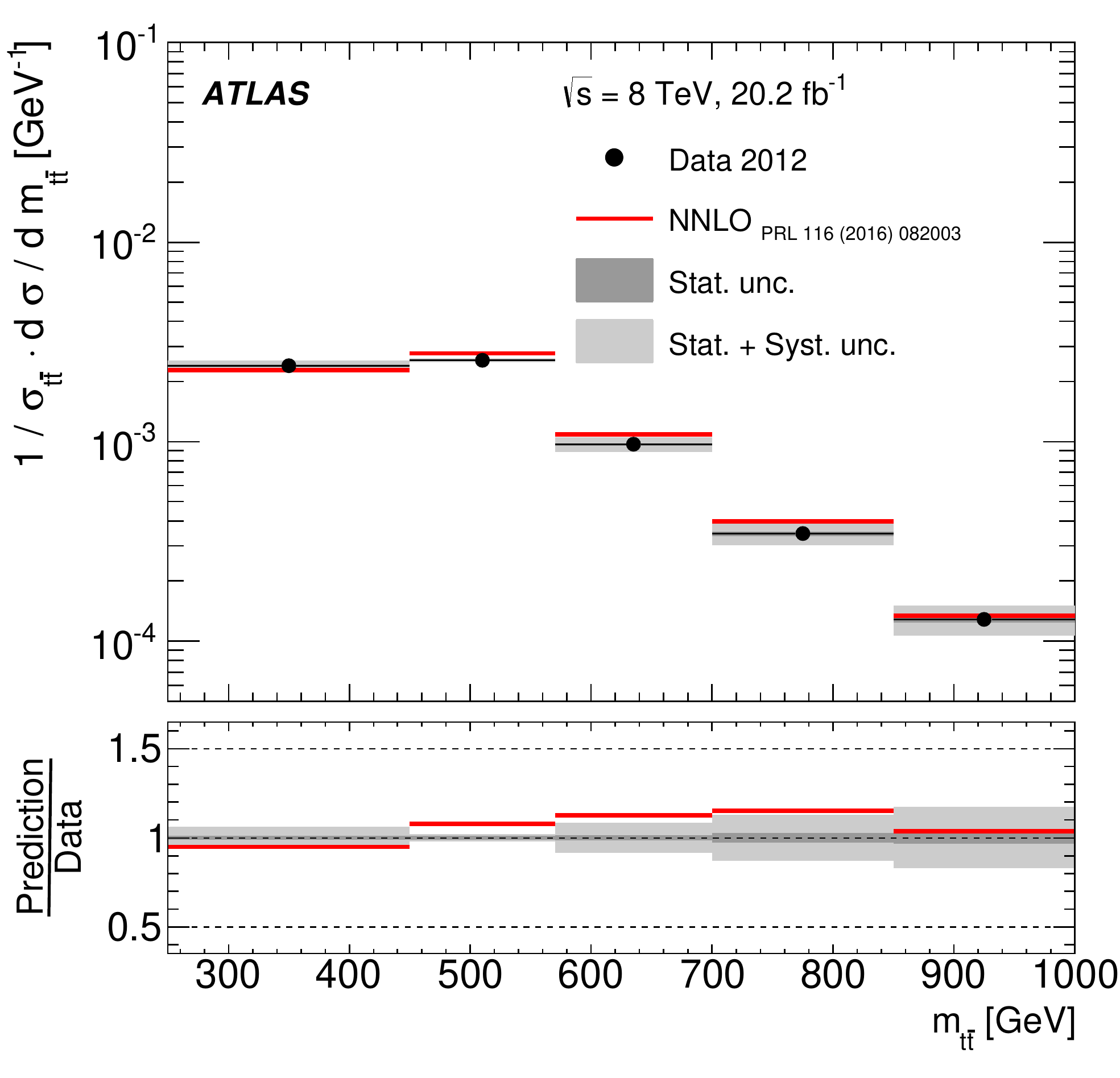} 
\label{fig:Normalized_NNLO_Mtt}
}
\hspace{8pt}
\subfloat[$\absyttbar$]{
\includegraphics[width=0.45\textwidth]{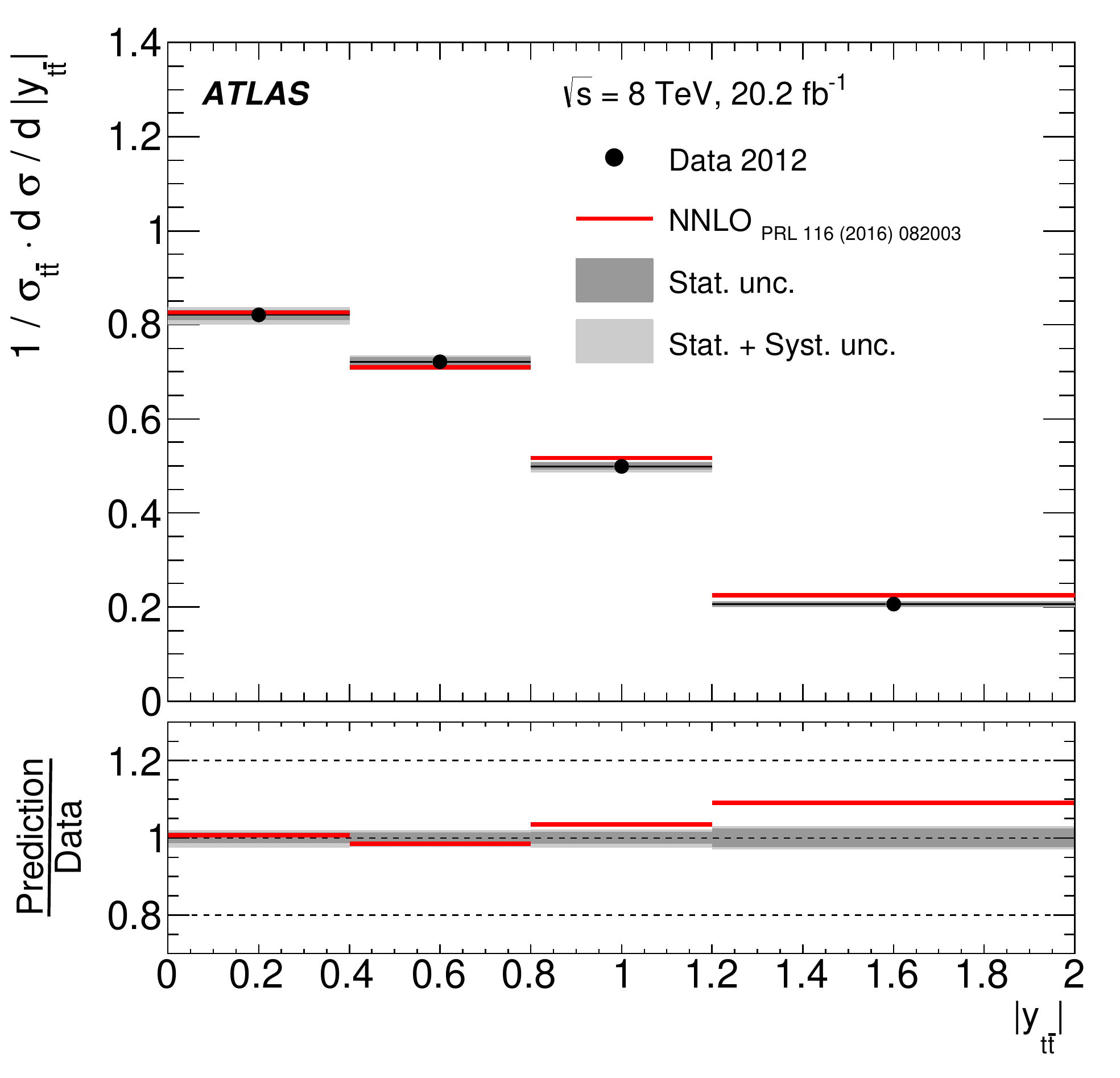} 
\label{fig:Normalized_NNLO_ytt} 
}
\caption{
  Normalized $\ttbar$ differential cross-sections as a function of the 
  \protect\subref{fig:Normalized_NNLO_Mtt} invariant mass ($\mttbar$)
  and \protect\subref{fig:Normalized_NNLO_ytt} absolute value of the rapidity ($\absyttbar$)
  of the $\ttbar$ system at $\sqrt{s}=8$ TeV 
  measured in the dilepton $\emu$ channel 
  compared with theoretical QCD calculations at full NNLO accuracy.
  The predictions are calculated using the MSTW2008nnlo PDF.
  The bottom panel shows the ratio of prediction to data. 
  The light (dark) gray band includes the total (data statistical) uncertainty in the data in each bin.
  The NNLO prediction does not cover the highest bins in $\mttbar$ and $|\yttbar|$.
}
\label{fig:Normalized_NNLO} 
\end{figure}

\clearpage

\begin{table}[htbp]
\centering
\sisetup{round-mode=places, round-precision=2, retain-explicit-plus=true, group-integer-digits=true, group-decimal-digits=false, group-four-digits=true}
\scalebox{0.9}
{
\begin{tabular}{c|cc|cc|cc}
 \toprule
 & \multicolumn{2}{c|}{$\mttbar$} & \multicolumn{2}{c|}{$\ptttbar$} & \multicolumn{2}{c}{$\absyttbar$} \\
 \midrule
 MC generator & {$\chi^2$/NDF} & {$p$-value} & {$\chi^2$/NDF }& {$p$-value} & {$\chi^2$/NDF} & {$p$-value} \\
 \midrule
PWG+PY6 CT10 $h_{\rm damp}=\mt$ & 
\numRP{4.69644/4}{1} & \num{0.319885} & 
\numRP{2.23623/3}{1} & \num{0.524847} & 
\numRP{1.31152/2}{1} & \num{0.519047} \\
PWG+PY6 CT10 $h_{\rm damp}=\infty$ & 
\numRP{4.39173/4}{1} & \num{0.355579} & 
\numRP{6.37951/3}{1} & \num{0.0945374} & 
\numRP{1.28319/2}{1} & \num{0.526452} \\
MC@NLO+HW CT10 AUET2 & 
\numRP{3.85227/4}{1} & \num{0.426368} & 
\numRP{0.775003/3}{1} & \num{0.855436} & 
\numRP{0.665769/2}{1} & \num{0.716853} \\
PWG+HW CT10 AUET2 & 
\numRP{9.07335/4}{1} & \num{0.0592921} & 
\numRP{1.85955/3}{1} & \num{0.602063} & 
\numRP{1.17449/2}{1} & \num{0.555856} \\
ALPGEN+HW CTEQ6L1 AUET2 & 
\numRP{4.27043/4}{1} & \num{0.370642} & 
\numRP{3.30766/3}{1} & \num{0.346578}    & 
\numRP{0.453504/2}{1} & \num{0.797118} \\
 \bottomrule
\end{tabular}
}
\caption{
Comparisons between the measured normalized cross-sections and the MC predictions at $\sqrt{s}$ = 7 TeV.
For each variable and prediction a $\chi^2$ and a $p$-value are calculated using the covariance matrix of each measured spectrum.
    The number of degrees of freedom is equal to one less than the number of bins $(N_{\rm b}-1)$. 
The abbreviations PWG, PY and HW correspond to {\sc Powheg}, {\sc Pythia} and {\sc Herwig} respectively. 
}
\label{tab:chi2_dxs_norm_ll}
\end{table}

\begin{table}[htbp]
 \begin{center}
  \sisetup{round-mode=places, round-precision=2, retain-explicit-plus=true, group-integer-digits=true, group-decimal-digits=false, group-four-digits=true}
  \scalebox{0.9}{
 \begin{tabular}{c|cc|cc|cc}
    \toprule
    & \multicolumn{2}{c|}{$\mttbar$} & \multicolumn{2}{c|}{$\ptttbar$} & \multicolumn{2}{c}{$\absyttbar$} \\
    \midrule
    MC generator & {$\chi^2$/NDF} & {$p$-value} & {$\chi^2$/NDF }& {$p$-value} & {$\chi^2$/NDF} & {$p$-value} \\
    \midrule
        PWG+PY6 CT10 $h_{\rm damp}=\mt$ & 
        \numRP{1.29354/5}{1} & \num{0.935595} & 
        \numRP{4.07882/6}{1} & \num{0.666011} & 
        \numRP{38.2019/4}{1} & \num{< 0.01}\\
        PWG+PY6 CT10 $h_{\rm damp}=\infty$ & 
        \numRP{1.14644/5}{1} & \num{0.949912} & 
        \numRP{16.6705/6}{1} & \num{0.0105737} & 
        \numRP{39.3264/4}{1} & \num{< 0.01}\\
        MC@NLO+HW CT10 AUET2 & 
        \numRP{1.98943/5}{1} & \num{0.850606} & 
        \numRP{0.424909/6}{1} & \num{0.998636} & 
        \numRP{29.8026/4}{1} & \num{< 0.01}\\
        PWG+HW CT10 AUET2 & 
        \numRP{1.19053/5}{1} & \num{0.945782} & 
        \numRP{3.32858/6}{1} & \num{0.766619} & 
        \numRP{36.9976/4}{1} & \num{< 0.01}\\
        \bottomrule
  \end{tabular}
 }
  \caption{
Comparisons between the measured normalized cross-sections and the MC predictions at $\sqrt{s}$ = 8 TeV.
For each variable and prediction a $\chi^2$ and a $p$-value are calculated using the covariance matrix of each measured spectrum.
    The number of degrees of freedom is equal to one less than the number of bins $(N_{\rm b}-1)$. 
The abbreviations PWG, PY and HW correspond to {\sc Powheg}, {\sc Pythia} and {\sc Herwig} respectively. 
    }
  \label{tab:chi2_norm}
 \end{center}
\end{table}

\begin{table}[htbp]
 \begin{center}
  \sisetup{round-mode=places, round-precision=2, retain-explicit-plus=true, group-integer-digits=true, group-decimal-digits=false, group-four-digits=true}
  \scalebox{0.9}{
 \begin{tabular}{c|cc|cc|cc}
    \toprule
    & \multicolumn{2}{c|}{$\mttbar$} & \multicolumn{2}{c|}{$\ptttbar$} & \multicolumn{2}{c}{$\absyttbar$} \\
    \midrule
    PDF & {$\chi^2$/NDF} & {$p$-value} & {$\chi^2$/NDF }& {$p$-value} & {$\chi^2$/NDF} & {$p$-value} \\
    \midrule
    CT10 NLO & 
    \numRP{1.98943/5}{1} & \num{0.850606} & 
    \numRP{0.424909/6}{1} & \num{0.998636} & 
    \numRP{29.8026/4}{1} & \num{< 0.01}\\
    MSTW2008nlo & 
    \numRP{2.13175/5}{1} & \num{0.830632} & 
    \numRP{0.583443/6}{1} & \num{0.99667} & 
    \numRP{11.5989/4}{1} & \num{0.0205972}\\
    NNPDF23nlo & 
    \numRP{2.2889/5}{1} & \num{0.807896} & 
    \numRP{0.421544/6}{1} & \num{0.998666} & 
    \numRP{3.1781/4}{1} & \num{0.528475}\\
    HERAPDF15NLO & 
    \numRP{2.37959/5}{1} & \num{0.794509} & 
    \numRP{2.27062/6}{1} & \num{0.893204} & 
    \numRP{5.63515/4}{1} & \num{0.228103}\\
    \bottomrule
  \end{tabular}
  }
  \caption{
Comparisons between the measured normalized cross-sections and the {\sc MC@NLO+Herwig} predictions with varied PDF sets at $\sqrt{s}$ = 8 TeV.
For each variable and prediction a $\chi^2$ and a $p$-value are calculated using the covariance matrix of each measured spectrum.
    The number of degrees of freedom is equal to one less than the number of bins $(N_{\rm b}-1)$. }
  \label{tab:chi2_norm_pdf}
 \end{center}
\end{table}

\begin{table}[htbp]
 \begin{center}
  \sisetup{round-mode=places, round-precision=2, retain-explicit-plus=true, group-integer-digits=true, group-decimal-digits=false, group-four-digits=true}
  \scalebox{0.9}{
 \begin{tabular}{c|cc|cc|cc}
    \toprule
    & \multicolumn{2}{c|}{$\mttbar$} & \multicolumn{2}{c|}{$\ptttbar$} & \multicolumn{2}{c}{$\absyttbar$} \\
    \midrule
    MC generator & {$\chi^2$/NDF} & {$p$-value} & {$\chi^2$/NDF }& {$p$-value} & {$\chi^2$/NDF} & {$p$-value} \\
    \midrule
    PWG+PY6 CT10 $h_{\rm damp}=\mt$ & 
        \numRP{1.29354/5}{1} & \num{0.935595} & 
        \numRP{4.07882/6}{1} & \num{0.666011} & 
        \numRP{38.2019/4}{1} & \num{< 0.01}\\
    PWG+PY6 CT10 $h_{\rm damp}=\mt$, $\mu=2\mt$ & 
    \numRP{0.933353/5}{1} & \num{0.96776} & 
    \numRP{14.4779/6}{1} & \num{0.0247303} & 
    \numRP{39.8503/4}{1} & \num{< 0.01}\\
    PWG+PY6 CT10 $h_{\rm damp}=2.0\mt$, $\mu=0.5\mt$ & 
    \numRP{1.64397/5}{1} & \num{0.895881} & 
    \numRP{9.69229/6}{1} & \num{0.138223} & 
    \numRP{33.8276/4}{1} & \num{< 0.01}\\
    \bottomrule
  \end{tabular}
  }
  \caption{
Comparisons between the measured normalized cross-sections and the {\sc Powheg+Pythia} predictions with different levels of radiation at $\sqrt{s}$ = 8 TeV.
For each variable and prediction a $\chi^2$ and a $p$-value are calculated using the covariance matrix of each measured spectrum.
    The number of degrees of freedom is equal to one less than the number of bins $(N_{\rm b}-1)$. 
   The abbreviations PWG and PY correspond to {\sc Powheg} and {\sc Pythia} respectively. 
    }
  \label{tab:chi2_norm_ifsr}
 \end{center}
\end{table}

\begin{table}[htbp]
\centering
\sisetup{round-mode=places, round-precision=2, retain-explicit-plus=true, group-integer-digits=true, group-decimal-digits=false, group-four-digits=true}
\scalebox{0.9}
{
\begin{tabular}{c|cc|cc}
 \toprule
 & \multicolumn{2}{c|}{$\mttbar$} & \multicolumn{2}{c}{$\ptttbar$} \\
 \midrule
 QCD calculation & {$\chi^2$/NDF} & {$p$-value} & {$\chi^2$/NDF }& {$p$-value} \\
 \midrule
NLO+NNLL ($\sqrt{s}$ = 7 TeV) & 
\numRP{4.99566/4}{1} & \num{0.287743} & 
\numRP{14.3025/3}{1} & \num{< 0.01} \\
NLO+NNLL ($\sqrt{s}$ = 8 TeV) & 
\numRP{5.86843/5}{1} & \num{0.319232} & 
\numRP{121.534/6}{1} & \num{< 0.01} \\
 \bottomrule
\end{tabular}
}
\caption{
Comparisons between the measured normalized cross-sections and the QCD NLO+NNLL calculations at $\sqrt{s}$ = 7 TeV and $\sqrt{s}$ = 8 TeV.
The NLO+NNLL predictions are calculated using the MSTW2008nnlo PDF.
For each variable and prediction a $\chi^2$ and a $p$-value are calculated using the covariance matrix of each measured spectrum.
    The number of degrees of freedom is equal to one less than the number of bins $(N_{\rm b}-1)$. 
}
\label{tab:chi2_dxs_norm_ll_th}
\end{table}

\clearpage

\begin{figure}[htbp]
\centering
\subfloat[$\mttbar$]{
\includegraphics[width=0.42\textwidth]{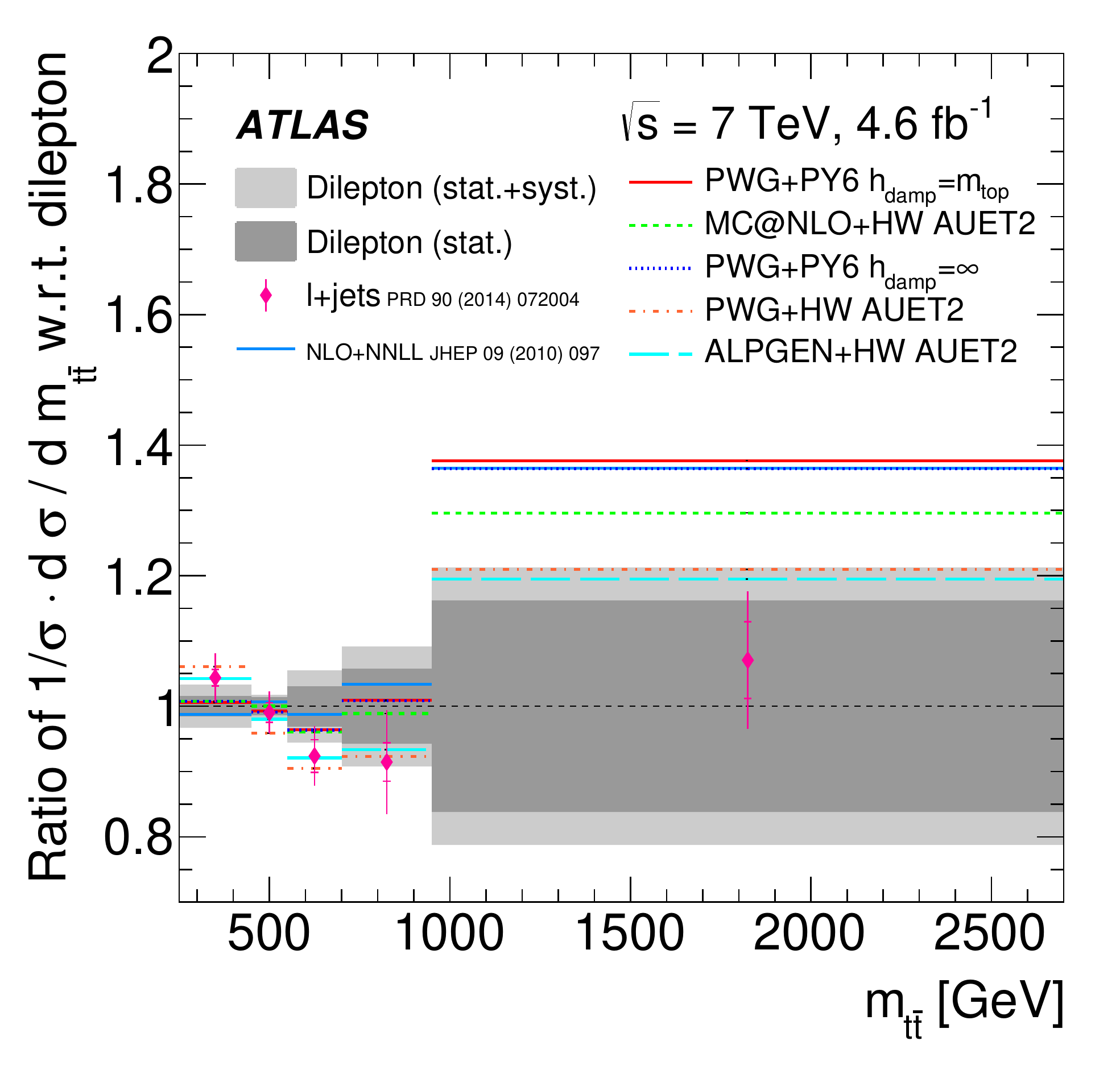} 
\label{fig:summary_ljll_dxs_norm_mtt}
}
\hspace{8pt}
\subfloat[$\ptttbar$]{
\includegraphics[width=0.42\textwidth]{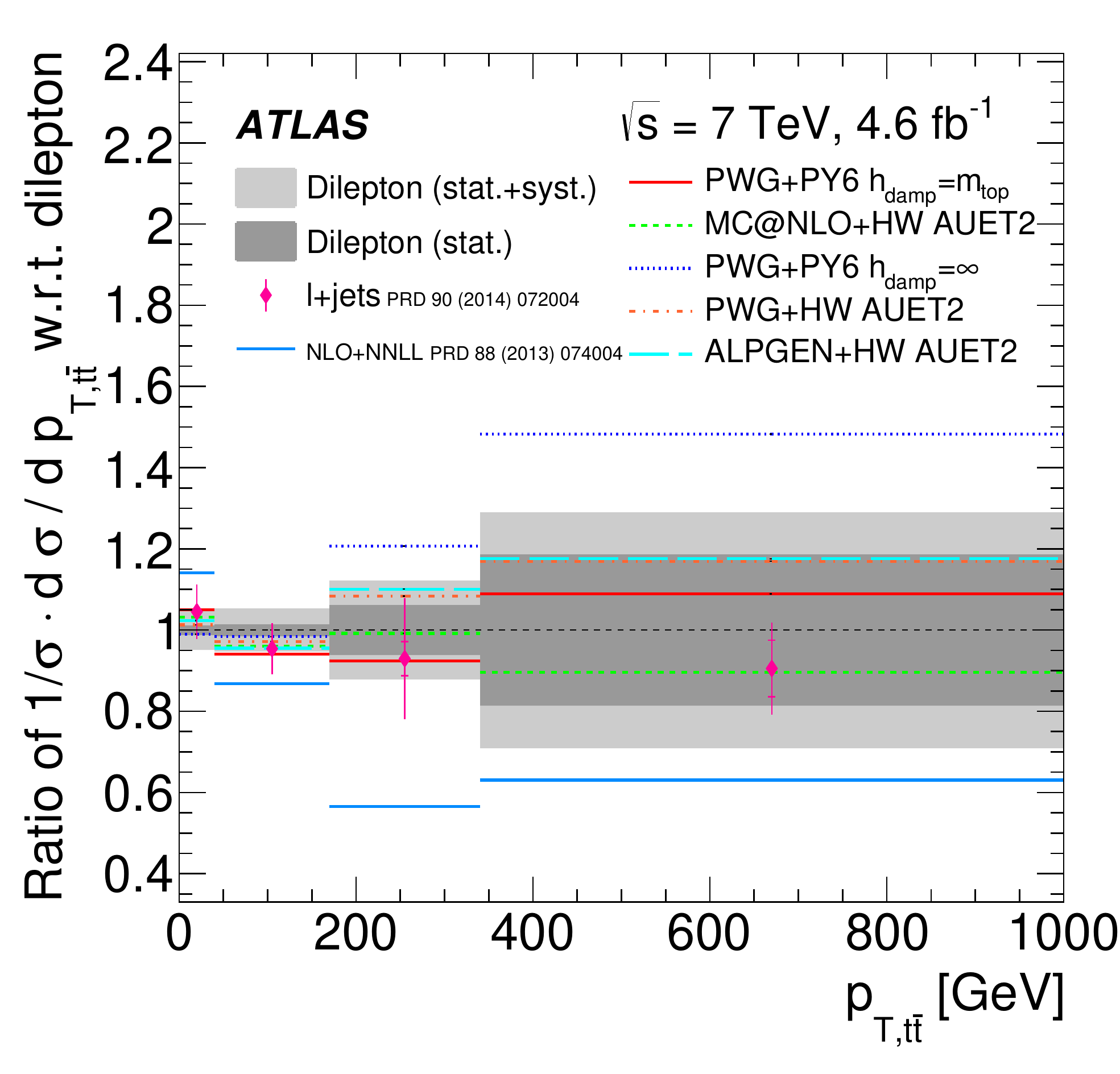} 
\label{fig:summary_ljll_dxs_norm_pttt}
}

\hspace{8pt}
\subfloat[$\absyttbar$]{
\includegraphics[width=0.42\textwidth]{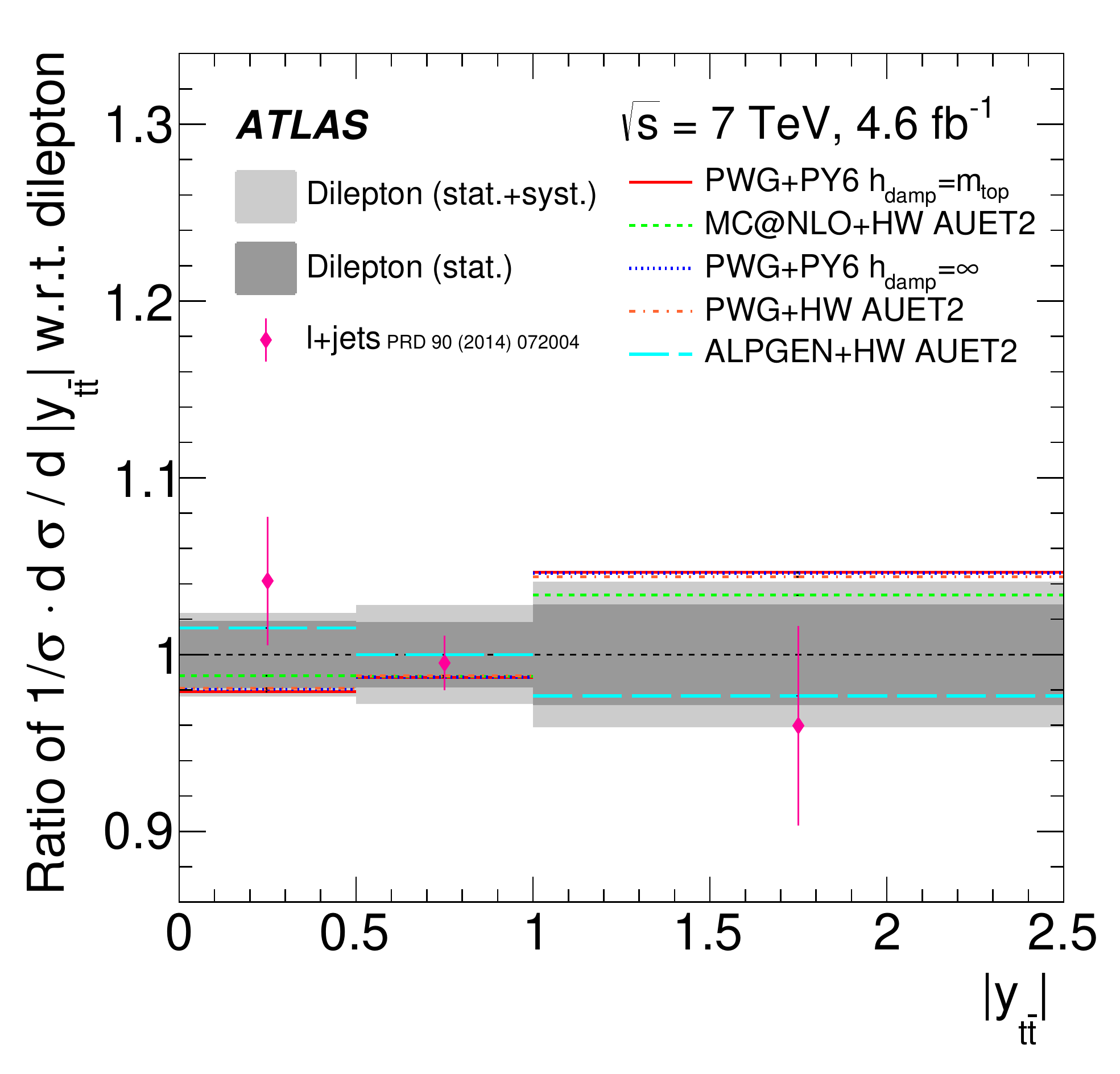} 
\label{fig:summary_ljll_dxs_norm_absytt}
}
\caption{
Ratio of different theoretical predictions and the lepton+jets measurement~\cite{TOPQ-2012-08} to the measurement of the normalized $\ttbar$ differential cross-sections in the dilepton channel for 
  \protect\subref{fig:summary_ljll_dxs_norm_mtt} invariant mass ($\mttbar$)
  \protect\subref{fig:summary_ljll_dxs_norm_pttt} transverse momentum ($\ptttbar$)
  and \protect\subref{fig:summary_ljll_dxs_norm_absytt} absolute value of the rapidity ($\absyttbar$)
  of the $\ttbar$ system at $\sqrt{s}=7$ TeV.
Theoretical QCD calculations at NLO+NNLL level 
are also included
in $\mttbar$
and $\ptttbar$.
All generators use the NLO CT10~\cite{CT10} PDF, except for {\sc Alpgen+Herwig} using the LO CTEQ6L1 PDF. 
The NLO+NNLL calculations use the MSTW2008nnlo PDF.
  The light (dark) gray band includes the total (data statistical) uncertainty in the data in each bin.
The uncertainties on the two data measurements do not account for the correlations of the systematic uncertainties between the two channels.
}
\label{fig:summary_all_dxs_norm}
\end{figure}

\FloatBarrier

\section{Conclusions}
\label{sec:conclusion}

Normalized differential $\ttbar$ production cross-sections
have been measured
as a function of the invariant mass, the transverse momentum, and the rapidity of the $\ttbar$ system
in \rts{} = 7 TeV and 8 TeV proton-proton collisions
using the dilepton channel.
The data correspond to an integrated luminosity of 4.6\,\ifb{} and 20.2\,\ifb{} 
for \rts{} = 7 TeV and 8 TeV, respectively, 
collected by the ATLAS detector at the CERN LHC. 
The results complement the other ATLAS measurements in the lepton+jets channel 
using the 7 TeV
and 8 TeV
datasets. 

The predictions from Monte Carlo and QCD calculations generally agree with data in a wide range of the kinematic distributions.
Most of the generators describe the $\mttbar$ spectrum fairly well in 7 TeV and 8 TeV data. 
The $\ptttbar$ spectrum  in both 7 TeV and 8 TeV data is well described by 
{\sc Powheg+Pythia} with $h_{\rm damp}=\mt$ and {\sc MC@NLO+Herwig}, 
but is particularly poorly described by {\sc Powheg+Pythia} with $h_{\rm damp}=\infty$. 
For $\absyttbar$, all of the generators predict higher cross-sections at large $\absyttbar$ 
than observed in data, 
and the level of agreement is improved when using 
NNPDF2.3 and HERAPDF1.5 PDF sets
instead of CT10.
The QCD calculation agrees well with data in the $\mttbar$ spectrum
at both NLO+NNLL and NNLO accuracy, 
while a large discrepancy for $\ptttbar$ is seen at NLO+NNLL accuracy for both $\sqrt{s}=7\TeV$ and $\sqrt{s}=8\TeV$. 
The results at both 7 TeV and 8 TeV are consistent with the other ATLAS measurements in the lepton+jets channel. 

\section*{Acknowledgements}

We honor the memory of our colleague Irene Vichou, who made a large contribution to this work, but died shortly before its completion.

We thank CERN for the very successful operation of the LHC, as well as the
support staff from our institutions without whom ATLAS could not be
operated efficiently.

We acknowledge the support of ANPCyT, Argentina; YerPhI, Armenia; ARC, Australia; BMWFW and FWF, Austria; ANAS, Azerbaijan; SSTC, Belarus; CNPq and FAPESP, Brazil; NSERC, NRC and CFI, Canada; CERN; CONICYT, Chile; CAS, MOST and NSFC, China; COLCIENCIAS, Colombia; MSMT CR, MPO CR and VSC CR, Czech Republic; DNRF and DNSRC, Denmark; IN2P3-CNRS, CEA-DSM/IRFU, France; GNSF, Georgia; BMBF, HGF, and MPG, Germany; GSRT, Greece; RGC, Hong Kong SAR, China; ISF, I-CORE and Benoziyo Center, Israel; INFN, Italy; MEXT and JSPS, Japan; CNRST, Morocco; FOM and NWO, Netherlands; RCN, Norway; MNiSW and NCN, Poland; FCT, Portugal; MNE/IFA, Romania; MES of Russia and NRC KI, Russian Federation; JINR; MESTD, Serbia; MSSR, Slovakia; ARRS and MIZ\v{S}, Slovenia; DST/NRF, South Africa; MINECO, Spain; SRC and Wallenberg Foundation, Sweden; SERI, SNSF and Cantons of Bern and Geneva, Switzerland; MOST, Taiwan; TAEK, Turkey; STFC, United Kingdom; DOE and NSF, United States of America. In addition, individual groups and members have received support from BCKDF, the Canada Council, CANARIE, CRC, Compute Canada, FQRNT, and the Ontario Innovation Trust, Canada; EPLANET, ERC, FP7, Horizon 2020 and Marie Sk{\l}odowska-Curie Actions, European Union; Investissements d'Avenir Labex and Idex, ANR, R{\'e}gion Auvergne and Fondation Partager le Savoir, France; DFG and AvH Foundation, Germany; Herakleitos, Thales and Aristeia programmes co-financed by EU-ESF and the Greek NSRF; BSF, GIF and Minerva, Israel; BRF, Norway; Generalitat de Catalunya, Generalitat Valenciana, Spain; the Royal Society and Leverhulme Trust, United Kingdom.

The crucial computing support from all WLCG partners is acknowledged gratefully, in particular from CERN, the ATLAS Tier-1 facilities at TRIUMF (Canada), NDGF (Denmark, Norway, Sweden), CC-IN2P3 (France), KIT/GridKA (Germany), INFN-CNAF (Italy), NL-T1 (Netherlands), PIC (Spain), ASGC (Taiwan), RAL (UK) and BNL (USA), the Tier-2 facilities worldwide and large non-WLCG resource providers. Major contributors of computing resources are listed in Ref.~\cite{ATL-GEN-PUB-2016-002}.

\printbibliography
\newpage \input{atlas_authlist} 

\end{document}

%% file: defs.tex
\newcommand{\yttbar}{\ensuremath{y_{\ttbar}}}

\newcommand{\mttbar}{\ensuremath{m_{\ttbar}}}
\newcommand{\mt}{\ensuremath{m_{t}}}

\newcommand{\ptttbar}{\ensuremath{p_{{\rm T},\ttbar}}} 
\newcommand{\Etmiss}{\ensuremath{\ET^{\rm miss}}}

\newcommand{\Et}{\ensuremath{E_{\rm T}}}

\newcommand{\absyttbar}{\ensuremath{|y_{\ttbar}|}}

\newcommand{\mll}{\mbox{$m_{\ell\ell}$}}

\newcommand{\AKT}{anti-$k_{t}$}

\newcommand{\sigmattbar}{\ensuremath{\sigma_{\ttbar}}}

\newcommand{\diel}{\ensuremath{ee}\xspace}  
\newcommand{\dimu}{\ensuremath{\mu\mu}\xspace}   
\newcommand{\emu}{\ensuremath{e\mu}\xspace}   

\def\pTZ{$p_{\mathrm{T}}^Z$}

\def\Put(#1,#2)#3{\leavevmode\makebox(0,0){\put(#1,#2){#3}}}

%% file: atlas_authlist.tex
\begin{flushleft}
{\Large The ATLAS Collaboration}

\bigskip

M.~Aaboud$^\textrm{\scriptsize 136d}$,
G.~Aad$^\textrm{\scriptsize 87}$,
B.~Abbott$^\textrm{\scriptsize 114}$,
J.~Abdallah$^\textrm{\scriptsize 65}$,
O.~Abdinov$^\textrm{\scriptsize 12}$,
B.~Abeloos$^\textrm{\scriptsize 118}$,
R.~Aben$^\textrm{\scriptsize 108}$,
O.S.~AbouZeid$^\textrm{\scriptsize 138}$,
N.L.~Abraham$^\textrm{\scriptsize 152}$,
H.~Abramowicz$^\textrm{\scriptsize 156}$
% ...
\bigskip
\\
% ...
$^{12}$ Institute of Physics, Azerbaijan Academy of Sciences, Baku, Azerbaijan\\
% ...
$^{65}$ University of Iowa, Iowa City IA, United States of America\\
% ...
$^{87}$ CPPM, Aix-Marseille Universit{\'e} and CNRS/IN2P3, Marseille, France\\
% ...
$^{108}$ Nikhef National Institute for Subatomic Physics and University of Amsterdam, Amsterdam, Netherlands\\
% ...
$^{114}$ Homer L. Dodge Department of Physics and Astronomy, University of Oklahoma, Norman OK, United States of America\\
% ...
$^{118}$ LAL, Univ. Paris-Sud, CNRS/IN2P3, Universit{\'e} Paris-Saclay, Orsay, France\\
% ...
$^{136}$ $^{(a)}$ Facult{\'e} des Sciences Ain Chock, R{\'e}seau Universitaire de Physique des Hautes Energies - Universit{\'e} Hassan II, Casablanca; $^{(b)}$ Centre National de l'Energie des Sciences Techniques Nucleaires, Rabat; $^{(c)}$ Facult{\'e} des Sciences Semlalia, Universit{\'e} Cadi Ayyad, LPHEA-Marrakech; $^{(d)}$ Facult{\'e} des Sciences, Universit{\'e} Mohamed Premier and LPTPM, Oujda; $^{(e)}$ Facult{\'e} des sciences, Universit{\'e} Mohammed V, Rabat, Morocco\\
% ...
$^{138}$ Santa Cruz Institute for Particle Physics, University of California Santa Cruz, Santa Cruz CA, United States of America\\
% ...
$^{152}$ Department of Physics and Astronomy, University of Sussex, Brighton, United Kingdom\\
% ...
$^{156}$ Raymond and Beverly Sackler School of Physics and Astronomy, Tel Aviv University, Tel Aviv, Israel\\
% ...
$^{a}$ Also at Department of Physics, King's College London, London, United Kingdom\\
$^{b}$ Also at Institute of Physics, Azerbaijan Academy of Sciences, Baku, Azerbaijan\\
$^{c}$ Also at Novosibirsk State University, Novosibirsk, Russia\\
$^{d}$ Also at TRIUMF, Vancouver BC, Canada\\
$^{e}$ Also at Department of Physics {\&} Astronomy, University of Louisville, Louisville, KY, United States of America\\
% ...
\end{flushleft}
